%% file: main.tex
\documentclass[preprint]{aastex631}

\newcommand{\revision}[1]{#1}
\newcommand{\revisionnew}[1]{#1}

\shorttitle{The EaRTH Disk Model}
\shortauthors{Grimble et al.}

\begin{document}

\title{The Empirical and Radiative Transfer Hybrid (EaRTH) Disk Model: Merging Analyses of Protoplanetary Dust Disk Mineralogy and Structure}

\author[0000-0002-8899-8914]{William Grimble}
\affiliation{Department of Physics and Astronomy, Vanderbilt University, Nashville, TN 37235, USA}
\affiliation{Frist Center for Autism and Innovation, Vanderbilt University, 2414 Highland Avenue, Suite 115, Nashville, TN 37212, USA}
\affiliation{Chester F. Carlson Center for Imaging Science, Rochester Institute of Technology, 54 Lomb Memorial Drive, Rochester, NY 14623, USA}

\author[0000-0002-3138-8250]{Joel Kastner}
\affiliation{Chester F. Carlson Center for Imaging Science, Rochester Institute of Technology, 54 Lomb Memorial Drive, Rochester, NY 14623, USA}

\author[0000-0001-5907-5179]{Christophe Pinte}
\affiliation{Monash Centre for Astrophysics (MoCA) and School of Physics and Astronomy, Monash University, Clayton, Vic 3800, Australia}

\author[0000-0001-9855-8261]{Beth Sargent}
\affiliation{Space Telescope and Science Institute (STScI), 3700 San Martin Drive, Baltimore, MD 21218, USA}
\affiliation{Center for Astrophysical Sciences, The William H. Miller III Department of Physics and Astronomy, Johns Hopkins University, Baltimore, MD 21218, USA}

\author[0000-0002-7939-377X]{David A. Principe}
\affiliation{MIT Kavli Institute for Astrophysics and Space Research, 77 Massachusetts Avenue, Cambridge, MA 02139, USA}

\author[0000-0002-4555-5144]{Annie Dickson-Vandervelde}
\affiliation{Physics and Astronomy Department, Vassar College, 124 Raymond Avenue, Poughkeepsie, NY 12604, USA}

\author{Aurora Bel\'{e}n Aguayo}
\affiliation{Departamento de Ciencias Fisicas, Facultad de Ciencias Exactas, Universidad Andres Bello. Av. Fernandez Concha 700, Las Condes, Santiago NAN, Chile}

\author[0000-0002-6617-3823]{Claudio Caceres}
\affiliation{Instituto de Astrof\'{\i}sica, Departmento de Ciencias Fisicas, Facultad de Ciencias Exactas, Universidad Andres Bello. Av. Fernandez Concha 700, Las Condes, Santiago, Chile}

\author[0000-0003-3903-8009]{Matthias R. Schreiber}
\affiliation{Departamento de F\'{\i}sica, Universidad Técnica Federico Santa María, Avenida España 1680, Valpara\'{\i}so, Chile}
\affiliation{Millennium Nucleus for Planet Formation, NPF, Valpara\'{\i}so, Chile}

\author[0000-0002-3481-9052]{Keivan G.\ Stassun}
\affiliation{Department of Physics and Astronomy, Vanderbilt University, Nashville, TN 37235, USA}

\begin{abstract}
 Our understanding of how exoplanets form and evolve relies on analyses of both the mineralogy of protoplanetary disks and their detailed structures; however, these key complementary aspects of disks are usually studied separately. We present initial results from a hybrid model that combines the empirical characterization of the mineralogy of a disk, as determined from its mid-infrared spectral features, with the MCFOST radiative transfer disk model, a combination we call the EaRTH Disk Model. With the results of the mineralogy detection serving as input to the radiative transfer model, we generate mid-infrared spectral energy distributions (SEDs) that reflect both the mineralogical and structural parameters of the corresponding disk. Initial fits of the SED output by the resulting integrated model to $Spitzer~Space~Telescope$ mid-infrared (IRS) spectra of the protoplanetary disk orbiting the nearby T Tauri star MP Mus demonstrate the potential advantages of this approach by revealing details like the dominance of \revision{$\mu m$-sized olivine} and \revision{$\mu m$-sized} forsterite in this dusty disk. The simultaneous insight into disk composition and structure provided by the EaRTH Disk methodology should be directly applicable to the interpretation of mid-infrared spectra of protoplanetary disks that will be produced by the James Webb Space Telescope.
\end{abstract}

\section{Introduction}\label{intro}
\input{introduction}

\section{Methodology}\label{models}
\input{methodology}

\section{Results: Application to the Protoplanetary Disk Orbiting MP Mus}\label{res}
\input{results}

\section{Conclusions}\label{conc}
\input{conclusion}

\section*{Acknowledgements}\label{ackn}

\revisionnew{We thank the anonymous referee for their careful review and helpful comments which helped to greatly improve this article.}

\revisionnew{This research was supported by NASA Astrophysics Data Analysis Program grant 80NSSC22K0625 and NASA Exoplanets Program grant 80NSSC19K0292 to RIT.}

\revisionnew{Partial support for this work was provided by Vanderbilt University's Frist Center for Autism \& Innovation.}

This research has made use of the NASA/ IPAC Infrared Science Archive, which is operated by
the Jet Propulsion Laboratory, California Institute of Technology, under contract with the National
Aeronautics and Space Administration.

This work is based in part on observations made with the $Spitzer~Space~Telescope$, obtained
from the NASA/ IPAC Infrared Science Archive, both of which are operated by the Jet Propulsion
Laboratory, California Institute of Technology under a contract with the National Aeronautics and
Space Administration.

This research makes use of data products from the Two Micron All Sky Survey, which is a joint
project of the University of Massachusetts and the Infrared Processing and Analysis Center/California Institute of Technology, funded by the National Aeronautics and Space Administration and the National Science Foundation.

This work has made use of data from the European Space Agency (ESA) mission
{\it Gaia} (\url{https://www.cosmos.esa.int/gaia}), processed by the {\it Gaia}
Data Processing and Analysis Consortium (DPAC,
\url{https://www.cosmos.esa.int/web/gaia/dpac/consortium}). Funding for the DPAC
has been provided by national institutions, in particular the institutions
participating in the {\it Gaia} Multilateral Agreement.

This research has made use of the SIMBAD database,
operated at CDS, Strasbourg, France.

This research has made use of the VizieR catalogue access tool, CDS,
Strasbourg, France (DOI : 10.26093/cds/vizier). The original description of the VizieR service was published in 2000, A$\&$AS 143, 23.

The online documentation for SciPy~\citep{2020SciPy-NMeth} was referenced for the utility of some Python functions, as well as sample code to see how the functions work, particularly \verb'scipy.interpolate.interp1d' for data interpolation. The program made significant use of NumPy~\citep{oliphant2006guide}~\citep{2011CSE....13b..22V} and AstroPy \citep{2013A&A...558A..33A}. Graph plotting was done using Matplotlib~\citep{Hunter:2007}.

This research made use of the Heidelberg - Jena - St.Petersburg - Database of Optical Constants (HJPDOC) for the retrieval of silicate optical constants \revision{\citep{1999A&AS..136..405H,2003JQSRT..79..765J}.}

This paper makes use of the following ALMA data: \url{ADS/JAO.ALMA#2017.1.01167.S}. ALMA is a partnership of ESO (representing its member states), NSF (USA) and NINS (Japan), together with NRC (Canada), MOST and ASIAA (Taiwan), and KASI (Republic of Korea), in cooperation with the Republic of Chile. The Joint ALMA Observatory is operated by ESO, AUI/NRAO and NAOJ.

The National Radio Astronomy Observatory is a facility of the National Science Foundation operated under cooperative agreement by Associated Universities, Inc.

Part of the data are retrieved from the JVO portal (\url{http://jvo.nao.ac.jp/portal}) operated by the NAOJ.

\revision{Images of MCFOST \citep{2022ascl.soft07023P} data were rendered by making use of pymcfost \citep{2022ascl.soft07024P}.}

C. Caceres acknowledges support by ANID BASAL project FB210003.

MRS acknowledges support by ANID, – Millennium Science Initiative Program – NCN19\_171, and FONDECYT (grant 1221059).

\appendix{\input{opacs}}\label{appe}

\bibliography{egbib}
\bibliographystyle{aasjournal}

\end{document}

%% file: introduction.tex
In recent years, exoplanets have been detected at a very high rate; it is widely recognized that these planets form in dusty circumstellar (aka protoplanetary) disks orbiting young stars \citep{2020ARA&A..58..483A}. Analysis of the dust in these disks is thus essential to the study of stellar and planetary evolution. We need to understand the relative amounts of constituents in protoplanetary disks and their temperatures in order to establish the disk compositions and physical structures that represent the initial conditions for planet formation. The same methods can be applied to study the composition of dust in a wide variety of circumstellar environments, including outflows from evolved stars and even disks orbiting massive stars \citep{2006ApJ...638L..29K, 2010AJ....139.1993K}.  

Protoplanetary disks are comprised mainly of gas and contain a small amount of dust ($\sim 1\%$ of the disk’s mass), but that dust is the main source of opacity in the disk, and this opacity manifests in distinguishable features in the infrared regime of the stellar spectra, such as a prominent peak at 10 $\micron$ and a lesser but still prominent peak at $\sim$20 $\micron$ \citep{2011ppcd.book...14C}. $Spitzer$ IRS spectroscopy of hundreds of classical T Tauri stars has established that properties of dust within the planet-forming ($\sim$0.5-5 au) regions of their disks span a wide range, in terms of degree of silicate crystallinity and grain size distributions. This mineralogical diversity among disks is poorly understood; disk lifetimes, levels of disk irradiation, internal shock waves, and grain transport within disks have all been considered as contributing factors \citep[e.g.][and references therein]{2019ApJ...882...33G,2016ApJS..226....8K,2009ApJ...690.1193S}. However, features of silicates and other minerals are easily and reliably distinguishable and have shown to sensibly model the shape of a disk's near- to mid-infrared spectrum \citep{2011ppcd.book..114H}. 

One useful empirical model in analyzing signatures of dust emission in IR spectra is the two-temperature empirical model, established by \citet{2009ApJ...690.1193S,2009ApJS..182..477S}. This approach has been used by many authors \citep[e.g.][]{2016ApJ...830...71F,2015ApJ...810...62R,2022ApJ...933...54L}. The successful use of crystalline materials in the model exposes the lack of clearly known mechanisms underlying their formation in the disk, though \citet{2009ApJ...690.1193S} postulate rapid grain transport through the disk. However, the simplifying assumption of two isothermal zones that underlies this modeling approach does not yield realistic descriptions of the disk structures where the various dust components reside. \revision{Alternatively, another empirical model, the two-layer temperature distribution, assumes that sections of the disk can be modeled based on the assumption that each section has a unique temperature distribution \citep{2009ApJ...695.1024J,2010ApJ...721..431J}. This model has likewise been used to model SEDs successfully \citep[e.g.][]{2023Natur.620..516P}, but struggles to fit spectra over a full mid-IR wavelength range primarily due to the presumption of only one population of dust in the optically thin atmosphere.}

In contrast, radiative transfer algorithms are designed to simulate a protoplanetary disk using input parameters corresponding to a set of (potentially overlapping) disk zones, each characterized by parameters such as scale height at some reference radius, flaring exponent, inner radius, viscosity \citep[which controls disk dust settling using treatments such as that by][]{1995Icar..114..237D}, and dust mass \citep{2006A&A...459..797P,2009A&A...498..967P}. However, the ``off the shelf" versions of such radiative transfer simulations of disk structure generally only incorporate a few simple mineralogies, and radiative transfer algorithms that have included additional mineralogical parameters to more accurately reproduce mid- to far-infrared spectra have generally incorporated limited sets of mineralogies that are fine-tuned to specific systems \citep[e.g.,][]{2012ApJ...759L..10M,2019A&A...623A.106L,2020A&A...642A.171R,2022A&A...666A.139A,2023A&A...672A..30K}.

This article demonstrates a prototype methodology that integrates the empirical mineralogical studies in these disks into the radiative transfer based structural analysis of the disks to create a single Empirical and Radiative Transfer Hybrid Disk Model (hereafter the EaRTH Disk Model). This hybrid modeling approach, like previous self-consistent (mineralogical+structural) disk models \citep[e.g.,][]{ 2012ApJ...759L..10M}, has the potential to generate highly accurate spectral energy distribution (SED) model outputs in the infrared regime, which we further improve by tuning the parameters of the disk structure, thus providing new insight into the compositions and structures of individual disks. We develop and test this model by fitting high-quality infrared spectral and radio interferometric data available for the nearby T Tauri star/disk system MP Muscae (MP Mus), to demonstrate the model’s potential.

%% file: methodology.tex
Given the impact of disk mineralogy on the infrared spectrum of the disk, it is intuitive that mineralogical model fits to data will provide important constraints on radiative transfer disk models, which depend on parameters like mineralogy to constrain the structure of the disk. Therefore, we develop a hybrid model \revision{based on the following methodology: we first apply a newly developed “two-zone temperature distribution” (TZTD) model to empirically fit the $Spitzer$ IRS spectrum, so as to establish both dust disk mineralogy and radial temperature structure; we use the resulting TZTD-constrained mineralogy as input to MCFOST and, in parallel, determine the MCFOST disk structural parameters that best match the TZTD radial temperature structure; and finally we fine-tune the MCFOST results to fit the $Spitzer$ IRS spectrum (primarily) and ALMA continuum mapping data (secondarily).}
This methodology forms the basis of the EaRTH Disk Model.

\subsection{Empirical Dust Composition Model: \revision{Two-Zone Temperature Distribution}}\label{2tmodel}
The first stage of the EaRTH Disk modeling methodology utilizes the empirical mineralogical analysis of the disk using \revision{what we call a two-zone temperature distribution model, which is based on both a two-layer temperature distribution and a two-temperature component model. The two-layer temperature distribution (TLTD) \citep{2009ApJ...695.1024J,2010ApJ...721..431J} is a method for mid-infrared spectral analysis that models the SED assuming that the protoplanetary disk can be partitioned into three sections: a puffed-up inner rim, the disk midplane, and the optically thin atmosphere. The hot inner rim and the optically thick midplane characterize the continuum along with stellar emission, while the silicate features manifest from the small dust grains in the optically thin atmosphere. The model is applied to an observed mid-IR spectrum, usually a preprocessed flux density function recorded by the $Spitzer~Space~Telescope$ Infrared Spectrograph (IRS) over a portion of the wavelength range of 5.2-38 $\micron$ (with the upper or lower bound typically set at 17 $\micron$), using mineralogy functions based on mineral opacities and blackbody functions integrated over corresponding disk temperatures. The results returned include the maximum temperatures $T_{max}$ corresponding to the disk components, assumed to decrease radially exponentially according to exponents $q$ corresponding to each component that are also returned, i.e. 
\begin{equation}
T(R)=T_{max}\left(\frac{R}{R_{inner/rim}}\right)^q,
\end{equation}
and constants corresponding to the mass fraction of each component, including the individual fractions of each mineralogical component in the disk atmosphere, which can be very helpful in surmising an approximate mineralogy-influenced disk structure. These parameters are estimated using a genetic algorithm \citep[{\tt pikaia};][]{1995ApJS..101..309C} to minimize the reduced $\chi^2$ value between the spectrum and the model constructed from the model fluxes of individual components,

\begin{equation}
\label{chisq}
{\chi}^{2}={\sum_{k}[{\frac{F_{\nu}({\lambda}_{k})^{irs}-F_{\nu}({\lambda}_{k})^{mod}}{{\Delta}F_{\nu}({\lambda}_{k})^{irs}}}}]^{2},
\end{equation}	
where $F_{\nu}({\lambda}_{k})_{irs}$ is the flux density of the disk measured by the $Spitzer~Space~Telescope$ at a given wavelength, ${\Delta}F_{\nu}({\lambda}_{k})_{irs}$ is the uncertainty of the flux density at that wavelength, and the modeled flux density is
\begin{equation}
\label{mod_flux0}
\begin{array}{l}
F_{\nu}({\lambda}_{k})^{mod}=\frac{\pi R_*^2}{d^2}B_{\nu}({\lambda}_{k},T_*)+D1\int_{T_{rim,max}}^{T_{rim,min}}\frac{2\pi}{d^2}B_{\nu}({\lambda}_{k},T)T^{\frac{2-q_{rim}}{q_{rim}}}dT\\
\indent+D2\int_{T_{mid,max}}^{T_{mid,min}}\frac{2\pi}{d^2}B_{\nu}({\lambda}_{k},T)T^{\frac{2-q_{mid}}{q_{mid}}}dT\\
\indent+\Sigma_{i=1}^{N}\Sigma_{i=1}^{M}D_{i,j}\kappa_{i,j}({\lambda}_{k})\int_{T_{atm,max}}^{T_{atm,min}}\frac{2\pi}{d^2}B_{\nu}({\lambda}_{k},T)T^{\frac{2-q_{atm}}{q_{atm}}}dT+\Sigma_{i=1}^{NP}C_iI_i^{PAH}.
\end{array}
\end{equation}
The first term represents an approximation of stellar photospheric emission, where $R_*$ is the stellar radius, $d$ is the distance to the star-disk system, and $B_{\nu}({\lambda}_{k}, T)$ is the blackbody intensity at wavelength ${\lambda}_{k}$ for effective stellar temperature $T_*$; the second term represents the puffed-up inner rim, where $D$ is the constant corresponding to the mass fraction of the component in question; the third term represents the disk midplane. These first three terms represent the continuum below the dust features. The fourth term represents the optically thin dusty atmosphere, where $\kappa_{i,j}({\lambda}_{k})$ is the opacity of mineral species $i$ and grain size $j$, $N$ is the total number of species and $M$ is the total number of sizes for a given species; the fifth and final term represents the contribution of polycyclic aromatic hydrocarbons (PAHs), where $I_i^{PAH}$ is the intensity function corresponding to PAH feature $i$, $C_i$ is the constant corresponding to the feature's contribution to the total flux, and $NP$ is the amount of PAH feature functions.}

\revision{The TLTD model is particularly effective at fitting observed {\it Spitzer} spectra over specific sections of the mid-infrared wavelength range, especially regions around the 10 $\mu$m and 20 $\mu$m silicate emission features. However, the model struggles to fit the full {\it Spitzer} wavelength range, primarily due to the assumption of a single atmospheric dust composition \citep{2009ApJ...695.1024J,2010ApJ...721..431J}}. 

\revision{The two-temperature (2-T) empirical model \citep{2009ApJ...690.1193S,2009ApJS..182..477S} mitigates this limitation, by approximating the mineralogy of a dusty disk under the assumption that the disk can be partitioned into two distinct temperature regimes of equilibrium dust temperature, each with its own distinct dust population.} We hereafter frequently refer to these two spatially distinct disk components as the ``cool and warm disk \revision{atmospheres}" since, as we later demonstrate, they can be modeled as spatially distinct physical structures, wherein the warm disk lies interior to the cool disk; these disks are representative estimates of the inner and outer regions of the disk \revision{atmosphere}, respectively. Within the two-temperature model, the two temperatures are estimated by setting a reasonable range given the materials whose spectral features appear in each disk component via a reduced $\chi^2$ fit between the spectrum and the model constructed from the model fluxes of individual components,

\begin{equation}
\label{mod_flux1}
\begin{array}{l}
F_{\nu}({\lambda}_{k})^{mod}=B_{\nu}({\lambda}_k,T_c)[{\Omega}_c+{\sum_i}{a_{c,i}}{\kappa}_i({\lambda}_k)]\\
\indent +B_{\nu}({\lambda}_k,T_w)[{\Omega}_w+{\sum_j}{a_{w,j}}{\kappa}_j({\lambda}_k)].
\end{array}
\end{equation}
Here the subscripts $c$ and $w$ denote the cold and warm disk flux densities, respectively, $B_{\nu}({\lambda}_{k}, T)$ is the blackbody intensity at wavelength ${\lambda}_{k}$ for temperature $T$, $\Omega$ is the blackbody solid angle, $a_{c,i}$ is the mass weight of a given disk component ($a_{c,i}$=$m_{c,i}/d^2$), and ${\kappa}_{i}({\lambda}_{k})$ is the opacity corresponding to the same component and wavelength. The $\chi^2$ fit is minimized for each temperature pair to extract the mass weight of each component from the model flux, removing most negative mass weights by setting them to zero until the model minimizing $\chi^2$ has no negative mass weights. \revision{Despite the few nonlinear variables, the spectral fits produced by the 2-T model are robust, well representing the atmosphere of a dusty disk via multiple dust populations. However, unlike the TLTD model, the 2-T model does not attempt to represent the realistic structure of a circumstellar, protoplanetary disk.}

\revision{To take advantage of the best aspects of both models, we have effectively combined them, yielding a hybrid hereafter referred to as the two-zone temperature distribution (TZTD) model. The TZTD model can be parameterized in the form
\begin{equation}
\label{mod_flux}
\begin{array}{l}
F_{\nu}({\lambda}_{k})^{mod}=\frac{\pi R_*^2}{d^2}B_{\nu}({\lambda}_{k},T_*)\\
\indent+(D1+\Sigma_{i=1}^{N}\Sigma_{j=1}^{M}Dw_{i,j}\kappa_{i,j}({\lambda}_{k}))\int_{T_{warm,min}}^{T_{warm,max}}\frac{2\pi}{d^2}B_{\nu}({\lambda}_{k},T)T^{\frac{2-q_{warm}}{q_{warm}}}dT\\
\indent+(D2+\Sigma_{i=1}^{N}\Sigma_{j=1}^{M}Dc_{i,j}\kappa_{i,j}({\lambda}_{k}))\int_{T_{cool,min}}^{T_{cool,max}}\frac{2\pi}{d^2}B_{\nu}({\lambda}_{k},T)T^{\frac{2-q_{cool}}{q_{cool}}}dT.
\end{array}
\end{equation}
where $B_\nu$ is again the Planck function, $D$ represents the constant corresponding to the mass fraction of the components, $\kappa_{i,j}({\lambda}_{k})$ is the opacity of mineral species $i$ and grain size $j$, $N$ is the total number of species, $M$ is the total number of sizes for a given species, and $q$ represents the temperature distribution exponent. In this model, the disk is split into a warm component and cool component with separate dust compositions, as in the 2-T model, but each component is assigned a distinct temperature distribution, in line with the TLTD model. In keeping with the 2-T model, we assume that each component has the same approximate minimum and maximum temperature, i.e., a range of 19 K to 1500 K, respectively, where the upper limit corresponds to a nominal dust sublimation temperature. The adoption of a temperature lower limit allows us to estimate a spectrum that incorporates the bulk of the disk surface without having to fit or estimate the radii corresponding to the outermost detectable emission from the two zones; we leave that estimation to the radiative transfer analysis (\S~\ref{mcfost}).}

\revision{In keeping with the TLTD model's use of a genetic algorithm to fit the nonlinear parameters, we use differential evolution \citep{1997JGOpt..11..341S}, a global optimization genetic algorithm of a stochastic nature, via the }\verb|scipy.optimize.differential_evolution|\revision{ function. We use this function to find the maximum temperatures as well as the radial temperature distribution exponents for both the warm and cool disk components and the dust grain size power law exponent of the whole disk; the allowed ranges for these exponents are $-$1.5 to $-$0.1 and $-$3.9 to $-$3.1, respectively. We permit the temperature search of each component, or zone, to span the full range from 19 K to 1500 K. The component with the higher maximum temperature is taken as the warm disk, and the one with the lower maximum temperature as the cool disk. }

\revision{For each set of the nonlinear parameters in Equation \ref{mod_flux}, we perform a linear fit of the component distribution constants to the stellar-subtracted (i.e. the first term of Equation \ref{mod_flux}) $Spitzer$ spectrum via the same approach used by the 2-T model.} The inherent degeneracies in dust composition as determined via this modeling methodology are discussed and explored in Appendix \ref{degeneracy}.

\subsection{Radiative Transfer Model: MCFOST}\label{mcfost}
We accomplish radiative transfer modelling in EaRTH Disk via the MCFOST code. MCFOST is a 3D continuum and line radiative transfer code that utilizes both Monte Carlo and ray-tracing methods \citep{2006A&A...459..797P,2009A&A...498..967P}. MCFOST performs these calculations based on a number of different parameters, such as mineralogy (mass fractions, mixtures, and sizes) and disk structure. Both the dust mineralogy (composition) and structural parameters --- i.e., inner and outer radii, scale height at a reference radius, and flaring exponent -- can be specified for multiple disk components. To specify dust mineralogy, input files specifying complex optical constants $n+i\kappa$ (distinct from opacity $\kappa$) of different minerals and wavelength grids such as that for $Spitzer$ can be provided.

As detailed in \citet{2006A&A...459..797P,2009A&A...498..967P}, MCFOST uses cylindrical grids, setting individual dust populations in each cell, with the resulting opacities, densities, and temperatures staying constant within the cell for the purposes of considering thermal emission. Opacities and scattering matrices are computed using \revision{either Mie theory \citep{1983uaz..rept.....B} or distribution of hollow spheres \citep[DHS;][]{2003A&A...404...35M}}. Individual ``photon packets" from photospheric and dust thermal emission are tracked as they move through each cell, with the distance of each movement calculated using optical depth. This permits both temperature throughout the disk structure and SEDs to be outputted from the program, as well as images at individual bands.

\subsection{EaRTH Disk Modeling Approach}\label{approach}
The EaRTH Disk Model utilizes a set of opacities to be used first in the empirical model equation fit, along with the corresponding parameters defining the opacity of a given mineral (optical indices and grain sizes) to be used with MCFOST. This set of opacities, discussed in Appendix \ref{opacsec}, \revision{is generated using the infrastructure of MCFOST itself for use with MCFOST after the empirical model returns the presence and amounts of these constituents; this union} is what effectively integrates the empirical and radiative transfer models into a single hybrid model, the EaRTH Disk Model. 

We first constrain the mineral \revision{compositions and mass fractions of the warm and cool disk components using our empirical model fit to the observed mid-IR spectrum. Next, we use the mass fractions of the detected minerals in the cool and warm disk atmosphere components to determine relative percentages corresponding to the dust population and total mass of each disk component.} This is a reasonable starting point for the optically thin regions of the disk, but may not well describe the optically thick regions.

We then use \revision{DHS, assuming $V_{max}=0.9$,} to compute the scattering and absorption opacity functions and matrices, based on mineralogy derived with \revision{our empirical} model fit. Simulation of \revision{DHS is implemented into MCFOST following the algorithm derived by \citet{2003A&A...404...35M}}; we use this MCFOST implementation to estimate the opacity of a mineral over a wide range of grain sizes \revision{based on the dust grain size power law. Since we have the opacity functions corresponding to individual grain sizes, we can calculate the opacity of each mineral for a given dust size power law exponent.} 

As minimum dust grain size has a negligible effect on the infrared SED of a disk \citep{2016A&A...586A.103W}, \revision{we set small grains, assumed to be sub-$\micron$ size, to range from 0.01-1 $\micron$, and we assume large grains to have a minimum size of 1 $\micron$; we set the maximum grain size for each large amorphous and crystalline mineral to 2 $\micron$ and 5 $\micron$, respectively, based on the opacities used by \citet{2010ApJ...721..431J}. The set of minerals we use are primarily based on those used with the the two-temperature model by \citet{2016ApJ...830...71F} but also has partial basis on the set used with the TLTD model by \citet{2010ApJ...721..431J}}. This opacity analysis is further detailed in Appendix \ref{opacsec}.

With the mineralogy of the disk and temperature distributions affecting the observed mid-IR spectrum constrained by the \revision{empirical} model fit, we vary \revision{seven} other parameters, outlined in Table \ref{initparams} and detailed in the following paragraphs, to \revision{find a MCFOST-generated disk temperature structure that matches the empirically fitted temperature distributions. The parameter values that minimize the normalized root mean squared error (NRMSE) --- i.e., the RMSE divided by the mean --- between the generated and observed temperature structures are taken as the initial representative parameters of the disk; we initially fit the temperature structure rather than SEDs to tie the radiative transfer model directly to the nonlinear parameter results from the empirical fit, and because this method is significantly faster, allowing many iterations to tune parameters to values that approach a strong fit between modeled and observed SEDs. We perform this fit using differential evolution, with bounds for each parameter. After establishing the set of parameters that minimizes the NRMSE between the generated temperature structures}, we fine-tune the values individually, as well as other parameters as necessary. This fine-tuning focuses on improving the fit to the mid-IR spectrum of the \revision{disk via minimization of the reduced $\chi^2$ value between the stellar-subtracted radiative transfer modeled SED and $Spitzer$ IRS spectrum; while MCFOST generates a photospheric curve, the generated curve is close to the empirically calculated photosphere within the mid-IR regime.}

The parameters we vary\revision{, as well as their allowed ranges,} are partially based on studies by \citet{2016A&A...586A.103W}, which indicate that certain parameters such as scale height and flaring exponent have a noticeable effect on the infrared SED of the disk, while other parameters like column density power law index have a negligible effect; we discuss and study this further in the model assessment in Section \ref{assess}. We set the \revision{range of the viscosity parameter $\alpha$ to 10$^{-4}$\revision{-10$^{-1}$}, which (respectively) constitute reasonable assumptions for the lower and upper limits for protoplanetary disks \citep[][and references therein]{2019MNRAS.486.4829R,2020A&A...642A.171R,2023ASPC..534..645P}}, assuming the treatment of dust settling by \citet{1995Icar..114..237D}. We alter the distribution of mass in the disk according to the proportional relation
\begin{equation}\label{dustdens}
\Sigma~\alpha~r^{p_1},
\end{equation}
where $p_1$ is the surface density exponent of the disk. 

\revision{To reduce the number of parameters fitted, we empirically constrain some parameters based on standard disk models \citep[e.g.,][]{2001ApJ...547.1077C,2001ApJ...560..957D}. \citet{2001ApJ...547.1077C} gives the disk flaring exponent as }
\begin{equation}
\label{flaring1}
\beta=\frac{3}{2}+\frac{1}{2}\frac{d\ln T}{d\ln r}.
\end{equation}
\revision{Since $q=d\ln{T}/d\ln{r}$ by definition (from our empirical fit),  we derive}
\begin{equation}
\label{flaring2}
\beta=\frac{q+3}{2}.
\end{equation}
\revision{However, since other parameters can affect the temperature distribution, we fit the flaring exponent of both disks. Nevertheless, this empirical flaring exponent should not be ignored; while the fitted parameter may not be exactly the same as the empirical parameter, we seek to ensure they are close in value.}

\revision{The scale height of the disk at the inner rim is given by}
\begin{equation}
\label{scale1}
h=\sqrt{\frac{kT_{rim}R_{rim}^3}{\mu m_p GM_*}},
\end{equation}
\revision{where $\mu$ is the mean molecular weight of the gas at the stellar surface \citep{2001ApJ...560..957D}. Following  \citet{1997ApJ...490..368C} and \citet{2001ApJ...560..957D}, we may generalize Eq.~\ref{scale1} as}
\begin{equation}
\label{scale2}
h(R)=\sqrt{\frac{kT_{max}(R/R_{in})^q R^3}{\mu m_p GM_*}}.
\end{equation}
\revision{We can use this function to derive a factor between the scale heights of the warm and cool disks,}
\begin{equation}
\label{scalefactor}
\frac{h_{warm}}{h_{cool}}(R)=\sqrt{\frac{T_{max,warm}(R/R_{in,warm})^{q_{warm}}}{T_{max,cool}(R/R_{in,cool})^{q_{cool}}}}.
\end{equation}
\revision{We fit the cool scale height at a reference radius of $R=100$ au. Using Eq.~\ref{scalefactor}, we can then empirically estimate the warm disk scale height as well as derive an approximate inner radius for the disk --- or, equivalently, the inner radius of the warm disk --- via Eq.~14 in \citet{2001ApJ...560..957D}, }
\begin{equation}
\label{inner_rad}
R_{in}=\sqrt{\frac{L_*}{4\pi T_{in}^4 \sigma}\left(1+\frac{H_{in}}{R_{in}}\right)}=R_*\left(\frac{T_*}{T_{max,warm}}\right)^2\sqrt{1+\frac{H_{in}}{R_{in}}},
\end{equation}
\revision{where $H_{in}$ is the full surface height of the disk inner rim, which is some factor $\chi$ of the warm disk scale height at the inner radius (where $\chi$ is greater than or equal to 1); we adopt the MCFOST default of $\chi=6.95$. 
Eq.~\ref{inner_rad} implicitly assumes isotropic luminosity and that the inner rim is at the derived maximum temperature of the warm disk.}

\revision{We fit the outer radius of the warm disk and the inner radius of the cool disk based on the value of $R_{in}$ determined from Eq.~\ref{inner_rad}, thus permitting the warm and cool disk radial zones to overlap or for a gap to be present between them.}

\revision{We derive and constrain some disk parameters --- namely the mass of the disk, the outer radius of the disk, and the approximate surface density exponent of the cool/outer disk --- based on available ALMA data. We further discuss these ALMA data, and the values for the parameters so constrained, in Section \ref{res}}.

\begin{deluxetable*}{lc}[htb!]
\deluxetablecaption{Initial values for the parameters used in the \revision{radiative transfer} model.\label{initparams}}
\tablewidth{0pt}
\tablehead{
\colhead{Parameter} & \colhead{Value Range}
}
\startdata
Warm Disk Flaring Exponent & \revision{0.75-1.5} \\
Warm Disk Surface Density Exponent & -1.5 - -0.5 \\
\revision{Warm Disk Outer Radius (au)} & \revision{1.1$\times$Inner Radius-11$\times$Inner Radius} \\
\revision{Cool Disk Inner Radius (au)} & \revision{1$\times$Inner Radius-11$\times$Inner Radius} \\
Cool Disk Scale Height (au) & 5-15 \\
Cool Disk Flaring Exponent & \revision{0.75-1.5} \\
Viscosity Parameter $\alpha$ & \revision{$10^{-4}-10^{-1}$} \\
\enddata
\tablecomments{The scale height in both sections of the disk is measured at a reference radius of R=100 au.}
\end{deluxetable*}

%% file: results.tex
To demonstrate the application of the EaRTH Disk Model to mid-IR spectroscopy data, we have fit the preprocessed $Spitzer$ IRS spectrum of the nearby star/disk system MP Mus \citep[][and references therein]{2010ApJ...723L.248K,2016ApJ...818L..15W}, observed by \citet{2004sptz.prop..148M}. We retrieved the spectrum (AORKEY 5198336) from CASSIS\footnote{The Combined Atlas of Sources with Spitzer IRS Spectra (CASSIS) is a product of the IRS instrument team, supported by NASA and JPL. CASSIS is supported by the "Programme National de Physique Stellaire" (PNPS) of CNRS/INSU co-funded by CEA and CNES and through the "Programme National Physique et Chimie du Milieu Interstellaire" (PCMI) of CNRS/INSU with INC/INP co-funded by CEA and CNES.} \citep{2011ApJS..196....8L} on 2020 May 21; the preprocessing of the $Spitzer$ IRS data is summarized in Appendix \ref{process}.

MP Mus is an actively accreting pre-main sequence star \citep{2007A&A...465L...5A} of spectral type K1 \citep{2006A&A...460..695T,2002AJ....124.1670M}, effective temperature 4920 K based on this spectral type \citep[see Table 6 of][]{2013ApJS..208....9P}, and stellar radius of $\revision{1.45}R_{\sun}$ based on the star's bolometric luminosity and effective temperature. The MP Mus star/disk system is of great interest for observation because it constitutes a rare example of an actively accreting T Tauri star with a molecule-rich protoplanetary disk within $\sim$100 pc of Earth \citep{2010ApJ...723L.248K,2014A&A...561A..42S}. As a member of the $\epsilon$ Cha association, MP Mus likely has an age of $\sim$5 Myr \citep{2021AJ....161...87D}. We adopt a distance of 97.89 pc, as obtained from the inverse of the $Gaia$ DR3 parallax \citep{2016A&A...595A...1G,2022arXiv220800211G,2021A&A...649A...2L} of MP Mus. 

MP Mus has been observed in the infrared and mm-range wavelengths to estimate disk parameters \revision{ \citep{2016ApJ...818L..15W}}. \citet{2005AJ....129.1049C} used the observed 1.2 mm continuum flux to estimate a total dust mass of $5\times10^{-5}M_{\sun}$. \citet{2009ApJ...697.1305C} resolved the disk using Hubble Space Telescope (HST)/NICMOS imaging and estimated an outer radius of 195 au (corrected for the $Gaia$ parallax distance) and disk inclination of $32\degr\pm5\degr$; the latter determination is consistent with that of \citet{2014AJ....148...59S}, who used HST/STIS coronagraphy to estimate a disk inclination of $27.3\degr\pm3.3\degr$. Additionally, based on a comparison between the MP Mus SED and the median Taurus SED, \citet{2009ApJ...697.1305C} detected a flux deficit in the 4-20 $\micron$ range, indicating either clearing in the disk or a shadow on the disk resulting from a puffed-up rim; however, \citet{2005A&A...431..165S} estimated an inner radius of 0.1 au, in line with the approximate dust sublimation radius, based on models of the mid-infrared SED, implying that the presence of a shadow is more likely than disk clearing.

Another set of parameter values come from a forthcoming Atacama Large Millimeter Array (ALMA)-based analysis of the disk by Aguayo, Caceres, et al. (private communication). These parameters include a total dust mass of $1.57\times10^{-4}M_{\sun}$ (or $1.72\times10^{-4}M_{\sun}$ with an additional dust ring in the disk from 80-100 au); a scale height of 5.5 au at a reference radius of 100 au (along with a section of dust with a scale height of 4 au at R=100 au that encompasses R=4.6 au to R=100 au); an inner radius of 0.03 au; an outer radius of 240 au; disk inclination $27.3\degr$ on the basis of HST/STIS imaging \citep{2014AJ....148...59S}; and a stellar radius of $1.65R_{\sun}$ based on the star's bolometric luminosity and effective temperature, estimated at 5035 K, which differs from the estimated effective temperature in this study. A recent analysis of these ALMA data by \citet{2023A&A...673A..77R} derived a dynamic stellar mass of 1.3 $M_{\sun}$, age 7-10 Myr based on this mass, dust disk mass $1.44\times10^{-4}M_{\sun}$, dust disk outer radius 60 au, gas disk outer radius 130 au (assuming small grain coupling with gas), and disk inclination $32\degr$, which we adopt as the inclination of our model disk \revision{along with the dust disk mass and dust disk outer radius}; furthermore, this study also indicates likely shadowing on the disk of MP Mus.

These sets of parameters indicate certain discrepancies in our knowledge of the disk, particularly in the wide range of estimated disk masses. The protoplanetary disk surrounding MP Mus therefore represents fertile ground for application of the EaRTH Disk Model, in an attempt to obtain a more robust set of fundamental disk structural parameters and to investigate disk dust composition.

\subsection{Disk Mineralogy}\label{mineralogy}

Our mineralogy analysis using the \revision{TZTD} model with the new opacities indicated dust continuum emission, represented here by very large grains and carbon, in the optically thick atmosphere making up \revision{essentially all of the total masses of the warm and cool dust disk components (99.95\%$\pm$0.01\% and 99.98\%$\pm$0.01\%, respectively)}. The model returns \revision{a dust temperature distribution parameterization} of \revision{$T_{warm}(r)=(1500~K)(\frac{r}{r_{in}})^{-1.5}$} and \revision{$T_{cool}(r)=(240~K)(\frac{r}{r_{in}})^{-0.9}$, and a dust grain size power law of 3.1}. A substantial amount of this optically thick dust is likely very large ($>$0.1 mm) silicates otherwise unrepresented in the fit, and thus does not represent the ratio of carbons and silicates. The results for the mineralogy of the disk \revision{excluding the continuum emission (representative of the optically thin portion of the atmosphere)} are presented in Figure \ref{mpmus_dust_plots} and in Table \ref{comp_table}\revision{, where we find the reduced $\chi^2$ value of the fit to be 2.63; the error within the $Spitzer$ IRS spectrum is taken to be the RMS error from the archival data}.

\begin{figure*}[htb!]
    \gridline{(a) \fig{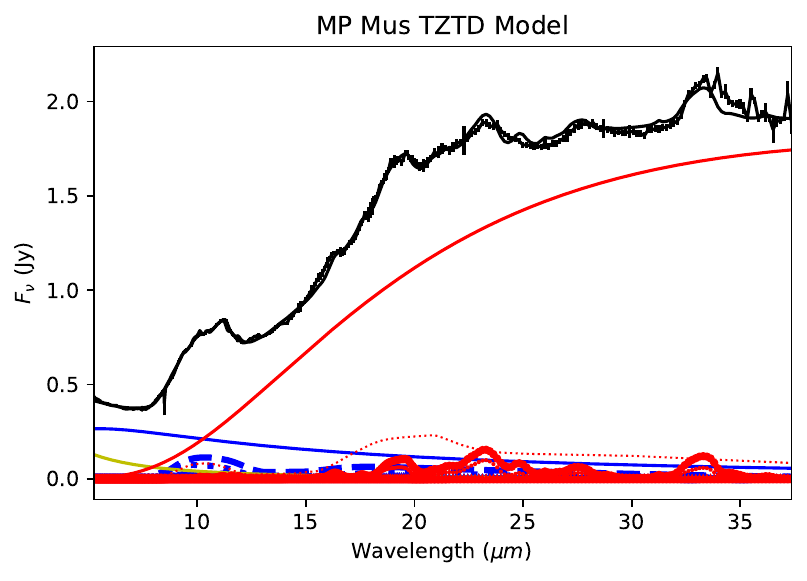}{0.5\textwidth}{}
              (b) \fig{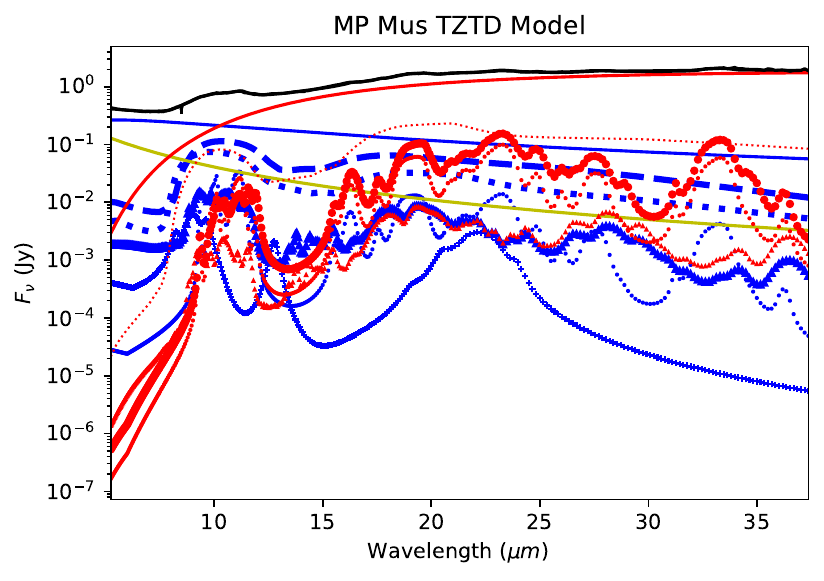}{0.5\textwidth}{}
              }
    \gridline{(c) \fig{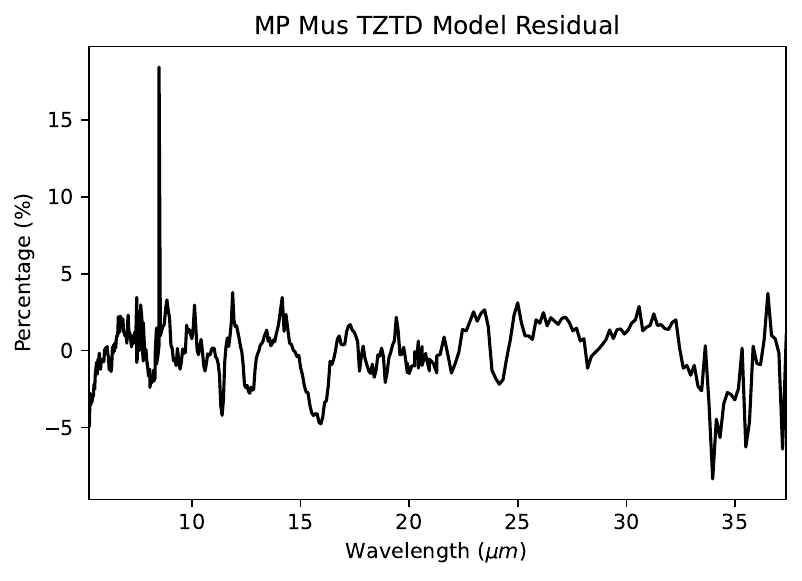}{0.5\textwidth}{}
              (d) \fig{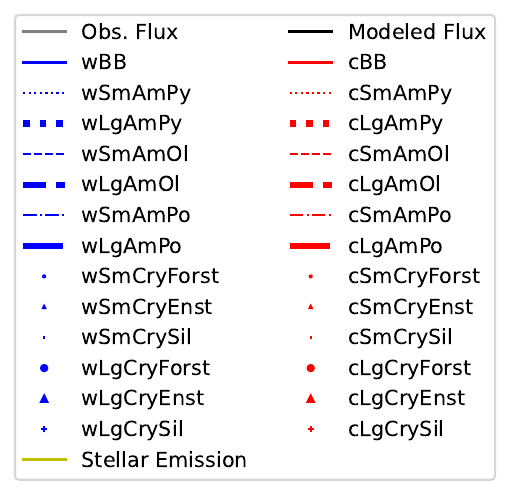}{0.39\textwidth}{}
              }
    \caption{Results of \revision{TZTD} empirical mineralogical analysis of MP Mus $Spitzer$ IRS spectrum; Top: Model fit plots, (a) Regular scaling to demonstrate the fit to the function, (b) Logarithmic scaling to demonstrate the contributions of each mineral; \revision{(c) Residual of mineralogical fit with $Spitzer$ IRS spectrum; (d)} Legend for the top plots; \revision{(e:)} Dust composition of optically thin portions of protoplanetary disk surrounding MP Mus, using values from Table \ref{comp_table}, which includes full names of abbreviated material names shown in plots; Red: Cool disk component constituents, Blue: Warm disk component constituents.}
    \label{mpmus_dust_plots}
\end{figure*}

\begin{figure*}[htb!]
    \figurenum{1}
    \gridline{(e) \fig{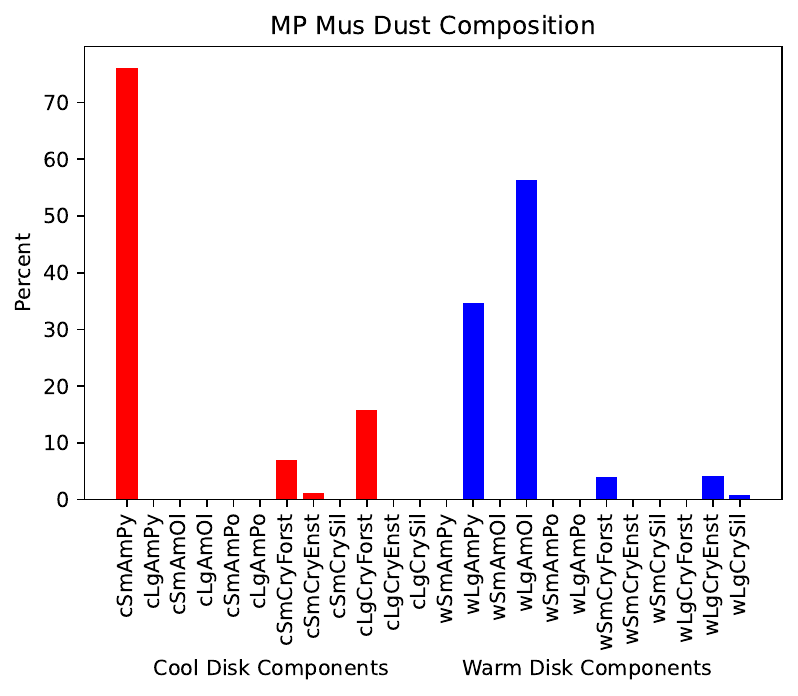}{0.5\textwidth}{}
             }
    \caption{(continued)}
\end{figure*}

\begin{deluxetable*}{lc}[htb!]
\deluxetablecaption{MP Mus Protoplanetary Disk: Dust Component Modeling.\label{comp_table}}
\tablewidth{0pt}
\tablehead{
\colhead{Dust Component} & \colhead{Percent}
}
\startdata
Warm Small Amorphous Pyroxene (wSmAmPy) & 0\revision{$\pm$5.3}\\
Warm Large Amorphous Pyroxene (wLgAmPy) & \revision{34.8$\pm$10.2}\\
Warm Small Amorphous Olivine (wSmAmOl) & \revision{0$\pm$6.8}\\
Warm Large Amorphous Olivine (wLgAmOl) & \revision{56.1$\pm$14.1}\\
Warm Small Amorphous Polivene (wSmAmPo) &  0\revision{$\pm$8.3}\\
Warm Large Amorphous Polivene (wLgAmPo) &  0\revision{$\pm$9.7}\\
\revision{Warm Small Crystalline Forsterite (wSmCryForst)} & \revision{4.0$\pm$5.3}\\
\revision{Warm Small Crystalline Enstatite (wSmCryEnst)} & \revision{0$\pm$4.8}\\
\revision{Warm Small Crystalline Silica (wSmCrySil)} & \revision{0$\pm$1.5}\\
\revision{Warm Large Crystalline Forsterite (wLgCryForst)} & \revision{0$\pm$12.3}\\
\revision{Warm Large Crystalline Enstatite (wLgCryEnst)} & \revision{4.3$\pm$6.4}\\
\revision{Warm Large Crystalline Silica (wLgCrySil)} & \revision{0.8$\pm$3.5}\\
\hline
\enddata
\tablecomments{Includes full names and abbreviations of materials shown in composition plots in Figure \ref{mpmus_dust_plots}; see Appendix \ref{uncert} for process of uncertainty derivation}
\end{deluxetable*}

\begin{deluxetable*}{lc}[htb!]
\tablenum{2}
\deluxetablecaption{(continued)}
\tablewidth{0pt}
\tablehead{
\colhead{Dust Component} & \colhead{Percent}
}
\startdata
Cool Small Amorphous Pyroxene (cSmAmPy) & \revision{76.1$\pm$22.6}\\
Cool Large Amorphous Pyroxene (cLgAmPy) & 0\revision{$\pm$11.7}\\
Cool Small Amorphous Olivine (cSmAmOl) & 0\revision{$\pm$8.4}\\
Cool Large Amorphous Olivine (cLgAmOl) & 0\revision{$\pm$8.7}\\
Cool Small Amorphous Polivene (cSmAmPo) &  \revision{0$\pm$12.4}\\
Cool Large Amorphous Polivene (cLgAmPo) &  0\revision{$\pm$12.1}\\
\revision{Cool Small Crystalline Forsterite (cSmCryForst)} & \revision{6.9$\pm$5.7}\\
\revision{Cool Small Crystalline Enstatite (cSmCryEnst)} & \revision{1.2$\pm$7.4}\\
\revision{Cool Small Crystalline Silica (cSmCrySil)} & \revision{0$\pm$5.4}\\
\revision{Cool Large Crystalline Forsterite (cLgCryForst)} & \revision{15.8$\pm$7.8}\\
\revision{Cool Large Crystalline Enstatite (cLgCryEnst)} & \revision{0$\pm$8.1}\\
\revision{Cool Large Crystalline Silica (cLgCrySil)} & \revision{0$\pm$7.1}\\
\hline
Reduced \revision{$\chi^2$} Value & \revision{2.63}\\
\enddata
\tablecomments{Includes full names and abbreviations of materials shown in composition plots in Figure \ref{mpmus_dust_plots}; see Appendix \ref{uncert} for process of uncertainty derivation}
\end{deluxetable*}

\revision{The results indicate that both disks are predominantly composed of amorphous minerals, with large ($\micron$-sized) amorphous grains in the warm disk and small (sub-$\micron$) pyroxene grains in the cool disk, suggesting grain growth focused in the warmer regions of the disk \citep[consistent with the findings of][]{2009ApJS..182..477S}. The presence of crystalline silicates also indicates substantial grain processing and growth within the MP Mus disk \citep{2009ApJS..180...84W}; $\micron$-sized forsterite has a prominent presence in the cool disk, as indicated by features in the $Spitzer$ IRS spectrum at wavelengths $>20~\micron$, suggesting grain growth is not limited to the warmer regions of the disk.} We note that degeneracies may be present between model dust components; see Appendix \ref{degeneracy}, where we detail a degeneracy analysis on the new set of components and find significant degeneracies between cool amorphous silicates, such as pyroxene and olivine.

\subsection{Disk Structure}\label{struc}

We implement the detected materials and mass fractions using the new set of opacities into MCFOST accordingly; summing up the mass fraction constants of each disk component and comparing the totals yields an estimate that the cool disk \revision{encompasses almost all of the mass of the disk; in this sense, the warm disk can represent the puffed-up inner rim of the disk, while the cool disk represents the rest of the disk, including the midplane and atmosphere of the disk.}

We use the EaRTH Disk Model to calculate the disk structural parameters and list the resulting parameters, including the \revision{reduced $\chi^2$ value} of the MCFOST generated model and the $Spitzer$ IRS spectrum, found to be \revision{10.81}, in Table \ref{mpmusres}. \revision{This result was established by fine-tuning the parameters to fit the $Spitzer$ IRS spectrum after the fit to the temperature structure.} We demonstrate the result of the corresponding MCFOST model as compared to the observed spectrum in Figure \ref{mpmus_seds}, alongside the mineralogy fit generated by the \revision{TZTD} model, as well as the modeled temperature structure of the disk\revision{, which is compared to the empirically fitted temperature distribution parameterization,} in Figure \ref{mpmus_temp}. \revision{In Figure \ref{mpmus_comp}, we illustrate the contributions of various emission and scattering components to the best-fit radiative transfer model.} We further demonstrate how variations in these parameters alter the model mid-IR SED in Figure \ref{both_disk_effects}.

As can be seen from the best-fit SED and the \revision{reduced $\chi^2$ value}, upon using the mineralogical input from the \revision{TZTD} model, MCFOST produces a robust fit in the infrared regime; the fit well reproduces the 10-$\micron$ feature, the 15-$\micron$ valley, the $\sim$20-$\micron$ spike, and the overall pattern. However, as the comparison demonstrates, with a reduced $\chi^2$ value of 2.63, the \revision{TZTD} model fit alone more robustly captures these features, as well as smaller features in the longer-wavelength infrared that the combined model fails to capture, \revision{particularly the boundaries of the wavelength range}. This slight mismatch demonstrates the inherent tendency of the EaRTH Disk Model to propagate systematic errors in feature wavelengths that are present in the \revision{TZTD} model fit, given this fit serves as the MCFOST mineralogical input\revision{; for example, both the empirical and radiative transfer SEDs have a slight mismatch in the 12-16 $\micron$ range, likely the result of mispointing, which we account for as detailed in Appendix \ref{misp}}. 

\subsection{Model Assessment}\label{assess}
As a result of the model's tendency to propagate the errors from the \revision{TZTD} fit, as well as the presence of physical structural constraints in MCFOST (and by extension the EaRTH Disk Model) that are not present in the \revision{TZTD} model, the \revision{TZTD} model will always produce a more robust fit than the EaRTH Disk Model; we again stress that degeneracy between dust components must be considered, as is discussed in Appendix \ref{degeneracy}. However, despite the additional parameters, \revision{as demonstrated in Figure \ref{mpmus_seds},} the EaRTH Disk Model fit is visually very close to that of the \revision{TZTD} model fit; the \revision{reduced $\chi^2$ value} of the EaRTH Disk Model fit is still relatively low, which demonstrates the robustness of the model in fitting the infrared SED to both mineralogical and structural disk parameters.

Additionally, many of the values are similar to those predicted by other studies. For example, \revision{the inner radius, 0.08 au}, is consistent with the determinations by \revision{\citet{2005A&A...431..165S}, Aguayo, Caceres, et al. (in prep.) and \citet{2023A&A...673A..77R}. Also, while the outer radius of 60 au we adopted based on the ALMA study by \citet{2023A&A...673A..77R} was effective in constraining the structural fit, we later tuned it to 53 au to match up with the edge falloff observed in the ALMA image and radial profile, a value which remained an effective constraint on the fit. These agreements are a promising demonstration of the hybrid model's capabilities.}

\begin{figure*}[htb!]
    \gridline{\fig{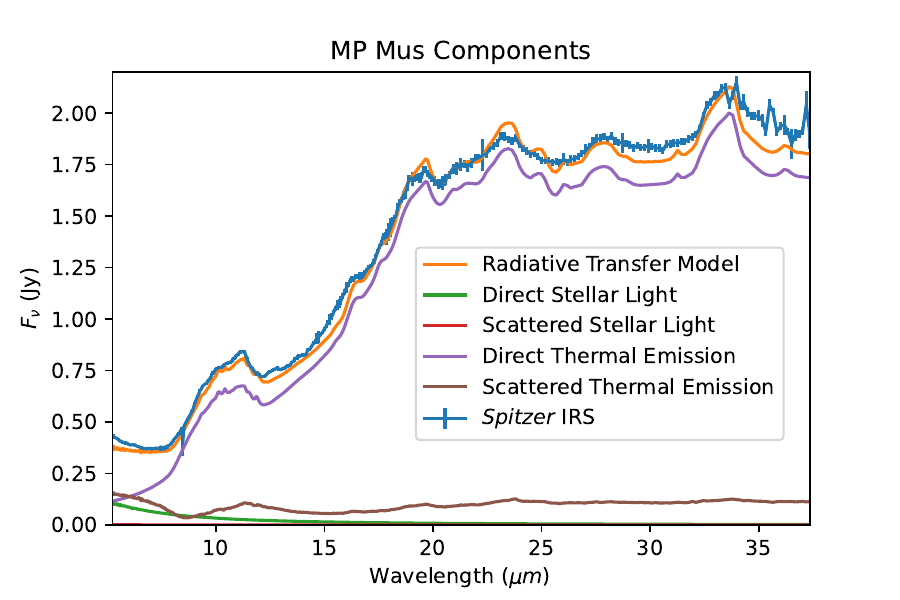}{\textwidth}{}}
    \caption{\revision{Contributions of light components to mid-IR SEDs; scattered stellar light does not noticeably contribute at mid-IR wavelengths.}}
    \label{mpmus_comp}
\end{figure*}

\revision{A complicating factor to be considered is that mass absorption coefficients are typically used as the opacity functions with mid-infrared empirical mineralogical models \citep[for example,][]{2009ApJ...695.1024J,2010ApJ...721..431J,2009ApJ...690.1193S,2009ApJS..182..477S}; such is the case with the TZTD model as well. These functions do not, and typically cannot, account for scattering opacity, and therefore implicitly assume that thermal emission, from both disk and star (which can be removed in advance and is removed in the EaRTH Disk Model; see Section \ref{2tmodel}), is the sole source of flux in the mid-infrared regime. Scattering is accounted for in MCFOST, however, and Figure \ref{mpmus_comp} demonstrates this; the dominant contribution to the mid-IR SED, as expected, is thermal emission from disk dust, and contributions from direct and scattered stellar photospheric radiation are negligible. However, while scattering of thermal radiation from the disk itself accounts for far less flux in the mid-infrared regime than direct thermal emission, the scattered dust emission component is not negligible, due to the presence of large grains in the mineralogical fit. This factor must be considered when using the EaRTH Disk Model for structural analysis, particularly if large grains are prevalent in the disk.}

Given the computational cost of our EaRTH Disk Model, we were not able to perform a Bayesian analysis to extract robust estimates of the errors on our parameter determination\revision{; \citet{2020A&A...642A.171R} indicates that full Bayesian analyses are not typically feasible with radiative transfer models due to the computation costs typically associated with them}. However, as an indication of the acceptable range of parameter values, Figure \ref{both_disk_effects} demonstrates the effects of each parameter on the mid-IR SED of MP Mus. \revision{We use these spectra to extract uncertainty estimates based on the $\Delta\chi^2$ approach outlined in Appendix \ref{uncert}; we empirically estimate the uncertainties of the inner radius and warm disk scale height using Equations \ref{inner_rad} and \ref{scalefactor} respectively along with these uncertainties and those derived from the TZTD model fit.} Based on this analysis, we surmise that the selected parameters for the fit are optimal for determining the structure of the disk; except for the surface density exponent of the warm inner disk \revision{(panel (g) of Figure \ref{both_disk_effects}), as reflected in the large uncertainty of the parameter,} altering the parameters highlighted in Table \ref{initparams} has a profound effect on the mid-infrared SED model. 

\revision{The viscosity parameter can be degenerate with some parameters, so it is necessary to be careful when using the mid-IR SED to determine the parameter value \citep{2023NewAR..9601674R}. For lower values of the viscosity parameter, grain mixing becomes less turbulent, causing larger grains to settle in the midplane \citep{1995Icar..114..237D,2016A&A...586A.103W}, which reduces their influence in the optically thin atmosphere, reducing mid-IR flux. Therefore, the assumption of the mixture of such grains in the TZTD model likely causes the proper viscosity parameter for the fit to be larger, i.e. a more turbulent, mixed disk; likewise, an empirical model that divides grains into separate zones by size would potentially require a weaker viscosity parameter, with weak turbulence and more settling to permit separation of grains in the disk by size. Thus, in future models using empirical results in radiative transfer modelling, an empirical model selection on the basis of known turbulence, or lack thereof, in the disk would be appropriate.}

With this set of parameters, the disk must have a \revision{very} small inner cavity, \revision{with only a radius of 0.08 au; furthermore, based on the temperature distributions from the TZTD model, the EaRTH Disk Model infers a potential gap within the disk at larger radii from $\sim$0.1 au up to $\sim$0.85 au, thereby creating a cavity with some hot dust, near the star, effectively creating a narrow, puffed up inner rim with most of its mass gathered at the inner radius to the point that its remaining mass is negligible at larger radii.} The ALMA image (Figure \ref{mpmus_mm}) and VLT/SPHERE image (Figure \ref{mpmus_sphere}) indicate a lack of an inner cavity down to their resolutions, as noted by \citet{2023A&A...673A..77R}. However, the resolution of the ALMA image is \revision{0.01} arcseconds, which translates to a resolution of $\sim$\revision{1} au at the $Gaia$ DR3 measured distance of 97.89 pc, so ALMA would not resolve such a cavity. Furthermore, the center of the disk within $\sim$0.1 arcseconds ($\sim$10 au) of the SPHERE image is occulted by the coronagraph, so such an inner cavity would not be imaged. Therefore, as the observed $Spitzer$ IRS spectrum and model indicate, it is possible that there is a small inner cavity in the disk of MP Mus that these images could not show.

\begin{figure*}[htb!]
    \gridline{\fig{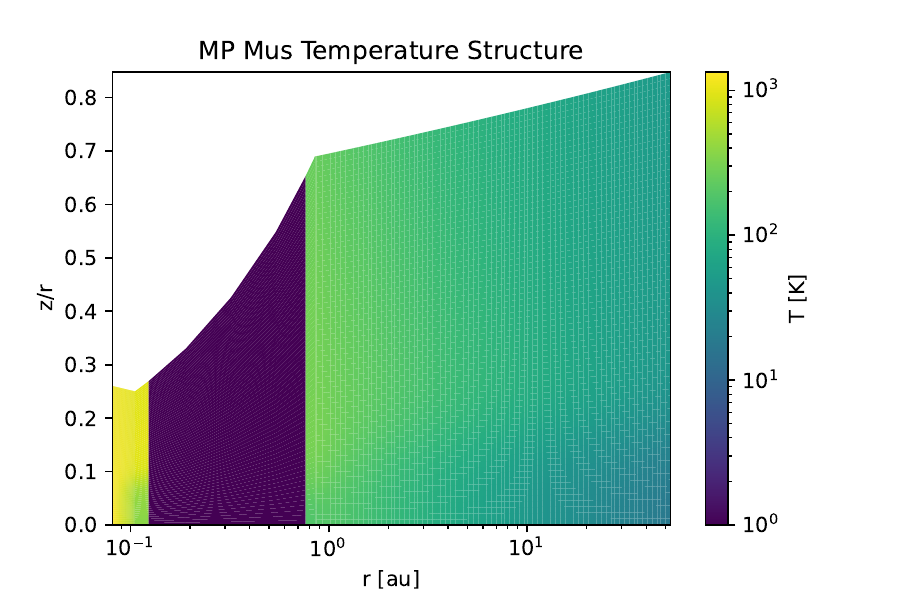}{0.7\textwidth}{}}
    \gridline{\fig{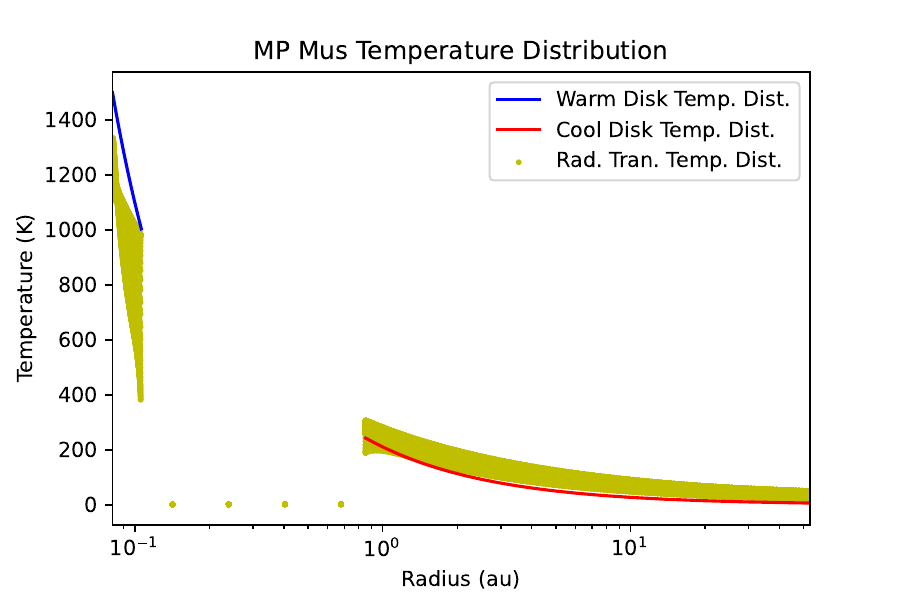}{0.7\textwidth}{}}
    \caption{\revision{Top:} MCFOST-simulated temperature structure of MP Mus protoplanetary disk; the discontinuity results from the \revision{gap formed between} the disk sections. The temperatures of the warm and cool parts of the disk are similar to the \revision{empirical model-fitted temperature distribution} predictions. \revision{Bottom: MCFOST-simulated temperature plotted against radius, compared with empirically fitted warm and cool disk temperature distributions.}}
    \label{mpmus_temp}
\end{figure*}

This hypothesis is further supported by the temperature structure of the disk (Figure \ref{mpmus_temp}). The maximum temperature of the dust in the disk is $\sim$\revision{1350 K, close to the warm maximum temperature of 1500 K returned by the TZTD model.} Variations in the inner radius would either introduce hotter dust or remove dust at reasonable temperatures; this further indicates that \revision{a puffy inner rim occupying an inner radius of $\sim$0.1 au} corresponding to a small cavity undetected by ALMA and SPHERE is a reasonable possibility\revision{, along with a main dust disk distribution starting at a radius of $\sim$0.85 au.}

\subsection{Model Images}\label{modimgs}

\begin{figure*}[htb!]
    \gridline{\fig{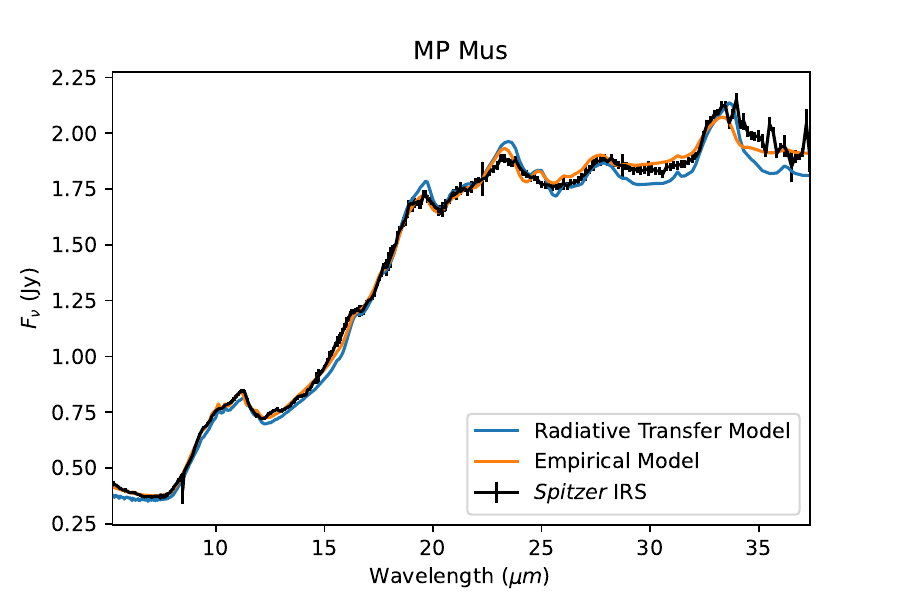}{0.7\textwidth}{}}
    \gridline{\fig{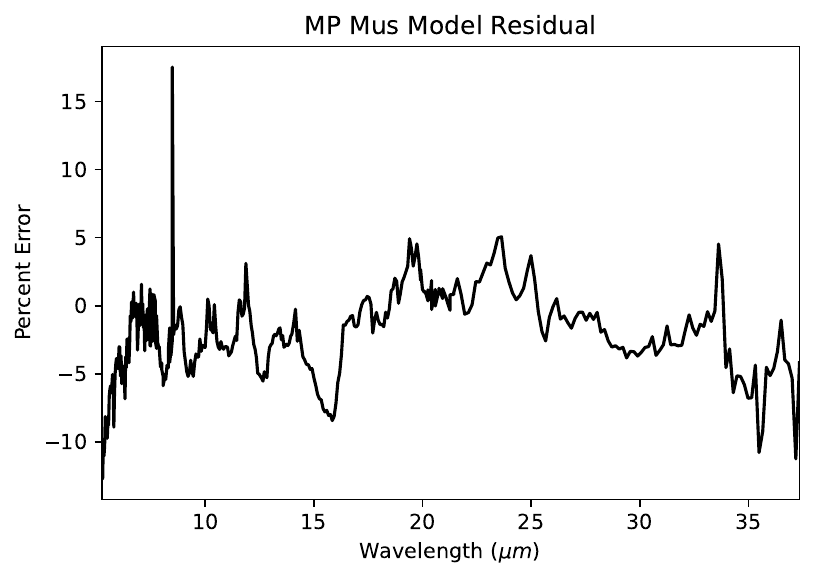}{0.7\textwidth}{}}
    \caption{\revision{Top:} Comparison between MCFOST-simulated spectrum, the \revision{TZTD} model fit, and the observed $Spitzer$ IRS spectrum of MP Mus; \revision{Bottom: Residual of radiative transfer model fit with $Spitzer$ IRS spectrum}}
    \label{mpmus_seds}
\end{figure*}

\begin{deluxetable*}{lc}[htb!]
\deluxetablecaption{MP Mus parameters determined by the EaRTH Disk Model.\label{mpmusres}}
\tablewidth{0pt}
\tablehead{
\colhead{Parameter} & \colhead{Value}
}
\startdata
Warm Disk Flaring Exponent & \revision{0.850$\pm$0.007} \\
Warm Disk Surface Density Exponent & \revision{-0.57$\pm$0.39} \\
Warm Disk Outer Radius (au) & \revision{(1.30$\pm$0.11)$\times$Inner Radius} \\
Cool Disk Inner Radius (au) & \revision{(10.5$\pm$0.4)$\times$Inner Radius} \\
Cool Disk Scale Height (au) & \revision{12.5$\pm$0.4} \\
Cool Disk Flaring Exponent & \revision{1.05$\pm$0.09} \\
Viscosity Parameter $\alpha$ & \revision{$1\times10^{-2}\pm7\times10^{-3}$} \\
\hline
Inner Radius (au) & \revision{0.08$\pm$0.25} \\
Warm Disk Scale Height (au) & \revision{1.30$\pm$1.18} \\
\hline
Cool Disk Surface Density Exponent & \revision{-0.642} \\
Cool Disk Outer Radius (au) & \revision{53} \\
Total Dust Mass ($M_{\sun}$) & \revision{$1.44\times10^{-4}$} \\
\revision{Reduced $\chi^2$ value} & \revision{10.81} \\
\enddata
\tablecomments{The scale height in both sections of the disk is measured at a reference radius of R=100 au\revision{; the top section of the table indicates the parameters directly fitted by the model, the middle section indicates the parameters indirectly fitted, empirically calculated using the above parameters, and the bottom section indicates the parameters constrained using ALMA archival data.}}
\end{deluxetable*}

\begin{figure*}[htb!]
    \gridline{
              (a) \fig{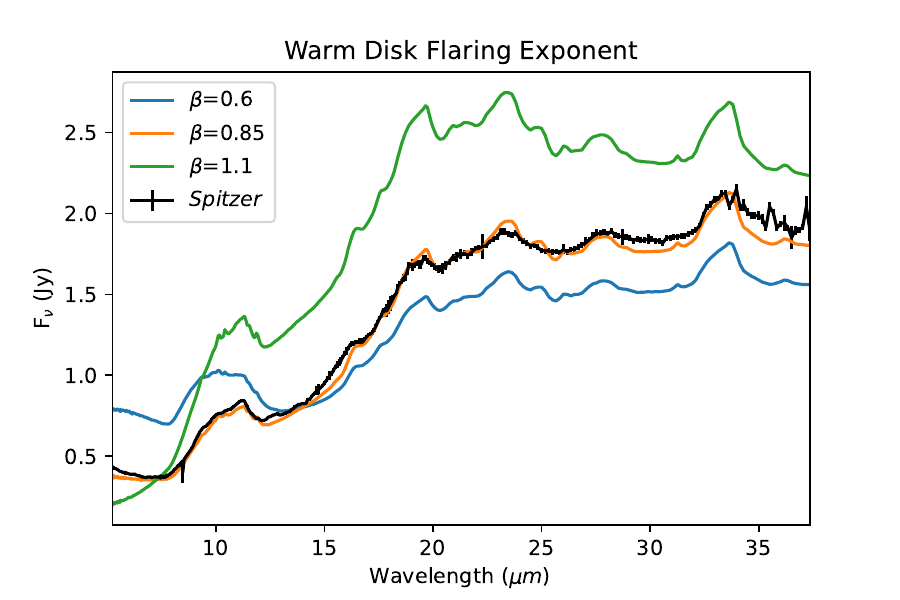}{0.49\linewidth}{}
              (b) \fig{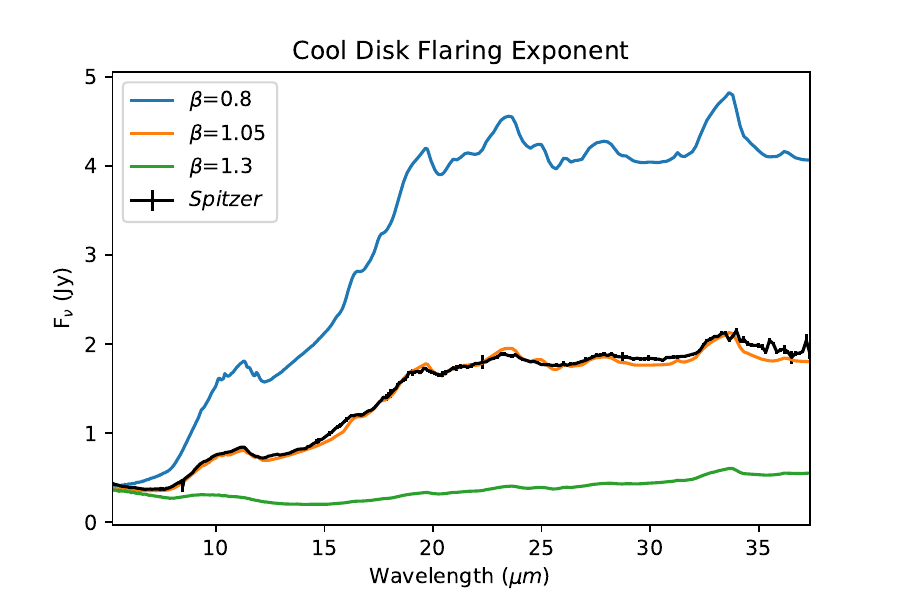}{0.49\linewidth}{}
             }
    \gridline{
              (c) \fig{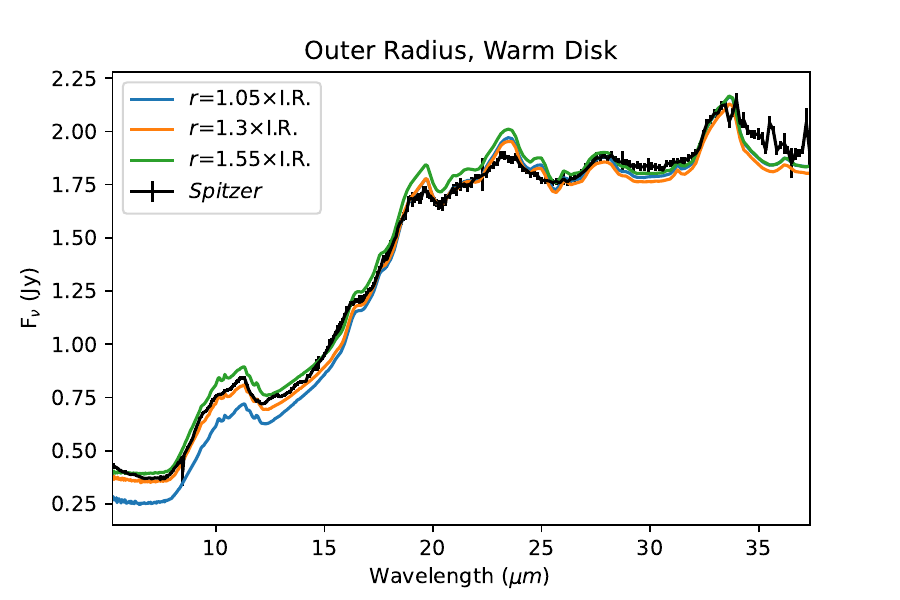}{0.49\linewidth}{}
              (d) \fig{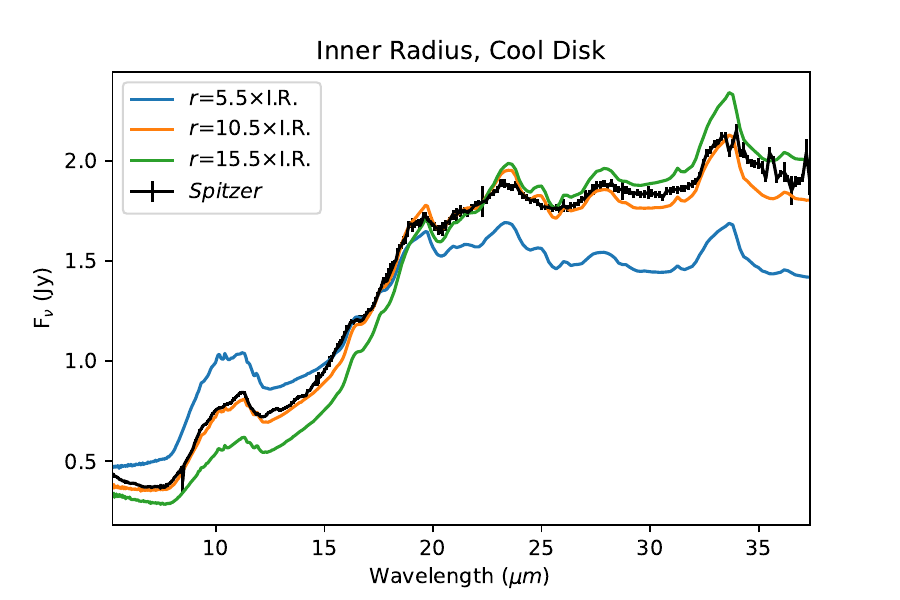}{0.49\linewidth}{}
             }
    \gridline{
              (e) \fig{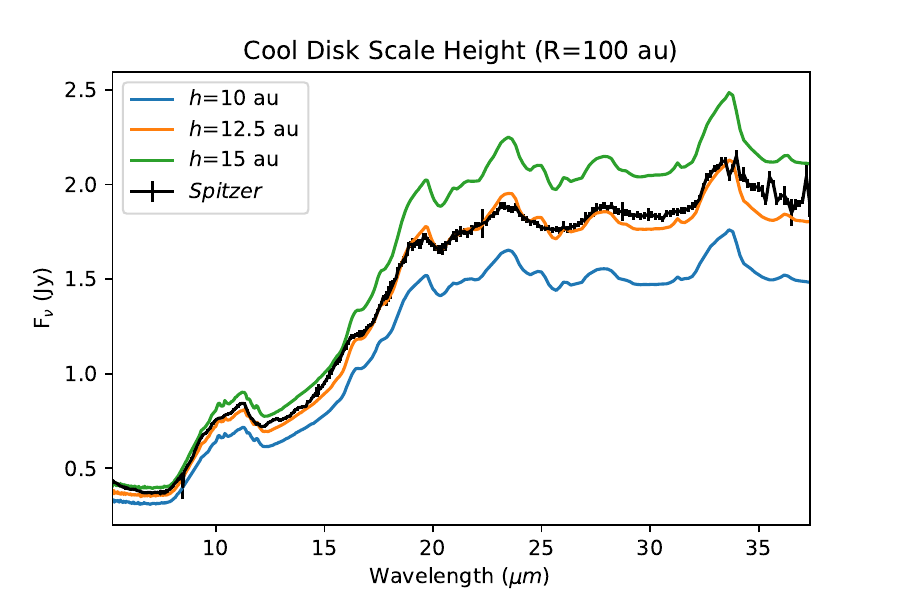}{0.49\linewidth}{}
              (f) \fig{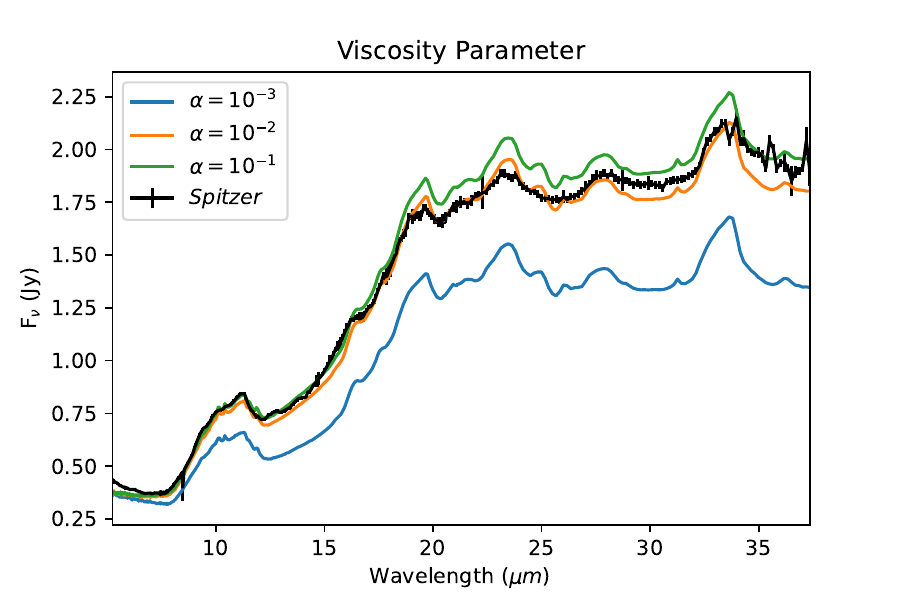}{0.49\linewidth}{}
              }
    \caption{Influence of individual parameters on the MCFOST-generated mid-IR SED models of the disk; titles indicate parameters in effect, and legends indicate values tested.}
    \label{both_disk_effects}
\end{figure*}

\begin{figure*}[htb!]
     \gridline{
              (g) \fig{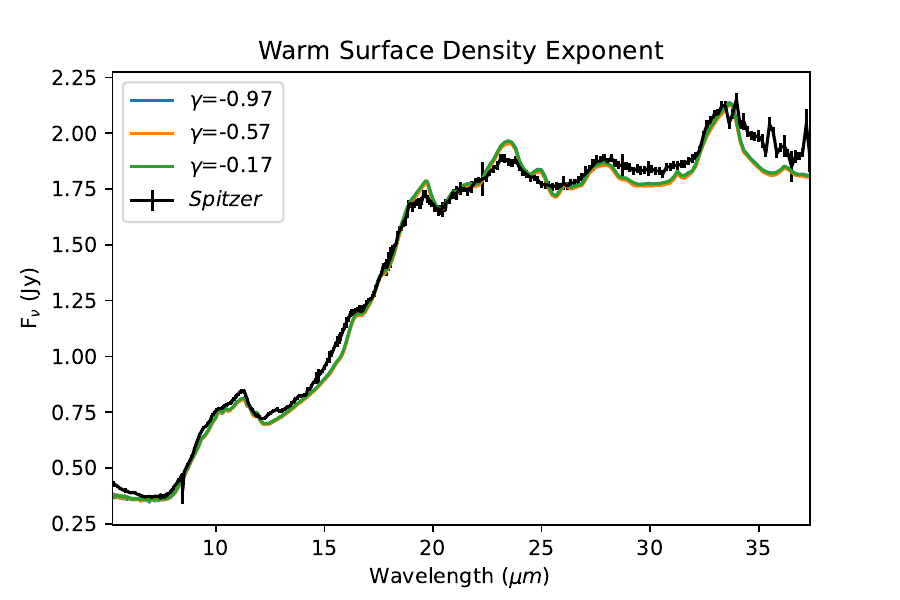}{0.49\linewidth}{}
              }
    \figurenum{5}
    \caption{(Continued)}
\end{figure*}

\begin{figure*}[htb!]
    \gridline{\fig{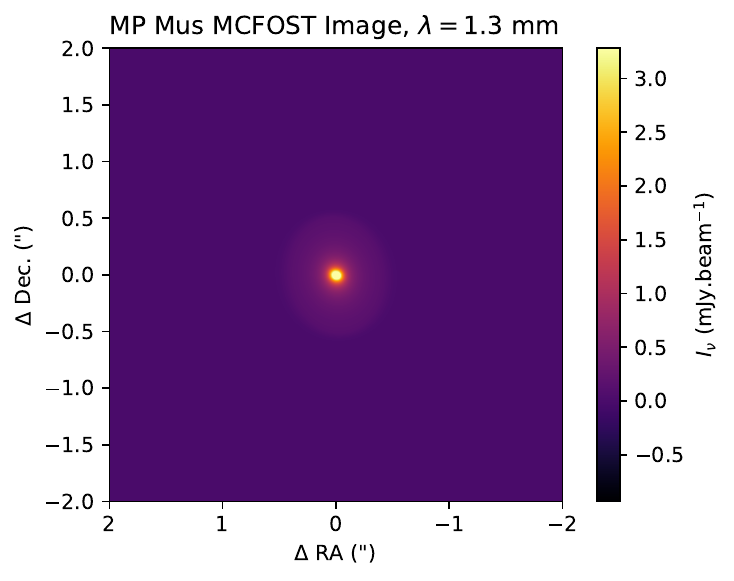}{0.45\textwidth}{}
              \fig{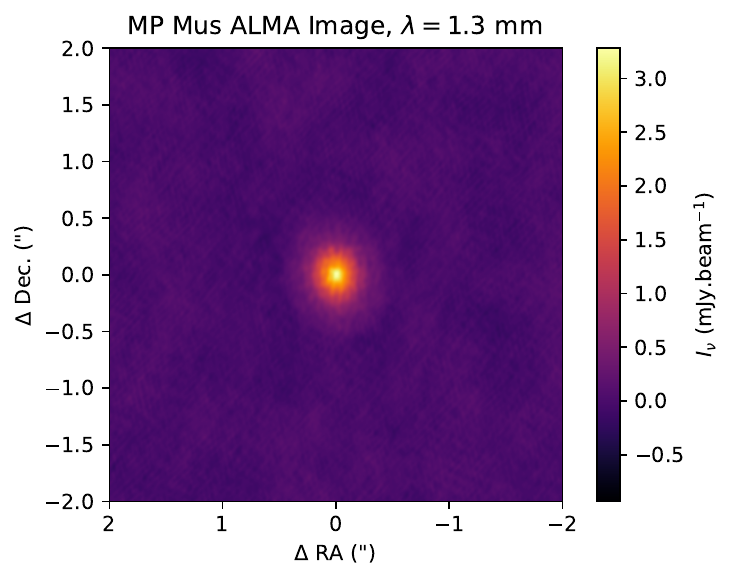}{0.45\textwidth}{}}
    \gridline{\fig{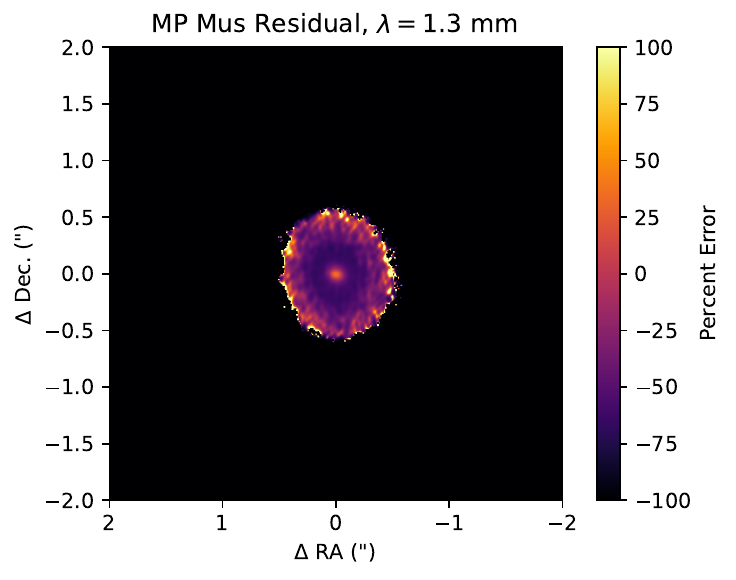}{0.45\textwidth}{}
              \fig{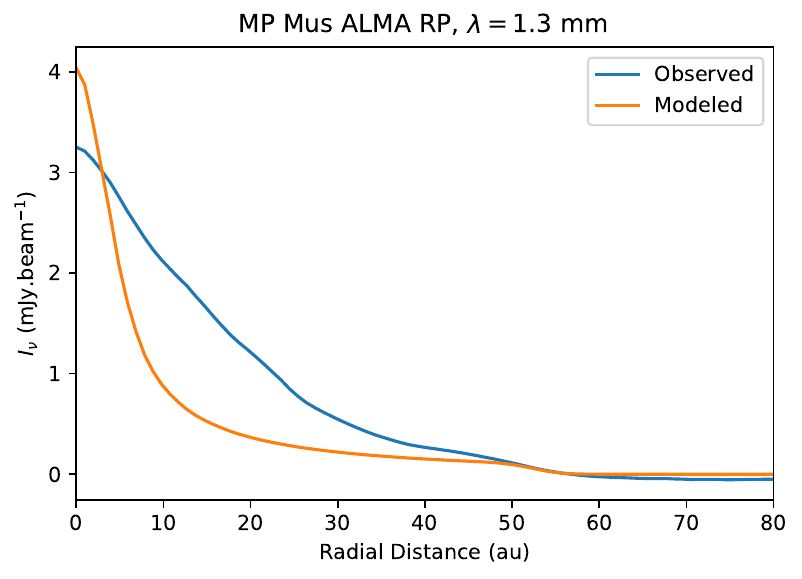}{0.45\textwidth}{}}
    \caption{Comparison of MCFOST-simulated image of MP Mus with archival ALMA data; Top Left: Simulated image at 1.3 mm; Top Right: 1.3 mm, observed image from ALMA; Bottom \revision{Left: Residual of simulated image and observed ALMA image at 1.3 mm. Bottom right}: Radial profile of simulated image and observed ALMA image at 1.3 mm.}
    \label{mpmus_mm}
\end{figure*}

We use the MCFOST model described in Section \ref{struc} (see Figure \ref{mpmus_seds} and Table \ref{mpmusres}) to generate simulated images at wavelengths of the $H$ band (i.e. 1.662 $\micron$) and 1.3 mm for comparison with these VLT/SPHERE and ALMA images, respectively. The rendered 1.3 mm image is compared to archival continuum mapping data obtained by the ALMA radio telescope at Band 6 \revision{(Program ID 2017.1.01167.S)}, with frequencies of \revision{229.607-247.819} GHz and a \revision{69} mas $\times$ \revision{50} mas Gaussian beam with position angle \revision{$-$11.88$\degr$}; the simulated 1.3 mm image is blurred with a Gaussian beam of the same parameters. As can be seen in Figure \ref{mpmus_mm}, the simulated disk image and radial profile well reproduce the observed image and radial profile \revision{within a radial angular size of 0.6 arcseconds}, with a NRMSE of \revision{1.02} mJy beam$^{-1}$ between the observed and simulated radial profile \revision{and a NRMSE of 1.90 mJy beam$^{-1}$ between the observed and simulated image. However, the flux radially decreases much more slowly in the ALMA image than the EaRTH Disk-simulated image, despite the outer disk's surface density exponent being estimated based on that of the ALMA image; this is because the real disk does not precisely radially decrease according to a surface density exponent. This is a limitation caused by some of the simplified assumptions of the EaRTH Disk Model, as well as the fact that the model is primarily focused on fitting the mid-IR SED rather than the image; to better fit the image beyond the model, a third intermediate overlapping zone may be necessary, potentially only consisting of very large dust grains}.

In Figure \ref{mpmus_sphere} we show the comparison of the normalized $Q_{\phi}$ model image rendered at the $H$ band to the normalized $Q_{\phi}$ image of MP Mus recorded by the infra-red dual imaging and spectrograph (IRDIS) instrument \citep{2008SPIE.7014E..3LD} of the Spectro-Polarimetic High contrast imager for Exoplanets REsearch (SPHERE) using the European Southern Observatory's Very Large Telescope (ESO VLT) \citep{2019A&A...631A.155B}, on 2016 March 16\revision{; this image was produced by sending the data through the IRDAP pipeline \citep[IRDIS Data reduction for Accurate Polarimetry, version 1.3.5,][]{2020A&A...633A..64V}}. We mask the central $\sim$120 mas of the image to approximate the occultation of the star from the coronagraphic imaging as seen in the original SPHERE image. However, besides the significantly narrower flux density peak in the central disk in the model compared to the SPHERE image, the latter demonstrates that MP Mus has a ring, located about 80 au from the central star, which is visible at the $H$ band, that is undetected by ALMA in the mm-wave regime, as Figure \ref{mpmus_mm} confirms; our model, \revision{restricted by the outer radius estimated using ALMA data}, also fails to capture this ring. This ring, detected by \citet{2016ApJ...818L..15W}, is either the result of the presence of a radially confined over-density of small grains (peaking near $\sim$80 au) \citep{2018ApJ...863...44A}, or is caused by shadowing of the disk at radii $<$80 au \citep{2023A&A...673A..77R}.

\begin{figure*}[htb!]
    \gridline{(a) \fig{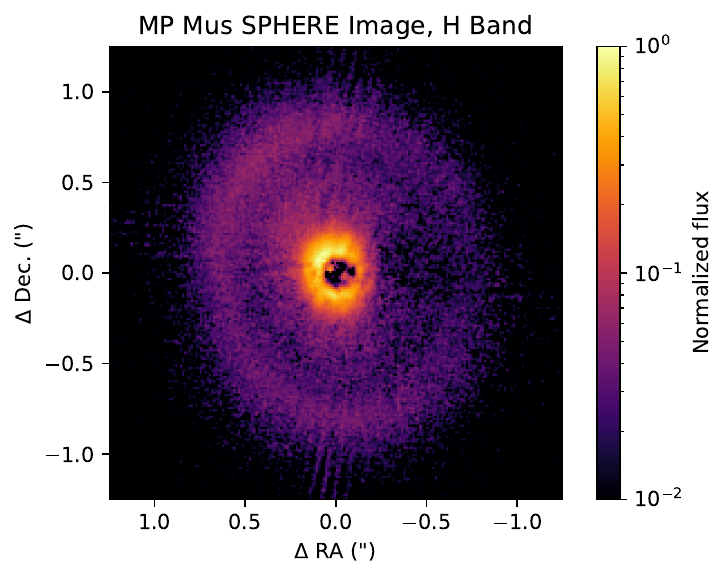}{0.45\textwidth}{}
              (b) \fig{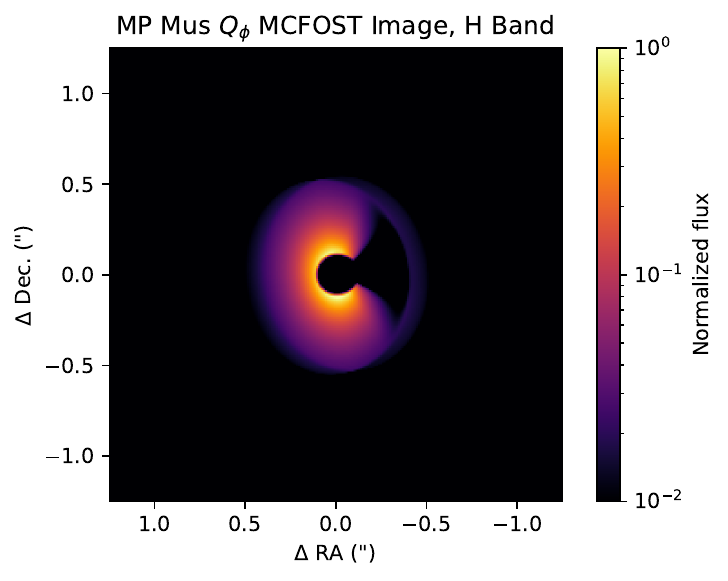}{0.45\textwidth}{}
              }
    \gridline{(c) \fig{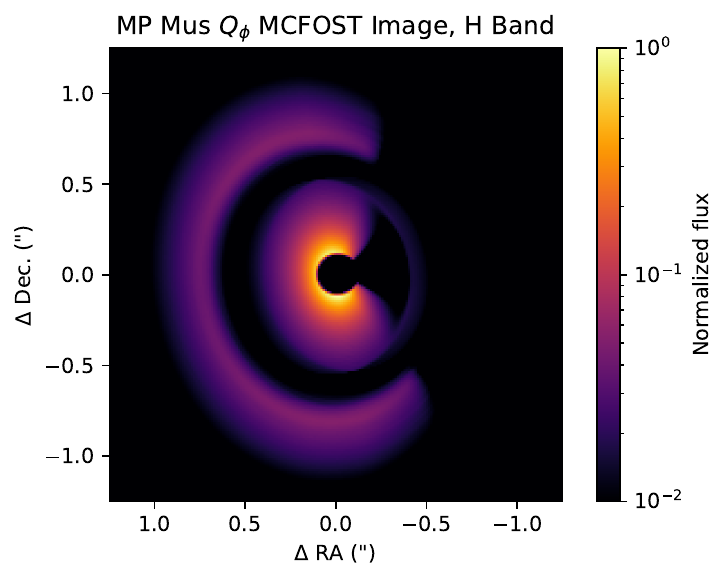}{0.45\textwidth}{}
            (d) \fig{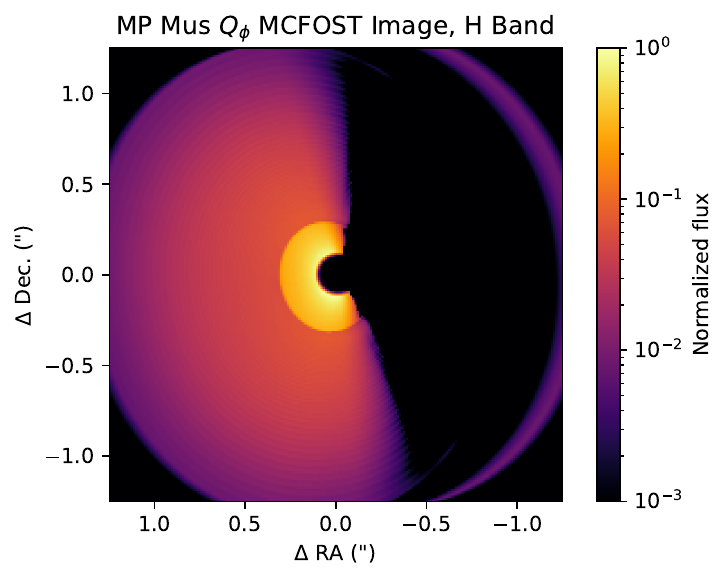}{0.45\textwidth}{}
              }
    \caption{Different visualizations of MP Mus disk at the $H$ band; Top and Middle: Normalized $Q_{\phi}$ images of MP Mus at the $H$ band; (a) Image observed by SPHERE; (b) Image of best fit model; (c) Image with an outer ring; (d) Image with shadowed disk; (e) $Q_{\phi} \times r^2$ radial profiles of simulated images and observed SPHERE image at the $H$ band normalized at peak intensity, plotted to emphasize the outer ring; (f) Comparison between MCFOST-simulated spectrum of original best fit, the best fit with a small-grain ring, optimal fits involving a shadowed disk, and the observed $Spitzer$ IRS spectrum of MP Mus.}
    \label{mpmus_sphere}
\end{figure*}
\begin{figure*}[htb!]
    \gridline{(e) \fig{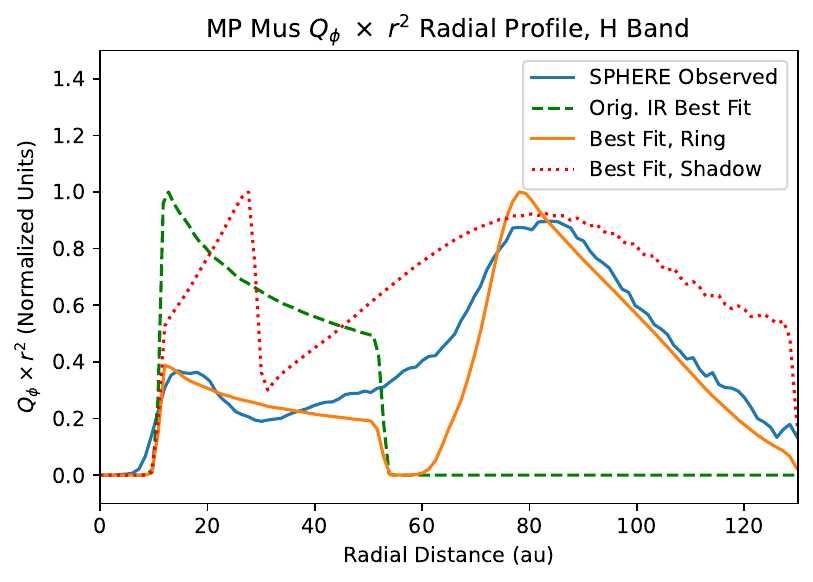}{0.45\textwidth}{}
              (f) \fig{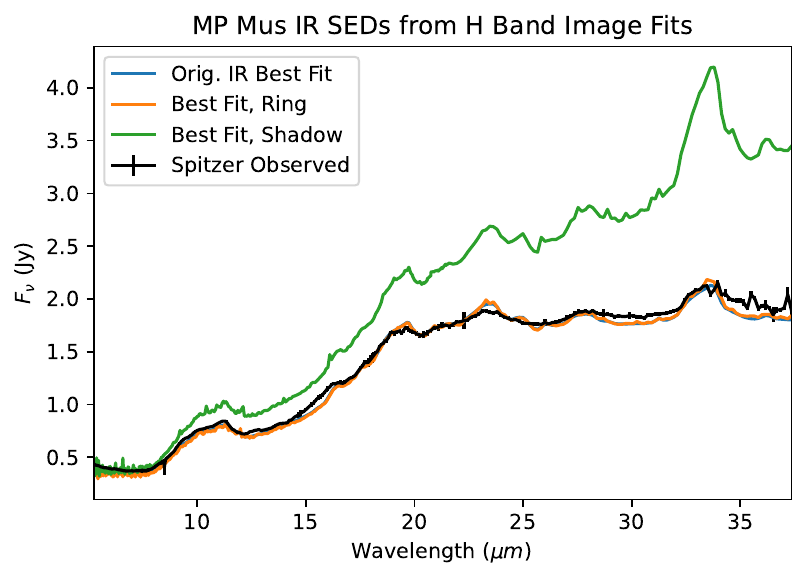}{0.45\textwidth}{}
              }
    \figurenum{7}
    \caption{(Continued)}
\end{figure*}

To address these possibilities, we first include a small grain ring in an amended model. This ring is permitted a tapered-edge structure, which has a density distribution of 

\begin{equation}\label{dustdens2}
\Sigma~\alpha~r^{-\gamma}\exp(-\left(\frac{r}{R_c}\right)^{2-\gamma_{\exp}}),
\end{equation}

\noindent where $\gamma$ is similar to the surface density exponent in Equation \ref{dustdens} but with a negative value, $R_c$ is the critical radius of the disk, and $\gamma_{\exp}$ affects the tapering rate with respect to the critical radius. The radial distribution of the ring is set from \revision{R=80 au to R=130 au}, with an edge parameter of \revision{4.5} (i.e. a density drop outside of these radii following a Gaussian with \revision{$\sigma=$4.5}) and \revision{$R_c=$100} au; we assign it a scale height of \revision{5.25} au at R=100 au, flaring exponent of 1.25, dust mass of $7\times10^{-5}M_{\sun}$, \revision{surface density exponent} 0.5, and $\gamma_{\exp}$=3.0. Furthermore, the very large amorphous carbon grains are removed, leaving only small dust in the $\micron$ and sub-$\micron$ size range. As before, we render images simulating the appearance of MP Mus with the ring at the $H$ band in Figure \ref{mpmus_sphere} along with the simulated image of MP Mus without the observed outer ring; we also render a mid-IR SED corresponding to this model, where it becomes obvious that the inclusion of this ring makes little difference. However, the radial profile well represents that of the SPHERE image, with NRMSE=\revision{0.99} normalized $Q_{\phi}$ units.

Alternatively, as proposed by \citet{2023A&A...673A..77R}, the ring could be the result of shadowing of the intermediate regions of the disk by the inner disk; we hence tested a model to simulate this shadowing, this time without a small grain ring. We accomplish this by \revision{cutting off the initially modeled disk at 30 au (with mass $\sim$1.1$\times~10^{-4}~M_{\sun}$), then setting a new disk from 30-130 au with $\sim$3.44$\times~10^{-5}~M_{\sun}$ of very large grains and $7\times~10^{-5}~M_{\sun}$ of small grains, scale height of 12.5 au at R=100 au, flaring exponent of 1.3, and surface density exponent -0.642; the small grains are modeled as a tapered-edge disk with edge parameter 4.5 and $\gamma_{\exp}$=3.0, co-spatial with the non-tapered-edge disk of large grains.} We end up with results that capture the general shape and peaks of the radial profile but contain an excess of flux in the disk\revision{, with NRMSE=0.99 normalized $Q_{\phi}$ units; while this implies a similar degree of success in simulating the SPHERE image as the addition of the ring, qualitatively, the image of the ring appears closer to reality (comparing panels (c) and (d) of Figure \ref{mpmus_sphere}) while also producing a mid-IR SED that more closely fits the $Spitzer$ IRS spectrum.} This is indicative of the inherent degeneracy in the parameters that must be considered when generating representative models of a disk; however, using multiple data sources, such as a comparison of $Spitzer$ and SPHERE data, can expose degenerate models while also indicating which model is closer to reality.

Both of these disk models maintain the estimated inner radius of $\sim$\revision{0.08} au. The dust mass in both cases was \revision{initialized} to the mass estimated by \citet{2023A&A...673A..77R}\revision{, with an additional $7~\times~10^{-5}~M_{\sun}$ of dust, either in the ring or the general outer area of the disk}. The low value of $\gamma$ in both disks is necessary to permit the distribution of mass in the disk such that the infrared flux from the outer part of the disk is as large as it is; otherwise, most of the mass will be contained at smaller radii and reduce flux from the outer portions of the disk. The values of $\gamma_{\exp}$ capture the large-radii falloff from the R=80 au flux peak. In both cases, setting the outer radius \revision{at 130} au\revision{, the outer radius of the gas/small grain disk measured by \citet{2023A&A...673A..77R}, results in models that match the data}.

%% file: conclusion.tex
We have presented a new modeling framework, the Empirical and Radiative Transfer (EaRTH) Disk Model, which integrates two complementary modeling approaches to elucidate the compositions and density structures of the dust components of circumstellar, protoplanetary disks orbiting young stars. \revision{This model operates on the following methodology:\\
1) apply a “two-zone temperature distribution” (TZTD) model we developed to empirically fit the $Spitzer$ IRS spectrum, so as to establish both dust disk mineralogy and radial temperature structure\\
2) use the resulting TZTD-constrained mineralogy as input to MCFOST and, in parallel, determine the MCFOST disk structural parameters that best match the TZTD radial temperature distribution\\
3) fine-tune the MCFOST results to fit the $Spitzer$ IRS spectrum (primarily) and ALMA continuum mapping data (secondarily).}

We use the EaRTH Disk Model to analyze archival Spitzer IRS and ALMA mm-wave data for the nearby pre-MS star/disk system MP Mus as a demonstration of the model’s promise to yield integrated mineralogy and structural parameters for protoplanetary disks. The modeling indicates MP Mus has a dense inner disk (scale height to radius ratio ranging from \revision{0.036-0.037), effectively a puffed-up inner rim, which} is composed primarily of \revision{large} amorphous silicates (dominated by amorphous olivine) but includes a significant ($\sim$\revision{8}$\%$) crystalline silicate (\revision{small} forsterite \revision{and large enstatite}) component. The warm inner disk extends from $\sim$\revision{0.08} au to $\sim$\revision{0.1} au, inside of which is a small, thus far unobserved cavity; this disk then gives way to a \revision{gap or otherwise area of negligible dust up to a radius of 0.85 au, followed by an} outer disk that is \revision{more} forsterite-rich, with scale height to radius ratio ranging from \revision{0.1 to 0.12} at the outermost radii of the disk. The total disk dust mass of \revision{1.44}$\times10^{-4}~M_{\sun}$ and outer disk radius of $\sim$\revision{53} au used within our EaRTH Disk modeling are \revision{based on previous ALMA estimates and enhance the analysis}.

Challenges remain to improve the model; the mineralogical output can be improved to better capture finer features in different mid-infrared SEDs, so different realistic mineralogical distributions can be tested for improved fits and reduced degeneracy. The model is also currently \revision{somewhat} slow, requiring many radiative transfer models to be generated in sequence and requires further fine-tuning after the fact; if realistic and less complex mineralogy distributions can be produced that improve the mineralogical fit to the mid-IR spectrum, this would speed up the model, potentially removing the need for fine-tuning and/or permitting the use of more parameters in the fit, such as those outlined in Section \ref{res}. \revision{Furthermore, if other nonlinear variables can be removed and still produce a good fit, the model can be made to run faster; for example, by constraining the minimum temperature of the warm disk to the maximum temperature of the cool disk, while removing the ability to produce a gap or overlap, the model would not need to fit for the outer radius of the warm disk or the inner radius of the cool disk.} A similar result could be achieved with a complex distribution with a pre-generated grid of models; however, due to the number of parameters that can be altered, such a grid would be massive and would take a long time to generate, and then would be tied to a single set of minerals. \revisionnew{The mass distribution can also be adjusted in a reasonable way to improve the model, such as constraining the warm and cool disk mass percentages to within an approximate radius where dust thermal emission is predominantly in the mid-infrared while maintaining the overall mass of the disk. Furthermore, the linearly fitted constants corresponding to the dust composition in the TZTD model as well as the original TLTD model contain factors corresponding to the fitted disk parameters $T_{max}$ and $q$, as well as $r_{in}$ \citep{2009ApJ...695.1024J}, which were disregarded as part of the constant since they do not affect the non-linear parameters; however, they can affect the linearly fitted dust component constants, which, while not affecting the composition, can affect the relative mass percentages of the warm and cool disk.}

The Mid-Infrared Instrument (MIRI) Medium Resolution Spectrometer (MRS) aboard the James Webb Space Telescope (JWST) \citep{2015PASP..127..646W} will soon produce spectra at sensitivity and spectral resolution far exceeding those of $Spitzer$/IRS. The EaRTH Disk model methodology described here should offer a particularly effective means to analyze JWST/MRS spectroscopy of dusty disks, enabling optimal extraction of the information that is inherent in such data concerning both disk composition and structure.

%% file: opacs.tex
\section{Mineral Opacities}\label{opacsec}
\subsection{Amorphous Grains}\label{amorph}
The amorphous materials considered in this model consist of olivine, pyroxene, and a material with intermediate stoichiometry between the two, which \citet{2016ApJ...830...71F} dubs ``polivene"; this term is used in this article to refer to this mineral. We assume a density of 3.3 $g~cm^{-3}$ for all amorphous minerals, in line with the study by \citet{2006ApJ...645..395S}. Amorphous carbon is also considered, but is detailed later.

\begin{figure*}[htb!]
    \gridline{\fig{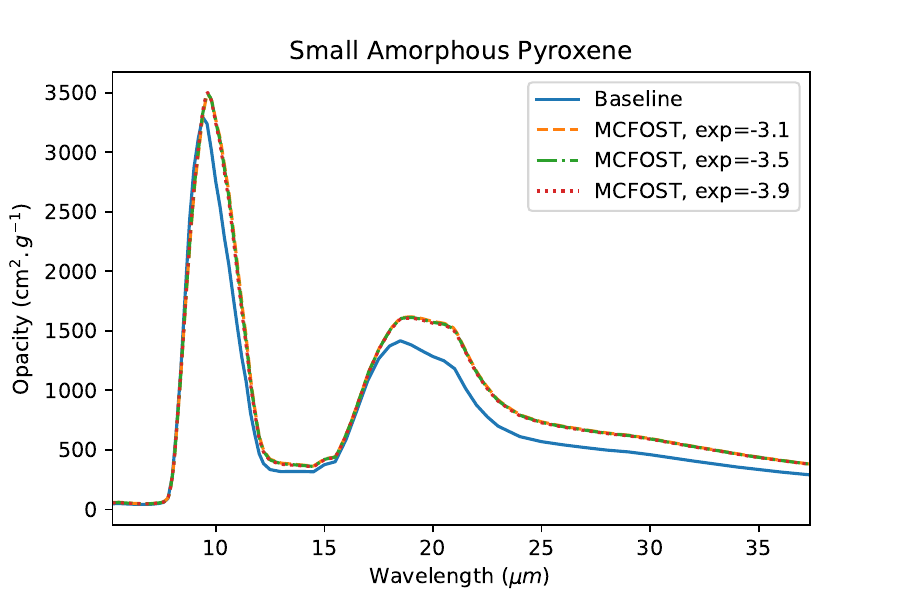}{0.3\textwidth}{}
              \fig{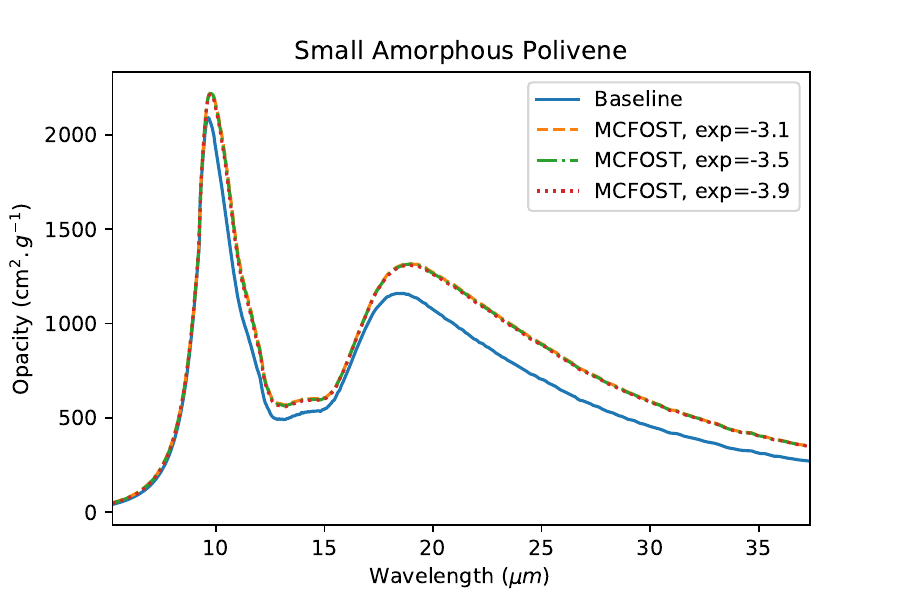}{0.3\textwidth}{}
              \fig{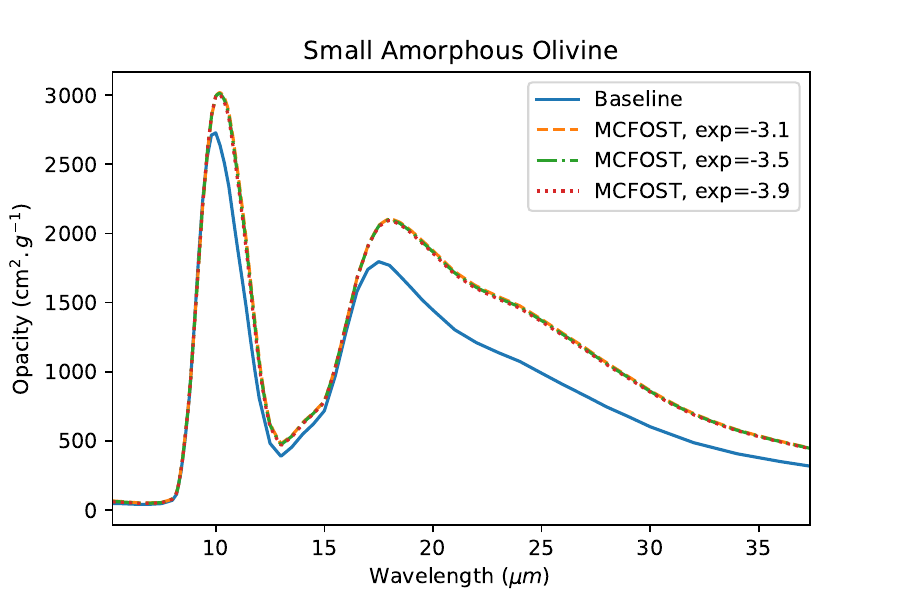}{0.3\textwidth}{}
              }
    \gridline{\fig{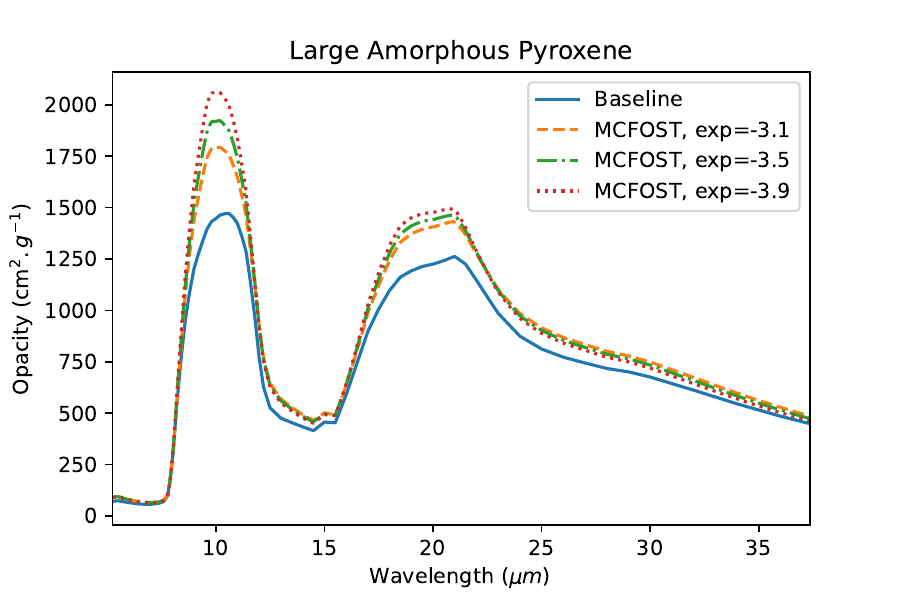}{0.3\textwidth}{}
              \fig{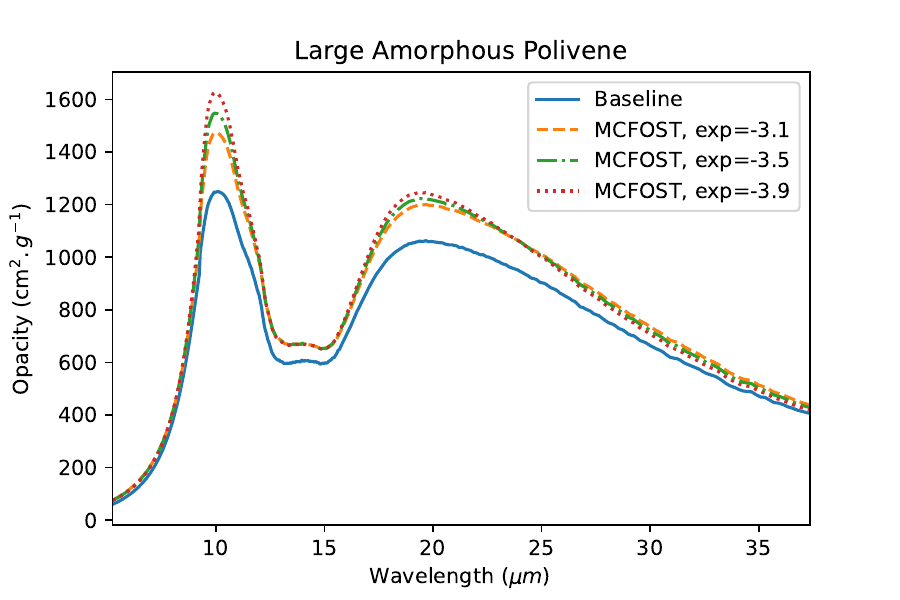}{0.3\textwidth}{}
              \fig{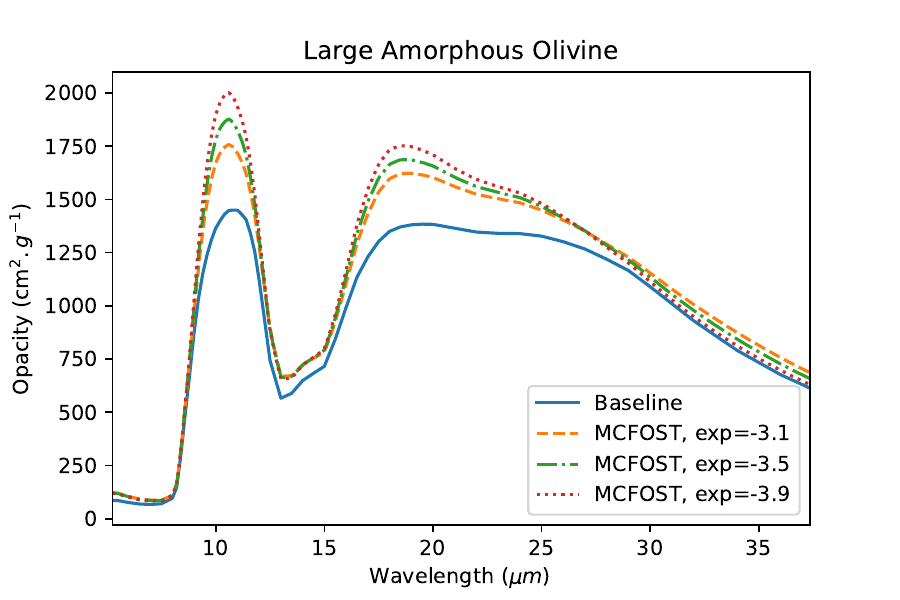}{0.3\textwidth}{}
              }
    \caption{Comparison between simulated opacities of amorphous silicates; \revision{Solid} curves: Baseline opacity functions by \citet{2016ApJ...830...71F}, \revision{Non-solid} curves: New opacity functions for use in the EaRTH Disk Model\revision{, varied by dust size power law exponent}; Top: Small (sub-$\micron$ sizes), Bottom: Large ($\micron$ sizes); Left: \revision{Pyroxene}; Middle: \revision{Polivene}; Right: Olivine.}
    \label{amorphcompare}
\end{figure*}

The baseline small amorphous mineral opacities \revision{from} the two-temperature model, received in a private communication with Shane Fogerty \revision{and outlined in Table \ref{2ttable}}, were calculated assuming CDE2 \citep{2001A&A...378..228F}; however, this distribution is not programmed into MCFOST, so it is necessary to approximate new opacities for the EaRTH Disk Model using \revision{DHS}. We use the same optical constants; under the assumption that these small amorphous minerals are sub-$\micron$, we use these optical constants and grain size range 0.01-1 $\micron$ with \revision{DHS} calculations to calculate the new opacities to use, as noted in Table \ref{mcfost_table}. Figure \ref{amorphcompare} demonstrates a comparison between the baseline small amorphous mineral opacities and the new opacities for use in the EaRTH Disk Model given different dust size power law exponents; in each case, the match between opacity calculations is very close up to the lower typical silicate peak at about 20 $\micron$, at which point the \revision{DHS} estimation becomes slightly larger. \revision{Furthermore, there is little variance between the size distributions due to the comparatively limited grain size range.}

In the baseline opacity set, calculations of the large amorphous mineral opacities are done using Mie theory, assuming a grain size of 5 $\micron$ and a porosity of 60\% using Bruggeman EMT, and by scaling the outputted absorption cross-section and dividing by the mass of a porous dust grain (see Equation \ref{opaceq} below), which is the mass of a non-porous dust grain scaled by $\rho_0(a/0.1 \micron)^{D-3}$ \citep{1998ApJ...496..971L}; D=2.766 is used by \citet{2006ApJ...645..395S}. \revisionnew{While} MCFOST can simulate grain porosity, \revisionnew{we do not include porosity within our DHS-modeled opacities.} While the general shapes of the distribution and distinguishing features are well preserved, \revision{the values of the DHS-modeled opacities are larger than those of the Mie theory-modeled opacities}; this is likely because the baseline opacities for the large amorphous minerals assume a single size, while the new opacities developed for the EaRTH Disk Model cover a range of sizes according to the dust size power law, meaning they are likely closer to reality. \revision{This power law exponent varies the large amorphous opacity functions, particularly at the 10- and 20-$\micron$ features. The more negative the dust size power law exponent is, the more prominent the features are; this is because more negative power law exponents indicate the presence of more small grains, which have opacities with more prominent silicate features, like those seen with the small amorphous grain opacities.}

\subsection{Crystalline Grains}\label{crystal}
In the case of the crystalline minerals, only the baseline opacities were given from the original sources; given how opacities are calculated, it is very difficult if not impossible to calculate optical constants from opacity functions alone. It was therefore necessary to retrieve substitute optical constants and derive opacities from those on the basis of \revision{DHS}. However, due to the crystalline nature of these minerals, 3 sets of optical constants were needed for each mineral, one for each crystallographic axis. This raises the question of how to implement crystalline minerals into MCFOST. \citet{2001A&A...378..228F} highlight two formulas, one that demonstrates how the extinction cross-section of a dust grain is related to opacity

\begin{equation}
\label{opaceq}
\kappa=\frac{C_{ext}}{V\rho},
\end{equation}
and how the cross-section of a crystalline dust grain is approximately the average of the cross-sections of the three crystallographic axes

\begin{equation}
\label{avg_c}
C_{ext}=\frac{1}{3}(C_{ext}(\epsilon_x)+C_{ext}(\epsilon_y)+C_{ext}(\epsilon_z)),
\end{equation}
where $\epsilon$ indicates the dielectric functions of the mineral along each of the crystallographic axes, as the cross-section is dependent on these dielectric functions. These two equations indicate that the opacity of a crystalline dust grain can be similarly calculated by the average of the opacities along the crystallographic axes; dividing both sides of Equation \ref{avg_c} by the mass of a dust grain ($V\rho$), assuming densities of \revision{3.26} $g~cm^{-3}$ for enstatite \citep{2003A&A...401...57J}, 3.23 $g~cm^{-3}$ for forsterite \citep{2006A&A...451..357S}, and 2.21 $g~cm^{-3}$ for silica \citep{2000A&A...364..282F}, yields

\begin{equation}
\label{avgopac}
\kappa=\frac{1}{3}(\kappa_x+\kappa_y+\kappa_z).
\end{equation}

\begin{figure*}[htb!]
    \gridline{\fig{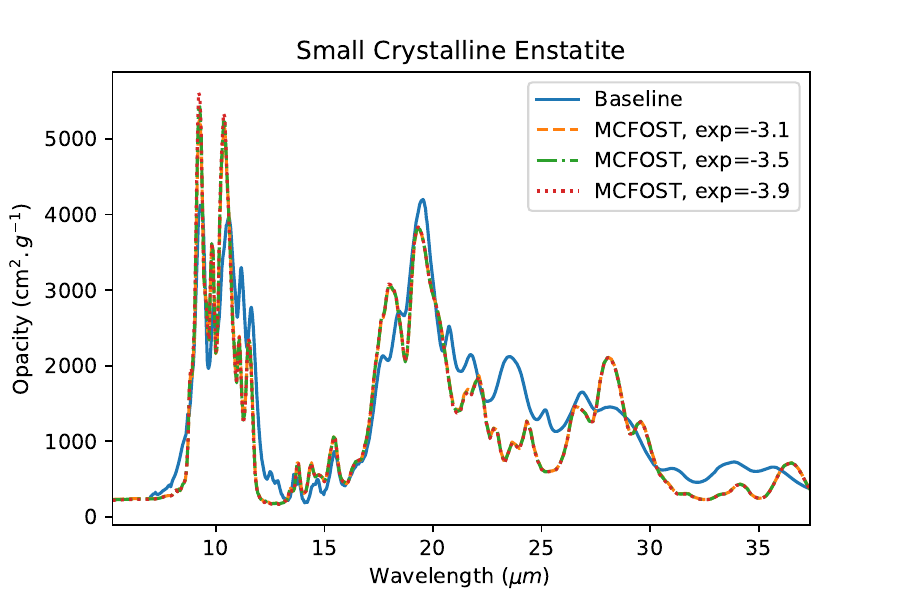}{0.3\textwidth}{}
              \fig{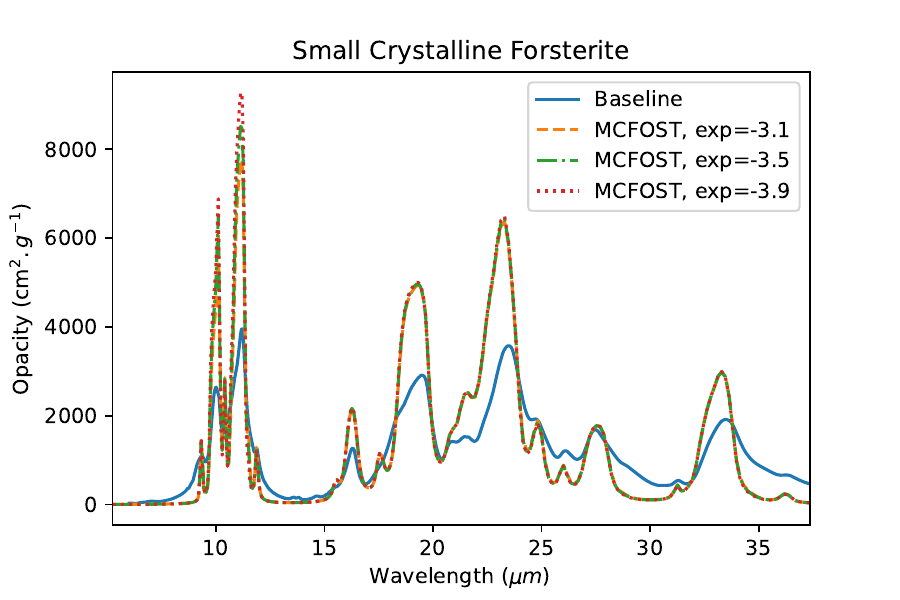}{0.3\textwidth}{}
              \fig{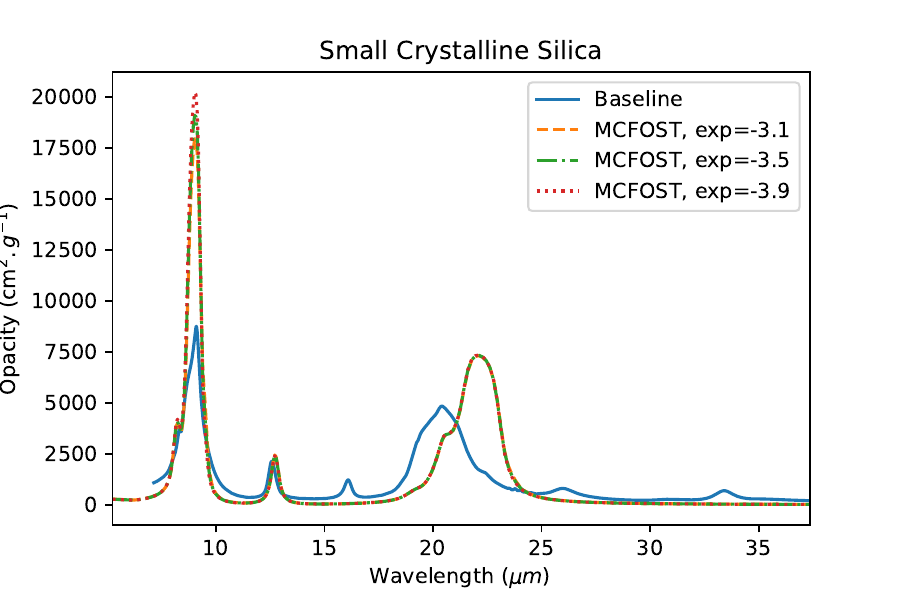}{0.3\textwidth}{}
              }
    \gridline{\fig{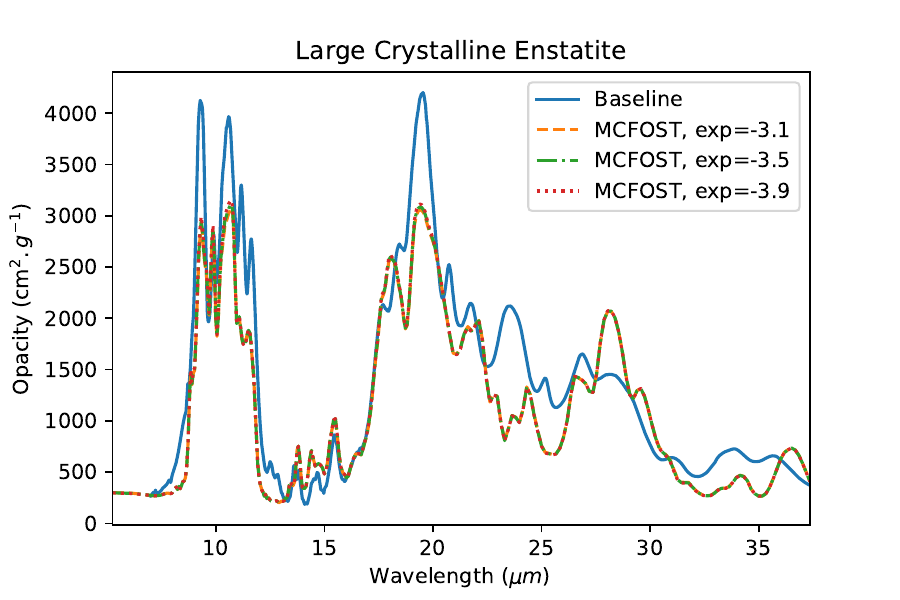}{0.3\textwidth}{}
              \fig{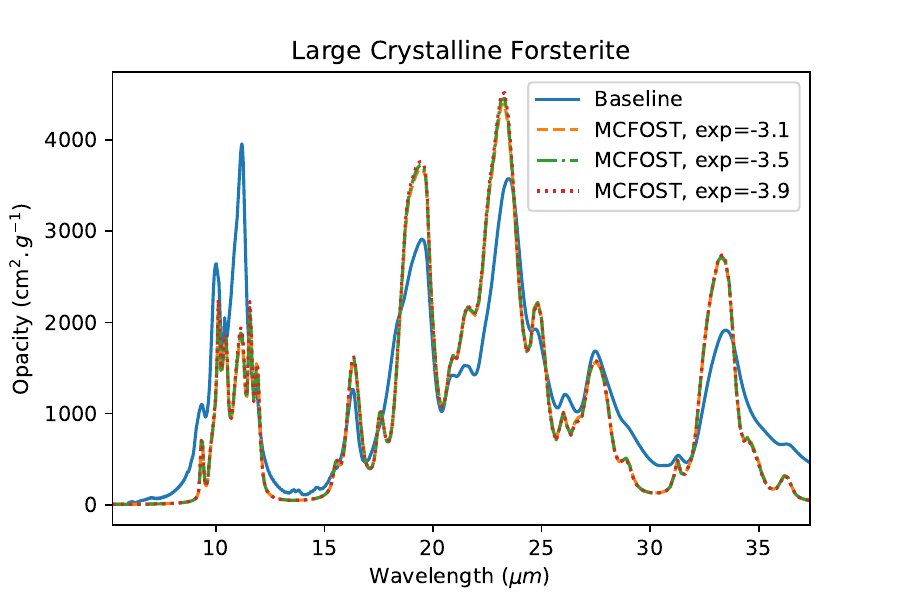}{0.3\textwidth}{}
              \fig{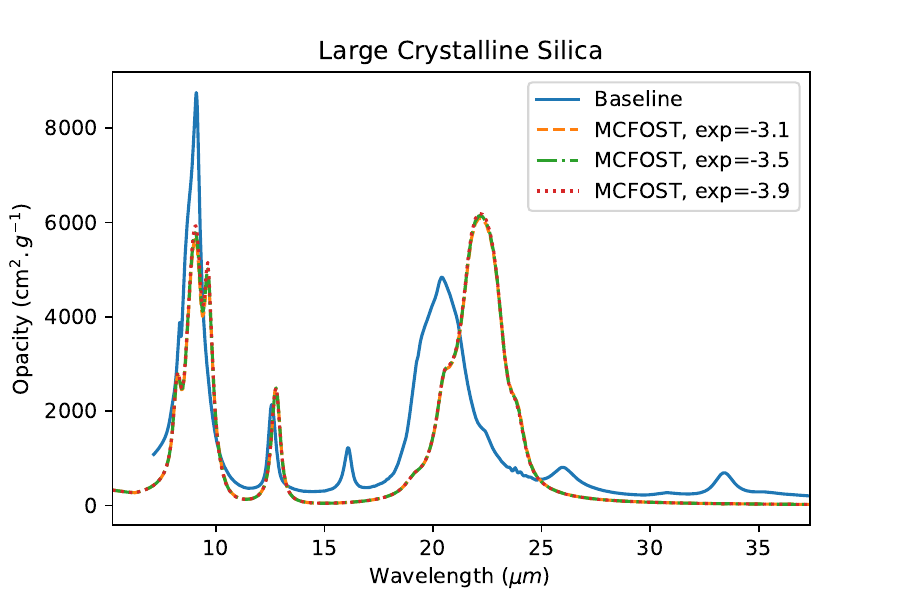}{0.3\textwidth}{}
              }
    \caption{Direct comparisons between opacities of crystalline minerals cited in Table \ref{2ttable} and those calculated using sets of optical constants corresponding to crystallographic axes cited in Table \ref{mcfost_table} assuming \revision{DHS} across a range of assumed sizes shared between the axes; \revision{Solid} curves: Baseline opacity functions by \citet{2016ApJ...830...71F}, \revision{Non-solid} curves: New opacity functions for use in the EaRTH Disk Model\revision{, varied by dust size power law exponent; Top: Small (sub-$\micron$ sizes), Bottom: Large ($\micron$ sizes), though in the baseline set, only one opacity was used for each crystalline mineral, so that one is used for comparison in both rows;} Left: Enstatite, Middle: Forsterite, Right: Silica.}
    \label{crystalcompare}
\end{figure*}

Therefore, we converted each set of individual constants that we retrieved to opacity using the implementation of \revision{DHS} in MCFOST assuming a single size \revision{range} for all 3 crystallographic axes and \revision{calculating} the average opacity of these axes. \revision{As with amorphous materials, we permit small and large crystalline size ranges; small grains are assumed to be sub-micron (i.e. 0.01-1 $\micron$). Large grains are assumed to range from 1-2 $\micron$; we select this different upper bound as \citet{2010ApJ...721..431J} found no evidence for the presence of crystalline grains larger than that within the dusty disks they studied.} We reference the sets of constants selected in Table \ref{mcfost_table} alongside the size ranges used for the DHS approximation and compare the resulting opacities with the opacities from Table \ref{2ttable} graphically in Figure \ref{crystalcompare}. While the crystalline opacities \revision{do not} line up as well as the amorphous opacities, the significant features of each spectrum are well represented. This allows the results of the \revision{empirical} model to be implemented into the EaRTH Disk Model for crystalline as well as amorphous grains.

We implement the consideration of averaging opacities along crystallographic axes to find the opacity of a crystalline mineral into MCFOST accordingly. If a crystalline mineral is detected in a disk spectrum, we divide the relative amount of that mineral in that portion of the disk (i.e. the percentage) into thirds; we apply this resulting value to the set of optical constants for each crystallographic axis, and all three sets are included in the mineralogy parameters with sizes set accordingly.

\subsection{Amorphous Carbon}\label{carbon}
The \revision{empirical} model also fits \revision{results representing continuum emission from the disk}; this works as a bias term to improve the fit, but it also represents very large grains and amorphous carbon grains. Experimental results, such as those detailed in this article, indicate that these grains of featureless \revision{mid-IR} emission can make up \revision{a significant portion} of the mass of the dust in a protoplanetary disk. This emission can suppress all but the most prominent features in the optically thin atmosphere of the disk, even though such large dust tends to settle into the midplane or \revision{rim} of the disk, and cannot be ignored. Therefore, as these terms represent very large grains and amorphous grains, we represent this bias in MCFOST as very large amorphous carbon grains, \revision{modeled using Mie theory \citep{1983uaz..rept.....B}} with a size range of \revision{5} mm to 30 mm and a porous vacuum volume fraction of 60\%. This generated opacity has very weak wavelength dependence in the infrared regime, and thus remains at an approximate constant value of \revision{1.025} $cm^{2}~g^{-1}$. This bias term in the \revision{empirical} model is then appropriately normalized upon implementation in the MCFOST fit.

{\catcode`\&=11
\gdef\citea{\cite{2003A&A...408..193J}}}
{\catcode`\&=11
\gdef\citeb{\cite{1995A&A...300..503D}}}
{\catcode`\&=11
\gdef\citec{\cite{1994A&A...292..641J}}}
{\catcode`\&=11
\gdef\cited{\cite{2002A&A...391..267C}}}
{\catcode`\&=11
\gdef\citee{\cite{1998A&A...339..904J}}}
{\catcode`\&=11
\gdef\citef{\cite{2000A&A...364..282F}}}
{\catcode`\&=11
\gdef\citeg{\citep{2001A&A...378..228F}}}
{\catcode`\&=11
\gdef\citeh{\cite{2006A&A...451..357S}}}
{\catcode`\&=11
\gdef\citei{\cite{2013A&A...553A..81Z}}}

\begin{deluxetable*}{lcc}[htb!]
\deluxetablecaption{Baseline opacities, mostly the same as those from \citet{2016ApJ...830...71F}.\label{2ttable}}
\tablewidth{0pt}
\tablehead{
\colhead{Name} & \colhead{Description} & \colhead{Reference}
}
\startdata
Small Amorphous Pyroxene & Optical constants for amorphous pyroxene of cosmic composition$^a$ & 1\\
Large Amorphous Pyroxene & Optical constants for amorphous pyroxene of cosmic composition$^b$ & 1\\
Small Amorphous Polivene & Optical constants for amorphous silicate of Mg$_{1.5}$SiO$_{3.5}$ composition$^a$ & 2 \\
Large Amorphous Polivene & Optical constants for amorphous silicate of Mg$_{1.5}$SiO$_{3.5}$ composition$^b$ & 2\\
Small Amorphous Olivine & Optical constants for amorphous olivine MgFeSiO$_4$ composition$^a$ & 3\\
Large Amorphous Olivine & Optical constants for amorphous olivine MgFeSiO$_4$ composition$^b$ & 3\\
Enstatite & Opacity for clinoenstatite Mg$_{0.9}$Fe$_{0.1}$SiO$_3$ & 4\\
Forsterite & Opacity for forsterite Mg$_2$SiO$_4$ & 5\\
Silica & Opacity for silica SiO$_2$ annealed at 1220 K for 5 hours & 6\\
\enddata
\tablecomments{(a) Assuming CDE2 \citeg; (b) Using the Bruggeman EMT and Mie theory \citep{1983uaz..rept.....B} with a vacuum volume fraction of f = 0.6 for porous spherical grains of radius 5 $\micron$}
\tablerefs{(1) \citec; (2) \citea; (3) \citeb; (4) \cited; (5) \citee; (6) \citef}
\end{deluxetable*}

\begin{longrotatetable}
\begin{deluxetable*}{lccc}
\deluxetablecaption{New minerals and corresponding optical constants used to calculate opacities assuming \revision{DHS}.\label{mcfost_table}}
\tablehead{
\colhead{Name} & \colhead{Description of Optical Constants} & \colhead{Assumed Size ($\micron$)} & \colhead{Reference}
}
\startdata
Small Amorphous Pyroxene & Amorphous pyroxene of cosmic composition & 0.01-1 & 1\\
Large Amorphous Pyroxene & Amorphous pyroxene of cosmic composition & \revision{1-5} & 1\\
Small Amorphous Polivene & Amorphous silicate of Mg\ensuremath{_{1.5}}SiO\ensuremath{_{3.5}} composition & 0.01-1 & 2\\
Large Amorphous Polivene & Amorphous silicate of Mg\ensuremath{_{1.5}}SiO\ensuremath{_{3.5}} composition & \revision{1-5} & 2\\
Small Amorphous Olivine & Amorphous olivine MgFeSiO\ensuremath{_{4}} composition & 0.01-1 & 3\\
Large Amorphous Olivine & Amorphous olivine MgFeSiO\ensuremath{_{4}} composition & \revision{1-5} & 3\\
\revision{Small Crystalline Enstatite} & Natural clinoenstatite MgSiO\ensuremath{_{3}} along crystallographic axes (p,s1,s2) & 0.01-1 & 4\\
\revision{Large Crystalline Enstatite} & Natural clinoenstatite MgSiO\ensuremath{_{3}} along crystallographic axes (p,s1,s2) & \revision{1-2} & 4\\
\revision{Small Crystalline Forsterite} & Forsterite Mg\ensuremath{_{2}}SiO\ensuremath{_{4}} along crystallographic axes (1u,2u,3u) & 0.01-1 & 5\\
\revision{Large Crystalline Forsterite} & Forsterite Mg\ensuremath{_{2}}SiO\ensuremath{_{4}} along crystallographic axes (1u,2u,3u) & \revision{1-2} & 5\\
\revision{Small Crystalline Silica} & \ensuremath{\alpha}-quartz SiO\ensuremath{_{2}} at 955 K along crystallographic axes (\ensuremath{E_{para},E_{perp1},E_{perp2}}) & 0.01-1 & 6\\
\revision{Large Crystalline Silica} & \ensuremath{\alpha}-quartz SiO\ensuremath{_{2}} at 955 K along crystallographic axes (\ensuremath{E_{para},E_{perp1},E_{perp2}}) & \revision{1-2} & 6\\
\enddata
\tablerefs{(1) \citec; (2) \citea; (3) \citeb; (4) \citee; (5) \citeh; (6) \citei}
\end{deluxetable*}
\end{longrotatetable}

\subsection{Composition Uncertainty Analysis}\label{uncert}
\revision{We derive uncertainties for the TZTD fitted parameters based on a process described by \citet{2009ApJS..182..477S}, in which the Taylor expansion of the reduced $\chi^2$ formula is used to estimate the shift in parameter value that produces a unity increase in the reduced $\chi^2$ metric. This is based on findings by \citet{1976ApJ...210..642A} that uncertainties based on increments of $\chi^2$ are similar to those from Monte Carlo simulations; \citet{2009ApJS..182..477S} indicates that with a limited amount of constraints, these uncertainty estimations are more conservative than those from Monte Carlo-based error analyses, quickly returning greater than one-sigma uncertainty values. We tested the applicability of this assertion to our TZTD model by comparing uncertainties from a simple Monte Carlo analysis using the process outlined by \citet{2005A&A...437..189V}. 1000 spectra are generated by adding normally distributed noise with standard deviation corresponding to the uncertainty of each spectrum data point, and each are fitted using the TZTD model; the means of each of these results are used as the best fit parameters, and the standard deviations of the parameters are taken as the uncertainties. We compared the values of the Monte Carlo-based uncertainties against the $\Delta\chi^2$-based uncertainties (in the case of the dust component values, we compared the raw constant values from Equation \ref{mod_flux} rather than percentages) and found that in all but three cases, the $\Delta\chi^2$-based uncertainties were greater than the Monte Carlo-based uncertainties, confirming the effectiveness of conservative estimation of uncertainty based on $\Delta\chi^2$; we therefore recommend conservative $\Delta\chi^2$-based uncertainty analysis when using complex mineralogical models such as TZTD. Beyond being faster and more conservative than Monte Carlo-based analyses, certain dust components end up excluded from every spectral fit in the Monte Carlo simulation, returning a standard deviation of zero for the parameter; with $\Delta\chi^2$-based uncertainty measurements, every parameter, including absent dust components, returns an uncertainty measurement greater than zero, as noted in Table \ref{comp_table} and by \citet{2009ApJS..182..477S} in their study. The three parameters in which the Monte Carlo-based uncertainty was greater were $q_c$, the contribution of the continuum in the cool disk, and the contribution of small amorphous pyroxene in the cool disk; in the case of these three parameters, there is potential degeneracy within the fit, particularly in the case of the cool dust components, so care must be taken when estimating the uncertainty of parameters corresponding to cooler dust, as detailed further below.}

\subsection{Degeneracy in Model Components}\label{degeneracy}

\citet{2009ApJS..182..477S} outlines a process for modeling the degeneracy between modeled dust components in the two-temperature model, a process itself derived from \citet{1992nrfa.book.....P}. For the \revision{parameters} corresponding to the best reduced $\chi^2$ fit, another $\chi^2$ fit is performed, but all resulting components are kept, including those with negative mass weights. Next, based on these components, the covariance matrix $C$ is calculated, with the elements calculated using the formula

\begin{equation}
\label{corrmat}
C_{i,j}=[\sum_{k}{\frac{\kappa_{i,j}({\lambda}_{k})\int_{T_{warm,min}}^{T_{warm,max}}\frac{2\pi}{d^2}B_{\nu}({\lambda}_{k},T)T^{\frac{2-q_{warm}}{q_{warm}}}dT\kappa_{i,j}({\lambda}_{k})\int_{T_{cool,min}}^{T_{cool,max}}\frac{2\pi}{d^2}B_{\nu}({\lambda}_{k},T)T^{\frac{2-q_{cool}}{q_{cool}}}dT}{{\Delta}F_{\nu}({\lambda}_{k})^{irs}}}]^{-1},
\end{equation}	
where $i$ and $j$ represent the row and column corresponding to different components, respectively, and the other variables represent those from Equations \ref{chisq} and \ref{mod_flux} in Section \ref{models}. The off-diagonal elements of the matrix represent the covariances between dust components corresponding to given rows and columns, and the diagonal elements represent the variances of the dust components. The correlation matrix $R$ can be calculated from the covariance matrix using the equation

\begin{equation}
\label{correq}
R_{i,j}=\frac{C_{i,j}}{\sqrt{C_{i,i}C_{j,j}}}.
\end{equation}

The diagonal elements of $R$ are all equal to 1, and the off-diagonal elements correspond to the correlation coefficient $r$ of each dust component. Values of $r$ approaching 1 indicate highly non-degenerate components, with both components required for a good fit; $r$ around 0 indicate uncorrelated components; negative values of $r$ indicate degenerate components, with more negative coefficients corresponding to higher degeneracy, indicating the degree to which the components could replace each other in the fit for similar goodness of fit \citep{2009ASPC..414...77W}.

\begin{figure*}[htb!]
    \gridline{\fig{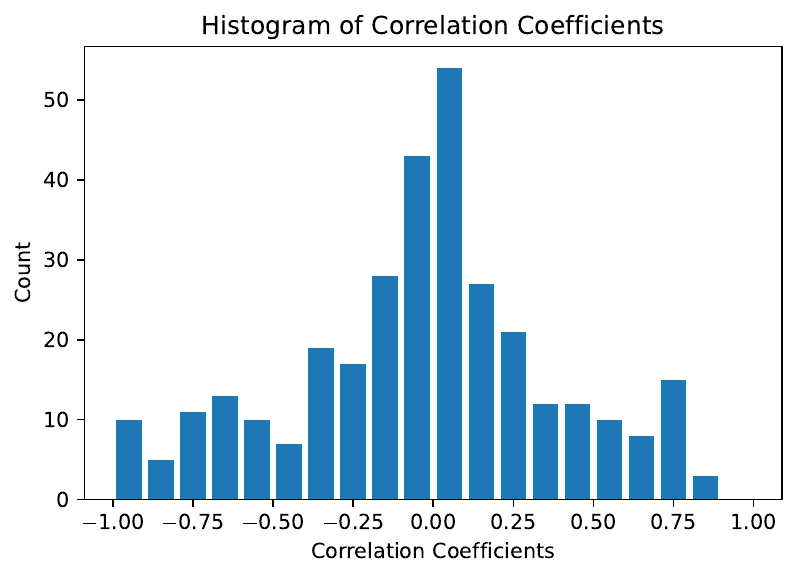}{\textwidth}{}}
    \caption{Distribution of correlation coefficients of component pairs; bins have width 0.1.}
    \label{correlation_histogram}
\end{figure*}

Including the dust continuum functions and all minerals in the warm and cool disks, there are \revision{26} components in the fit of MP Mus and \revision{325} total pairs. Table \ref{degen} lists the most degenerate dust component pairs, the \revision{22} with $r$ less than -0.75, and Figure \ref{correlation_histogram} shows a histogram of the distribution of correlation coefficients. Based on this histogram, most pairs are uncorrelated or negligibly correlated. However, there are several degenerate pairs.

\begin{deluxetable*}{lcc}[htb!]
\deluxetablecaption{Highly degenerate dust component pairs ($r<-0.75$).\label{degen}}
\tabletypesize{\scriptsize}
\tablewidth{0pt}
\tablehead{
\colhead{Dust Component 1} & \colhead{Dust Component 2} & \colhead{Correlation Coefficient}
}
\startdata
Warm Small Amorphous Polivene & Warm Large Amorphous Polivene & -0.991\\
Cool Small Amorphous Polivene & Cool Large Amorphous Polivene & -0.982\\
Cool Small Crystalline Enstatite & Cool Large Crystalline Enstatite & -0.972\\
Warm Small Amorphous Pyroxene & Warm Large Amorphous Pyroxene & -0.949\\
Cool Small Amorphous Olivine & Cool Large Amorphous Olivine & -0.936\\
Cool Small Crystalline Silica & Cool Large Crystalline Silica & -0.932\\
Cool Small Amorphous Pyroxene & Cool Large Amorphous Pyroxene & -0.930\\
Cool Small Crystalline Forsterite & Cool Large Crystalline Forsterite & -0.913\\
Cool Small Amorphous Olivine & Cool Small Amorphous Polivene & -0.904\\
Warm Small Crystalline Silica & Warm Large Crystalline Silica & -0.901\\
Warm Small Amorphous Olivine & Warm Large Amorphous Olivine & -0.888\\
Warm Small Amorphous Polivene & Cool Small Amorphous Polivene & -0.823\\
Warm Large Amorphous Olivine & Cool Small Amorphous Polivene & -0.810\\
Warm Small Crystalline Forsterite & Cool Small Crystalline Forsterite & -0.804\\
Warm Small Crystalline Enstatite & Cool Small Crystalline Enstatite & -0.803\\
Warm Large Amorphous Polivene & Cool Large Amorphous Polivene & -0.791\\
Warm Small Amorphous Pyroxene & Warm Small Amorphous Polivene & -0.788\\
Warm Small Amorphous Pyroxene & Warm Small Crystalline Silica & -0.784\\
Warm Small Amorphous Olivine & Warm Small Amorphous Polivene & -0.783\\
Warm Large Amorphous Pyroxene & Warm Large Crystalline Silica & -0.778\\
Warm Small Crystalline Enstatite & Warm Large Crystalline Enstatite & -0.772\\
Warm Small Amorphous Polivene & Warm Large Crystalline Silica & -0.767\\
\enddata
\tablecomments{Abbreviations of materials shown in composition plots in Figure \ref{mpmus_dust_plots} and Table \ref{comp_table}}
\end{deluxetable*}

In general, the cooler components are more highly degenerate. This is because cooler temperatures have a tendency to suppress more prominent silicate features, making the component functions similar. As \citet{2009ApJS..182..477S} note, functions corresponding to amorphous silicates, and especially at cooler temperatures, tend to be most degenerate, as these results confirm; amorphous silicate functions, particularly at cooler temperatures, have less unique features to distinguish them well from others in the infrared regime. More unique features distinguish components more easily and make them less degenerate, which is why degeneracy is less common and less prominent with crystalline minerals; this is also a reason why mineralogy is calculated over a wider mid-infrared spectral range,\revision{ the full $Spitzer$ IRS range}, which incorporates multiple silicate features \citep{2016ApJ...830...71F}. Still, degeneracies can occur if overall patterns and trends are somewhat similar, as with amorphous silicates and dust continua, or if the components share similar features, such as in instances where warm and cool components \revision{or small and large grains} of the same mineral are found to be degenerate. Furthermore, polivene is degenerate with pyroxene and olivine in several pairs, as polivene is stoichiometrically intermediate to pyroxene and olivine\revision{; while most crystalline silicates on this list are degenerate between different temperatures and sizes, silica is also somewhat degenerate with some amorphous materials due to the fact that it has relatively fewer features than the other crystalline silicates (see Figure \ref{crystalcompare})}. Overall, while relatively few pairs of components are found to be highly degenerate, this degeneracy must be kept in mind during mineralogical fits.

\section{Spitzer IRS Preprocessing}\label{process}
We preprocess the low-resolution $Spitzer$ IRS spectra retrieved from CASSIS before analysis. This preprocessing involved mispointing correction at 14 $\micron$ \revision{and} interstellar extinction correction.

\subsection{Extinction Correction}\label{extcorr}
We deredden the spectrum via an extinction correction obtained from the stellar $J$ and $K_s$ \citep[2MASS;][]{2006AJ....131.1163S} magnitudes, assuming an intrinsic $J-K$ color appropriate for the spectral type of the star being processed \revision{\citep[see Table 6 of][for example]{2013ApJS..208....9P}}. Utilizing extinction curves from \citet{2009ApJ...693L..81M}, we find an approximate extinction ratio for these bands to be $A_J/A_{K_s}=2.585$. This allows for a value of $R_{K_s}$ to be derived:

\begin{equation}
\label{exteq}
R_{K_s}=\frac{1}{\frac{A_J}{A_{K_s}}-1}=\frac{1}{2.585-1}=0.63
\end{equation}

\begin{figure*}[htb!]
    \gridline{\fig{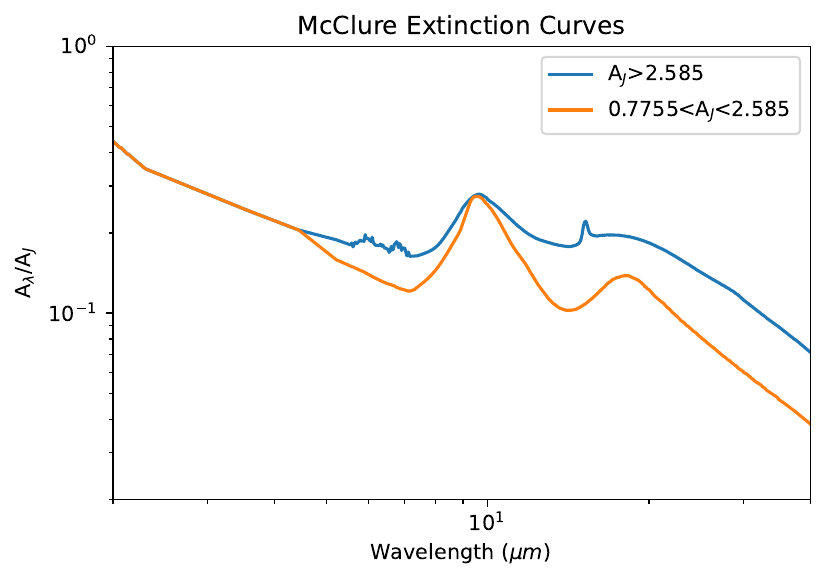}{\textwidth}{}}
    \caption{Curves used to correct extinction in observed spectra based on those by \citet{2009ApJ...693L..81M} and \citet{2001ApJ...548..296W} adjusted to demonstrate their relation to extinction at the $J$ band\revision{; note that these curves are similar in shape to those from Figure 1 of \citet{2009ApJ...693L..81M}}.}
    \label{ext_curve}
\end{figure*}

This yields the extinction formula $A_{K_s}=0.63E(J-K)$; this allows for a direct analysis of the necessary extinction curve, as this extinction law is directly based on the curve found by~\citet{2009ApJ...693L..81M}. If $A_{K_s}<0.3$, then no extinction correction is performed at all, rather than utilizing a different curve; this is done because such a low extinction value at the $K_s$ band implies negligible extinction in the \revision{$Spitzer$ IRS wavelength} region, so correction is not necessary. If $A_{K_s}~\underline{>}~0.3$, the appropriate extinction curve from~\citet{2009ApJ...693L..81M} is used for wavelengths 5-20 $\micron$ appended with the curve from~\citet{2001ApJ...548..296W} noted in the former paper for higher wavelengths. These extinction curves are demonstrated in Figure \ref{ext_curve}. \revision{Note that the extinction law corresponding to $A_{J}~\underline{>}~2.585$ represents a jump in extinction; \citet{2009ApJ...693L..81M} indicates that this is consistent with IRAC photometry analysis of molecular clouds by \citet{2009ApJ...690..496C}, who find that extinction law curves such as these increase and flatten for higher values of extinction, soon approaching the $R_V=5.5$ extinction curves derived by \citet{2001ApJ...548..296W}, which were derived on the assumption of larger maximum grain sizes.}

\subsection{Mispointing Correction}\label{misp}
In addition to extinction correction, mispointing correction is necessary; for a very large data set, adjusting each by eye is infeasible. Therefore, a new strategy proposed and used is to interpolate both portions of the spectrum just past the 14 $\micron$ point, find the value of each at 14 $\micron$, then increase whichever portion of the spectrum is at a lower value at the points just greater and just less than 14 $\micron$ (typically the short-low (SL) spectrum) by the amount of the change in values such that they are approximately equal at 14 $\micron$; if neither is less than the other at both points, then no change is made.

%% file: main.bbl
\begin{thebibliography}{}
\expandafter\ifx\csname natexlab\endcsname\relax\def\natexlab#1{#1}\fi
\providecommand{\url}[1]{\href{#1}{#1}}
\providecommand{\dodoi}[1]{doi:~\href{http://doi.org/#1}{\nolinkurl{#1}}}
\providecommand{\doeprint}[1]{\href{http://ascl.net/#1}{\nolinkurl{http://ascl.net/#1}}}
\providecommand{\doarXiv}[1]{\href{https://arxiv.org/abs/#1}{\nolinkurl{https://arxiv.org/abs/#1}}}

\bibitem[{{Andrews}(2020)}]{2020ARA&A..58..483A}
{Andrews}, S.~M. 2020, \araa, 58, 483, \dodoi{10.1146/annurev-astro-031220-010302}

\bibitem[{{Arabhavi} {et~al.}(2022){Arabhavi}, {Woitke}, {Cazaux}, {Kamp}, {Rab}, \& {Thi}}]{2022A&A...666A.139A}
{Arabhavi}, A.~M., {Woitke}, P., {Cazaux}, S.~M., {et~al.} 2022, \aap, 666, A139, \dodoi{10.1051/0004-6361/202141825}

\bibitem[{{Argiroffi} {et~al.}(2007){Argiroffi}, {Maggio}, \& {Peres}}]{2007A&A...465L...5A}
{Argiroffi}, C., {Maggio}, A., \& {Peres}, G. 2007, \aap, 465, L5, \dodoi{10.1051/0004-6361:20067016}

\bibitem[{{Astropy Collaboration} {et~al.}(2013){Astropy Collaboration}, {Robitaille}, {Tollerud}, {Greenfield}, {Droettboom}, {Bray}, {Aldcroft}, {Davis}, {Ginsburg}, {Price-Whelan}, {Kerzendorf}, {Conley}, {Crighton}, {Barbary}, {Muna}, {Ferguson}, {Grollier}, {Parikh}, {Nair}, {Unther}, {Deil}, {Woillez}, {Conseil}, {Kramer}, {Turner}, {Singer}, {Fox}, {Weaver}, {Zabalza}, {Edwards}, {Azalee Bostroem}, {Burke}, {Casey}, {Crawford}, {Dencheva}, {Ely}, {Jenness}, {Labrie}, {Lim}, {Pierfederici}, {Pontzen}, {Ptak}, {Refsdal}, {Servillat}, \& {Streicher}}]{2013A&A...558A..33A}
{Astropy Collaboration}, {Robitaille}, T.~P., {Tollerud}, E.~J., {et~al.} 2013, \aap, 558, A33, \dodoi{10.1051/0004-6361/201322068}

\bibitem[{{Avenhaus} {et~al.}(2018){Avenhaus}, {Quanz}, {Garufi}, {Perez}, {Casassus}, {Pinte}, {Bertrang}, {Caceres}, {Benisty}, \& {Dominik}}]{2018ApJ...863...44A}
{Avenhaus}, H., {Quanz}, S.~P., {Garufi}, A., {et~al.} 2018, \apj, 863, 44, \dodoi{10.3847/1538-4357/aab846}

\bibitem[{{Avni}(1976)}]{1976ApJ...210..642A}
{Avni}, Y. 1976, \apj, 210, 642, \dodoi{10.1086/154870}

\bibitem[{{Beuzit} {et~al.}(2019){Beuzit}, {Vigan}, {Mouillet}, {Dohlen}, {Gratton}, {Boccaletti}, {Sauvage}, {Schmid}, {Langlois}, {Petit}, {Baruffolo}, {Feldt}, {Milli}, {Wahhaj}, {Abe}, {Anselmi}, {Antichi}, {Barette}, {Baudrand}, {Baudoz}, {Bazzon}, {Bernardi}, {Blanchard}, {Brast}, {Bruno}, {Buey}, {Carbillet}, {Carle}, {Cascone}, {Chapron}, {Charton}, {Chauvin}, {Claudi}, {Costille}, {De Caprio}, {de Boer}, {Delboulb{\'e}}, {Desidera}, {Dominik}, {Downing}, {Dupuis}, {Fabron}, {Fantinel}, {Farisato}, {Feautrier}, {Fedrigo}, {Fusco}, {Gigan}, {Ginski}, {Girard}, {Giro}, {Gisler}, {Gluck}, {Gry}, {Henning}, {Hubin}, {Hugot}, {Incorvaia}, {Jaquet}, {Kasper}, {Lagadec}, {Lagrange}, {Le Coroller}, {Le Mignant}, {Le Ruyet}, {Lessio}, {Lizon}, {Llored}, {Lundin}, {Madec}, {Magnard}, {Marteaud}, {Martinez}, {Maurel}, {M{\'e}nard}, {Mesa}, {M{\"o}ller-Nilsson}, {Moulin}, {Moutou}, {Orign{\'e}}, {Parisot}, {Pavlov}, {Perret}, {Pragt}, {Puget}, {Rabou}, {Ramos}, {Reess}, {Rigal}, {Rochat}, {Roelfsema}, {Rousset},
  {Roux}, {Saisse}, {Salasnich}, {Santambrogio}, {Scuderi}, {Segransan}, {Sevin}, {Siebenmorgen}, {Soenke}, {Stadler}, {Suarez}, {Tiph{\`e}ne}, {Turatto}, {Udry}, {Vakili}, {Waters}, {Weber}, {Wildi}, {Zins}, \& {Zurlo}}]{2019A&A...631A.155B}
{Beuzit}, J.~L., {Vigan}, A., {Mouillet}, D., {et~al.} 2019, \aap, 631, A155, \dodoi{10.1051/0004-6361/201935251}

\bibitem[{{Bohren} \& {Huffman}(1983)}]{1983uaz..rept.....B}
{Bohren}, C.~F., \& {Huffman}, D.~R. 1983, {Absorption and scattering of light by small particles}, Tech. rep.

\bibitem[{{Calvet} \& {D'Alessio}(2011)}]{2011ppcd.book...14C}
{Calvet}, N., \& {D'Alessio}, P. 2011, {Protoplanetary Disk Structure and Evolution}, ed. P.~J.~V. {Garcia}, 14--54

\bibitem[{{Carpenter} {et~al.}(2005){Carpenter}, {Wolf}, {Schreyer}, {Launhardt}, \& {Henning}}]{2005AJ....129.1049C}
{Carpenter}, J.~M., {Wolf}, S., {Schreyer}, K., {Launhardt}, R., \& {Henning}, T. 2005, \aj, 129, 1049, \dodoi{10.1086/427131}

\bibitem[{{Chapman} {et~al.}(2009){Chapman}, {Mundy}, {Lai}, \& {Evans}}]{2009ApJ...690..496C}
{Chapman}, N.~L., {Mundy}, L.~G., {Lai}, S.-P., \& {Evans}, Neal~J., I. 2009, \apj, 690, 496, \dodoi{10.1088/0004-637X/690/1/496}

\bibitem[{{Charbonneau}(1995)}]{1995ApJS..101..309C}
{Charbonneau}, P. 1995, \apjs, 101, 309, \dodoi{10.1086/192242}

\bibitem[{{Chiang} \& {Goldreich}(1997)}]{1997ApJ...490..368C}
{Chiang}, E.~I., \& {Goldreich}, P. 1997, \apj, 490, 368, \dodoi{10.1086/304869}

\bibitem[{{Chiang} {et~al.}(2001){Chiang}, {Joung}, {Creech-Eakman}, {Qi}, {Kessler}, {Blake}, \& {van Dishoeck}}]{2001ApJ...547.1077C}
{Chiang}, E.~I., {Joung}, M.~K., {Creech-Eakman}, M.~J., {et~al.} 2001, \apj, 547, 1077, \dodoi{10.1086/318427}

\bibitem[{{Chihara} {et~al.}(2002){Chihara}, {Koike}, {Tsuchiyama}, {Tachibana}, \& {Sakamoto}}]{2002A&A...391..267C}
{Chihara}, H., {Koike}, C., {Tsuchiyama}, A., {Tachibana}, S., \& {Sakamoto}, D. 2002, \aap, 391, 267, \dodoi{10.1051/0004-6361:20020791}

\bibitem[{{Cortes} {et~al.}(2009){Cortes}, {Meyer}, {Carpenter}, {Pascucci}, {Schneider}, {Wong}, \& {Hines}}]{2009ApJ...697.1305C}
{Cortes}, S.~R., {Meyer}, M.~R., {Carpenter}, J.~M., {et~al.} 2009, \apj, 697, 1305, \dodoi{10.1088/0004-637X/697/2/1305}

\bibitem[{{Dickson-Vandervelde} {et~al.}(2021){Dickson-Vandervelde}, {Wilson}, \& {Kastner}}]{2021AJ....161...87D}
{Dickson-Vandervelde}, D.~A., {Wilson}, E.~C., \& {Kastner}, J.~H. 2021, \aj, 161, 87, \dodoi{10.3847/1538-3881/abd0fd}

\bibitem[{{Dohlen} {et~al.}(2008){Dohlen}, {Langlois}, {Saisse}, {Hill}, {Origne}, {Jacquet}, {Fabron}, {Blanc}, {Llored}, {Carle}, {Moutou}, {Vigan}, {Boccaletti}, {Carbillet}, {Mouillet}, \& {Beuzit}}]{2008SPIE.7014E..3LD}
{Dohlen}, K., {Langlois}, M., {Saisse}, M., {et~al.} 2008, in Society of Photo-Optical Instrumentation Engineers (SPIE) Conference Series, Vol. 7014, Ground-based and Airborne Instrumentation for Astronomy II, ed. I.~S. {McLean} \& M.~M. {Casali}, 70143L, \dodoi{10.1117/12.789786}

\bibitem[{{Dorschner} {et~al.}(1995){Dorschner}, {Begemann}, {Henning}, {Jaeger}, \& {Mutschke}}]{1995A&A...300..503D}
{Dorschner}, J., {Begemann}, B., {Henning}, T., {Jaeger}, C., \& {Mutschke}, H. 1995, \aap, 300, 503

\bibitem[{{Dubrulle} {et~al.}(1995){Dubrulle}, {Morfill}, \& {Sterzik}}]{1995Icar..114..237D}
{Dubrulle}, B., {Morfill}, G., \& {Sterzik}, M. 1995, \icarus, 114, 237, \dodoi{10.1006/icar.1995.1058}

\bibitem[{{Dullemond} {et~al.}(2001){Dullemond}, {Dominik}, \& {Natta}}]{2001ApJ...560..957D}
{Dullemond}, C.~P., {Dominik}, C., \& {Natta}, A. 2001, \apj, 560, 957, \dodoi{10.1086/323057}

\bibitem[{{Fabian} {et~al.}(2001){Fabian}, {Henning}, {J{\"a}ger}, {Mutschke}, {Dorschner}, \& {Wehrhan}}]{2001A&A...378..228F}
{Fabian}, D., {Henning}, T., {J{\"a}ger}, C., {et~al.} 2001, \aap, 378, 228, \dodoi{10.1051/0004-6361:20011196}

\bibitem[{{Fabian} {et~al.}(2000){Fabian}, {J{\"a}ger}, {Henning}, {Dorschner}, \& {Mutschke}}]{2000A&A...364..282F}
{Fabian}, D., {J{\"a}ger}, C., {Henning}, T., {Dorschner}, J., \& {Mutschke}, H. 2000, \aap, 364, 282

\bibitem[{{Fogerty} {et~al.}(2016){Fogerty}, {Forrest}, {Watson}, {Sargent}, \& {Koch}}]{2016ApJ...830...71F}
{Fogerty}, S., {Forrest}, W., {Watson}, D.~M., {Sargent}, B.~A., \& {Koch}, I. 2016, \apj, 830, 71, \dodoi{10.3847/0004-637X/830/2/71}

\bibitem[{{Gaia Collaboration} {et~al.}(2016){Gaia Collaboration}, {Prusti}, {de Bruijne}, {Brown}, {Vallenari}, {Babusiaux}, {Bailer-Jones}, {Bastian}, {Biermann}, {Evans}, {Eyer}, {Jansen}, {Jordi}, {Klioner}, {Lammers}, {Lindegren}, {Luri}, {Mignard}, {Milligan}, {Panem}, {Poinsignon}, {Pourbaix}, {Randich}, {Sarri}, {Sartoretti}, {Siddiqui}, {Soubiran}, {Valette}, {van Leeuwen}, {Walton}, {Aerts}, {Arenou}, {Cropper}, {Drimmel}, {H{\o}g}, {Katz}, {Lattanzi}, {O'Mullane}, {Grebel}, {Holland}, {Huc}, {Passot}, {Bramante}, {Cacciari}, {Casta{\~n}eda}, {Chaoul}, {Cheek}, {De Angeli}, {Fabricius}, {Guerra}, {Hern{\'a}ndez}, {Jean-Antoine-Piccolo}, {Masana}, {Messineo}, {Mowlavi}, {Nienartowicz}, {Ord{\'o}{\~n}ez-Blanco}, {Panuzzo}, {Portell}, {Richards}, {Riello}, {Seabroke}, {Tanga}, {Th{\'e}venin}, {Torra}, {Els}, {Gracia-Abril}, {Comoretto}, {Garcia-Reinaldos}, {Lock}, {Mercier}, {Altmann}, {Andrae}, {Astraatmadja}, {Bellas-Velidis}, {Benson}, {Berthier}, {Blomme}, {Busso}, {Carry}, {Cellino}, {Clementini},
  {Cowell}, {Creevey}, {Cuypers}, {Davidson}, {De Ridder}, {de Torres}, {Delchambre}, {Dell'Oro}, {Ducourant}, {Fr{\'e}mat}, {Garc{\'\i}a-Torres}, {Gosset}, {Halbwachs}, {Hambly}, {Harrison}, {Hauser}, {Hestroffer}, {Hodgkin}, {Huckle}, {Hutton}, {Jasniewicz}, {Jordan}, {Kontizas}, {Korn}, {Lanzafame}, {Manteiga}, {Moitinho}, {Muinonen}, {Osinde}, {Pancino}, {Pauwels}, {Petit}, {Recio-Blanco}, {Robin}, {Sarro}, {Siopis}, {Smith}, {Smith}, {Sozzetti}, {Thuillot}, {van Reeven}, {Viala}, {Abbas}, {Abreu Aramburu}, {Accart}, {Aguado}, {Allan}, {Allasia}, {Altavilla}, {{\'A}lvarez}, {Alves}, {Anderson}, {Andrei}, {Anglada Varela}, {Antiche}, {Antoja}, {Ant{\'o}n}, {Arcay}, {Atzei}, {Ayache}, {Bach}, {Baker}, {Balaguer-N{\'u}{\~n}ez}, {Barache}, {Barata}, {Barbier}, {Barblan}, {Baroni}, {Barrado y Navascu{\'e}s}, {Barros}, {Barstow}, {Becciani}, {Bellazzini}, {Bellei}, {Bello Garc{\'\i}a}, {Belokurov}, {Bendjoya}, {Berihuete}, {Bianchi}, {Bienaym{\'e}}, {Billebaud}, {Blagorodnova}, {Blanco-Cuaresma}, {Boch},
  {Bombrun}, {Borrachero}, {Bouquillon}, {Bourda}, {Bouy}, {Bragaglia}, {Breddels}, {Brouillet}, {Br{\"u}semeister}, {Bucciarelli}, {Budnik}, {Burgess}, {Burgon}, {Burlacu}, {Busonero}, {Buzzi}, {Caffau}, {Cambras}, {Campbell}, {Cancelliere}, {Cantat-Gaudin}, {Carlucci}, {Carrasco}, {Castellani}, {Charlot}, {Charnas}, {Charvet}, {Chassat}, {Chiavassa}, {Clotet}, {Cocozza}, {Collins}, {Collins}, {Costigan}, {Crifo}, {Cross}, {Crosta}, {Crowley}, {Dafonte}, {Damerdji}, {Dapergolas}, {David}, {David}, {De Cat}, {de Felice}, {de Laverny}, {De Luise}, {De March}, {de Martino}, {de Souza}, {Debosscher}, {del Pozo}, {Delbo}, {Delgado}, {Delgado}, {di Marco}, {Di Matteo}, {Diakite}, {Distefano}, {Dolding}, {Dos Anjos}, {Drazinos}, {Dur{\'a}n}, {Dzigan}, {Ecale}, {Edvardsson}, {Enke}, {Erdmann}, {Escolar}, {Espina}, {Evans}, {Eynard Bontemps}, {Fabre}, {Fabrizio}, {Faigler}, {Falc{\~a}o}, {Farr{\`a}s Casas}, {Faye}, {Federici}, {Fedorets}, {Fern{\'a}ndez-Hern{\'a}ndez}, {Fernique}, {Fienga}, {Figueras}, {Filippi},
  {Findeisen}, {Fonti}, {Fouesneau}, {Fraile}, {Fraser}, {Fuchs}, {Furnell}, {Gai}, {Galleti}, {Galluccio}, {Garabato}, {Garc{\'\i}a-Sedano}, {Gar{\'e}}, {Garofalo}, {Garralda}, {Gavras}, {Gerssen}, {Geyer}, {Gilmore}, {Girona}, {Giuffrida}, {Gomes}, {Gonz{\'a}lez-Marcos}, {Gonz{\'a}lez-N{\'u}{\~n}ez}, {Gonz{\'a}lez-Vidal}, {Granvik}, {Guerrier}, {Guillout}, {Guiraud}, {G{\'u}rpide}, {Guti{\'e}rrez-S{\'a}nchez}, {Guy}, {Haigron}, {Hatzidimitriou}, {Haywood}, {Heiter}, {Helmi}, {Hobbs}, {Hofmann}, {Holl}, {Holland }, {Hunt}, {Hypki}, {Icardi}, {Irwin}, {Jevardat de Fombelle}, {Jofr{\'e}}, {Jonker}, {Jorissen}, {Julbe}, {Karampelas}, {Kochoska}, {Kohley}, {Kolenberg}, {Kontizas}, {Koposov}, {Kordopatis}, {Koubsky}, {Kowalczyk}, {Krone-Martins}, {Kudryashova}, {Kull}, {Bachchan}, {Lacoste-Seris}, {Lanza}, {Lavigne}, {Le Poncin-Lafitte}, {Lebreton}, {Lebzelter}, {Leccia}, {Leclerc}, {Lecoeur-Taibi}, {Lemaitre}, {Lenhardt}, {Leroux}, {Liao}, {Licata}, {Lindstr{\o}m}, {Lister}, {Livanou}, {Lobel}, {L{\"o}ffler},
  {L{\'o}pez}, {Lopez-Lozano}, {Lorenz}, {Loureiro}, {MacDonald}, {Magalh{\~a}es Fernandes}, {Managau}, {Mann}, {Mantelet}, {Marchal}, {Marchant}, {Marconi}, {Marie}, {Marinoni}, {Marrese}, {Marschalk{\'o}}, {Marshall}, {Mart{\'\i}n-Fleitas}, {Martino}, {Mary}, {Matijevi{\v{c}}}, {Mazeh}, {McMillan}, {Messina}, {Mestre}, {Michalik}, {Millar}, {Miranda}, {Molina}, {Molinaro}, {Molinaro}, {Moln{\'a}r}, {Moniez}, {Montegriffo}, {Monteiro}, {Mor}, {Mora}, {Morbidelli}, {Morel}, {Morgenthaler}, {Morley}, {Morris}, {Mulone}, {Muraveva}, {Musella}, {Narbonne}, {Nelemans}, {Nicastro}, {Noval}, {Ord{\'e}novic}, {Ordieres-Mer{\'e}}, {Osborne}, {Pagani}, {Pagano}, {Pailler}, {Palacin}, {Palaversa}, {Parsons}, {Paulsen}, {Pecoraro}, {Pedrosa}, {Pentik{\"a}inen}, {Pereira}, {Pichon}, {Piersimoni}, {Pineau}, {Plachy}, {Plum}, {Poujoulet}, {Pr{\v{s}}a}, {Pulone}, {Ragaini}, {Rago}, {Rambaux}, {Ramos-Lerate}, {Ranalli}, {Rauw}, {Read}, {Regibo}, {Renk}, {Reyl{\'e}}, {Ribeiro}, {Rimoldini}, {Ripepi}, {Riva}, {Rixon},
  {Roelens}, {Romero-G{\'o}mez}, {Rowell}, {Royer}, {Rudolph}, {Ruiz-Dern}, {Sadowski}, {Sagrist{\`a} Sell{\'e}s}, {Sahlmann}, {Salgado}, {Salguero}, {Sarasso}, {Savietto}, {Schnorhk}, {Schultheis}, {Sciacca}, {Segol}, {Segovia}, {Segransan}, {Serpell}, {Shih}, {Smareglia}, {Smart}, {Smith}, {Solano}, {Solitro}, {Sordo}, {Soria Nieto}, {Souchay}, {Spagna}, {Spoto}, {Stampa}, {Steele}, {Steidelm{\"u}ller}, {Stephenson}, {Stoev}, {Suess}, {S{\"u}veges}, {Surdej}, {Szabados}, {Szegedi-Elek}, {Tapiador}, {Taris}, {Tauran}, {Taylor}, {Teixeira}, {Terrett}, {Tingley}, {Trager}, {Turon}, {Ulla}, {Utrilla}, {Valentini}, {van Elteren}, {Van Hemelryck}, {van Leeuwen}, {Varadi}, {Vecchiato}, {Veljanoski}, {Via}, {Vicente}, {Vogt}, {Voss}, {Votruba}, {Voutsinas}, {Walmsley}, {Weiler}, {Weingrill}, {Werner}, {Wevers}, {Whitehead}, {Wyrzykowski}, {Yoldas}, {{\v{Z}}erjal}, {Zucker}, {Zurbach}, {Zwitter}, {Alecu}, {Allen}, {Allende Prieto}, {Amorim}, {Anglada-Escud{\'e}}, {Arsenijevic}, {Azaz}, {Balm}, {Beck}, {Bernstein},
  {Bigot}, {Bijaoui}, {Blasco}, {Bonfigli}, {Bono}, {Boudreault}, {Bressan}, {Brown}, {Brunet}, {Bunclark}, {Buonanno}, {Butkevich}, {Carret}, {Carrion}, {Chemin}, {Ch{\'e}reau}, {Corcione}, {Darmigny}, {de Boer}, {de Teodoro}, {de Zeeuw}, {Delle Luche}, {Domingues}, {Dubath}, {Fodor}, {Fr{\'e}zouls}, {Fries}, {Fustes}, {Fyfe}, {Gallardo}, {Gallegos}, {Gardiol}, {Gebran}, {Gomboc}, {G{\'o}mez}, {Grux}, {Gueguen}, {Heyrovsky}, {Hoar}, {Iannicola}, {Isasi Parache}, {Janotto}, {Joliet}, {Jonckheere}, {Keil}, {Kim}, {Klagyivik}, {Klar}, {Knude}, {Kochukhov}, {Kolka}, {Kos}, {Kutka}, {Lainey}, {LeBouquin}, {Liu}, {Loreggia}, {Makarov}, {Marseille}, {Martayan}, {Martinez-Rubi}, {Massart}, {Meynadier}, {Mignot}, {Munari}, {Nguyen}, {Nordlander}, {Ocvirk}, {O'Flaherty}, {Olias Sanz}, {Ortiz}, {Osorio}, {Oszkiewicz}, {Ouzounis}, {Palmer}, {Park}, {Pasquato}, {Peltzer}, {Peralta}, {P{\'e}turaud}, {Pieniluoma}, {Pigozzi}, {Poels}, {Prat}, {Prod'homme}, {Raison}, {Rebordao}, {Risquez}, {Rocca-Volmerange}, {Rosen},
  {Ruiz-Fuertes}, {Russo}, {Sembay}, {Serraller Vizcaino}, {Short}, {Siebert}, {Silva}, {Sinachopoulos}, {Slezak}, {Soffel}, {Sosnowska}, {Strai{\v{z}}ys}, {ter Linden}, {Terrell}, {Theil}, {Tiede}, {Troisi}, {Tsalmantza}, {Tur}, {Vaccari}, {Vachier}, {Valles}, {Van Hamme}, {Veltz}, {Virtanen}, {Wallut}, {Wichmann}, {Wilkinson}, {Ziaeepour}, \& {Zschocke}}]{2016A&A...595A...1G}
{Gaia Collaboration}, {Prusti}, T., {de Bruijne}, J.~H.~J., {et~al.} 2016, \aap, 595, A1, \dodoi{10.1051/0004-6361/201629272}

\bibitem[{{Gaia Collaboration} {et~al.}(2022){Gaia Collaboration}, {Vallenari}, {Brown}, {Prusti}, {de Bruijne}, {Arenou}, {Babusiaux}, {Biermann}, {Creevey}, {Ducourant}, {Evans}, {Eyer}, {Guerra}, {Hutton}, {Jordi}, {Klioner}, {Lammers}, {Lindegren}, {Luri}, {Mignard}, {Panem}, {Pourbaix}, {Randich}, {Sartoretti}, {Soubiran}, {Tanga}, {Walton}, {Bailer-Jones}, {Bastian}, {Drimmel}, {Jansen}, {Katz}, {Lattanzi}, {van Leeuwen}, {Bakker}, {Cacciari}, {Casta{\~n}eda}, {De Angeli}, {Fabricius}, {Fouesneau}, {Fr{\'e}mat}, {Galluccio}, {Guerrier}, {Heiter}, {Masana}, {Messineo}, {Mowlavi}, {Nicolas}, {Nienartowicz}, {Pailler}, {Panuzzo}, {Riclet}, {Roux}, {Seabroke}, {Sordo{\o}rcit}, {Th{\'e}venin}, {Gracia-Abril}, {Portell}, {Teyssier}, {Altmann}, {Andrae}, {Audard}, {Bellas-Velidis}, {Benson}, {Berthier}, {Blomme}, {Burgess}, {Busonero}, {Busso}, {C{\'a}novas}, {Carry}, {Cellino}, {Cheek}, {Clementini}, {Damerdji}, {Davidson}, {de Teodoro}, {Nu{\~n}ez Campos}, {Delchambre}, {Dell'Oro}, {Esquej},
  {Fern{\'a}ndez-Hern{\'a}ndez}, {Fraile}, {Garabato}, {Garc{\'\i}a-Lario}, {Gosset}, {Haigron}, {Halbwachs}, {Hambly}, {Harrison}, {Hern{\'a}ndez}, {Hestroffer}, {Hodgkin}, {Holl}, {Jan{\ss}en}, {Jevardat de Fombelle}, {Jordan}, {Krone-Martins}, {Lanzafame}, {L{\"o}ffler}, {Marchal}, {Marrese}, {Moitinho}, {Muinonen}, {Osborne}, {Pancino}, {Pauwels}, {Recio-Blanco}, {Reyl{\'e}}, {Riello}, {Rimoldini}, {Roegiers}, {Rybizki}, {Sarro}, {Siopis}, {Smith}, {Sozzetti}, {Utrilla}, {van Leeuwen}, {Abbas}, {{\'A}brah{\'a}m}, {Abreu Aramburu}, {Aerts}, {Aguado}, {Ajaj}, {Aldea-Montero}, {Altavilla}, {{\'A}lvarez}, {Alves}, {Anders}, {Anderson}, {Anglada Varela}, {Antoja}, {Baines}, {Baker}, {Balaguer-N{\'u}{\~n}ez}, {Balbinot}, {Balog}, {Barache}, {Barbato}, {Barros}, {Barstow}, {Bartolom{\'e}}, {Bassilana}, {Bauchet}, {Becciani}, {Bellazzini}, {Berihuete}, {Bernet}, {Bertone}, {Bianchi}, {Binnenfeld}, {Blanco-Cuaresma}, {Blazere}, {Boch}, {Bombrun}, {Bossini}, {Bouquillon}, {Bragaglia}, {Bramante}, {Breedt},
  {Bressan}, {Brouillet}, {Brugaletta}, {Bucciarelli}, {Burlacu}, {Butkevich}, {Buzzi}, {Caffau}, {Cancelliere}, {Cantat-Gaudin}, {Carballo}, {Carlucci}, {Carnerero}, {Carrasco}, {Casamiquela}, {Castellani}, {Castro-Ginard}, {Chaoul}, {Charlot}, {Chemin}, {Chiaramida}, {Chiavassa}, {Chornay}, {Comoretto}, {Contursi}, {Cooper}, {Cornez}, {Cowell}, {Crifo}, {Cropper}, {Crosta}, {Crowley}, {Dafonte}, {Dapergolas}, {David}, {David}, {de Laverny}, {De Luise}, {De March}, {De Ridder}, {de Souza}, {de Torres}, {del Peloso}, {del Pozo}, {Delbo}, {Delgado}, {Delisle}, {Demouchy}, {Dharmawardena}, {Di Matteo}, {Diakite}, {Diener}, {Distefano}, {Dolding}, {Edvardsson}, {Enke}, {Fabre}, {Fabrizio}, {Faigler}, {Fedorets}, {Fernique}, {Fienga}, {Figueras}, {Fournier}, {Fouron}, {Fragkoudi}, {Gai}, {Garcia-Gutierrez}, {Garcia-Reinaldos}, {Garc{\'\i}a-Torres}, {Garofalo}, {Gavel}, {Gavras}, {Gerlach}, {Geyer}, {Giacobbe}, {Gilmore}, {Girona}, {Giuffrida}, {Gomel}, {Gomez}, {Gonz{\'a}lez-N{\'u}{\~n}ez},
  {Gonz{\'a}lez-Santamar{\'\i}a}, {Gonz{\'a}lez-Vidal}, {Granvik}, {Guillout}, {Guiraud}, {Guti{\'e}rrez-S{\'a}nchez}, {Guy}, {Hatzidimitriou}, {Hauser}, {Haywood}, {Helmer}, {Helmi}, {Sarmiento}, {Hidalgo}, {Hilger}, {H{\l}adczuk}, {Hobbs}, {Holland}, {Huckle}, {Jardine}, {Jasniewicz}, {Jean-Antoine Piccolo}, {Jim{\'e}nez-Arranz}, {Jorissen}, {Juaristi Campillo}, {Julbe}, {Karbevska}, {Kervella}, {Khanna}, {Kontizas}, {Kordopatis}, {Korn}, {K{\'o}sp{\'a}l}, {Kostrzewa-Rutkowska}, {Kruszy{\'n}ska}, {Kun}, {Laizeau}, {Lambert}, {Lanza}, {Lasne}, {Le Campion}, {Lebreton}, {Lebzelter}, {Leccia}, {Leclerc}, {Lecoeur-Taibi}, {Liao}, {Licata}, {Lindstr{\o}m}, {Lister}, {Livanou}, {Lobel}, {Lorca}, {Loup}, {Madrero Pardo}, {Magdaleno Romeo}, {Managau}, {Mann}, {Manteiga}, {Marchant}, {Marconi}, {Marcos}, {Marcos Santos}, {Mar{\'\i}n Pina}, {Marinoni}, {Marocco}, {Marshall}, {Polo}, {Mart{\'\i}n-Fleitas}, {Marton}, {Mary}, {Masip}, {Massari}, {Mastrobuono-Battisti}, {Mazeh}, {McMillan}, {Messina}, {Michalik},
  {Millar}, {Mints}, {Molina}, {Molinaro}, {Moln{\'a}r}, {Monari}, {Mongui{\'o}}, {Montegriffo}, {Montero}, {Mor}, {Mora}, {Morbidelli}, {Morel}, {Morris}, {Muraveva}, {Murphy}, {Musella}, {Nagy}, {Noval}, {Oca{\~n}a}, {Ogden}, {Ordenovic}, {Osinde}, {Pagani}, {Pagano}, {Palaversa}, {Palicio}, {Pallas-Quintela}, {Panahi}, {Payne-Wardenaar}, {Pe{\~n}alosa Esteller}, {Penttil{\"a}}, {Pichon}, {Piersimoni}, {Pineau}, {Plachy}, {Plum}, {Poggio}, {Pr{\v{s}}a}, {Pulone}, {Racero}, {Ragaini}, {Rainer}, {Raiteri}, {Rambaux}, {Ramos}, {Ramos-Lerate}, {Re Fiorentin}, {Regibo}, {Richards}, {Rios Diaz}, {Ripepi}, {Riva}, {Rix}, {Rixon}, {Robichon}, {Robin}, {Robin}, {Roelens}, {Rogues}, {Rohrbasser}, {Romero-G{\'o}mez}, {Rowell}, {Royer}, {Ruz Mieres}, {Rybicki}, {Sadowski}, {S{\'a}ez N{\'u}{\~n}ez}, {Sagrist{\`a} Sell{\'e}s}, {Sahlmann}, {Salguero}, {Samaras}, {Sanchez Gimenez}, {Sanna}, {Santove{\~n}a}, {Sarasso}, {Schultheis}, {Sciacca}, {Segol}, {Segovia}, {S{\'e}gransan}, {Semeux}, {Shahaf}, {Siddiqui}, {Siebert},
  {Siltala}, {Silvelo}, {Slezak}, {Slezak}, {Smart}, {Snaith}, {Solano}, {Solitro}, {Souami}, {Souchay}, {Spagna}, {Spina}, {Spoto}, {Steele}, {Steidelm{\"u}ller}, {Stephenson}, {S{\"u}veges}, {Surdej}, {Szabados}, {Szegedi-Elek}, {Taris}, {Taylo}, {Teixeira}, {Tolomei}, {Tonello}, {Torra}, {Torra}, {Torralba Elipe}, {Trabucchi}, {Tsounis}, {Turon}, {Ulla}, {Unger}, {Vaillant}, {van Dillen}, {van Reeven}, {Vanel}, {Vecchiato}, {Viala}, {Vicente}, {Voutsinas}, {Weiler}, {Wevers}, {Wyrzykowski}, {Yoldas}, {Yvard}, {Zhao}, {Zorec}, {Zucker}, \& {Zwitter}}]{2022arXiv220800211G}
{Gaia Collaboration}, {Vallenari}, A., {Brown}, A.~G.~A., {et~al.} 2022, arXiv e-prints, arXiv:2208.00211, \dodoi{10.48550/arXiv.2208.00211}

\bibitem[{{Giacalone} {et~al.}(2019){Giacalone}, {Teitler}, {K{\"o}nigl}, {Krijt}, \& {Ciesla}}]{2019ApJ...882...33G}
{Giacalone}, S., {Teitler}, S., {K{\"o}nigl}, A., {Krijt}, S., \& {Ciesla}, F.~J. 2019, \apj, 882, 33, \dodoi{10.3847/1538-4357/ab311a}

\bibitem[{{Henning} {et~al.}(1999){Henning}, {Il'In}, {Krivova}, {Michel}, \& {Voshchinnikov}}]{1999A&AS..136..405H}
{Henning}, T., {Il'In}, V.~B., {Krivova}, N.~A., {Michel}, B., \& {Voshchinnikov}, N.~V. 1999, \aaps, 136, 405, \dodoi{10.1051/aas:1999222}

\bibitem[{{Henning} \& {Meeus}(2011)}]{2011ppcd.book..114H}
{Henning}, T., \& {Meeus}, G. 2011, {Dust Processing and Mineralogy in Protoplanetary Accretion Disks}, ed. P.~J.~V. {Garcia}, 114--148

\bibitem[{Hunter(2007)}]{Hunter:2007}
Hunter, J.~D. 2007, Computing In Science \& Engineering, 9, 90, \dodoi{10.1109/MCSE.2007.55}

\bibitem[{{Jaeger} {et~al.}(1998){Jaeger}, {Molster}, {Dorschner}, {Henning}, {Mutschke}, \& {Waters}}]{1998A&A...339..904J}
{Jaeger}, C., {Molster}, F.~J., {Dorschner}, J., {et~al.} 1998, \aap, 339, 904

\bibitem[{{Jaeger} {et~al.}(1994){Jaeger}, {Mutschke}, {Begemann}, {Dorschner}, \& {Henning}}]{1994A&A...292..641J}
{Jaeger}, C., {Mutschke}, H., {Begemann}, B., {Dorschner}, J., \& {Henning}, T. 1994, \aap, 292, 641

\bibitem[{{J{\"a}ger} {et~al.}(2003{\natexlab{a}}){J{\"a}ger}, {Dorschner}, {Mutschke}, {Posch}, \& {Henning}}]{2003A&A...408..193J}
{J{\"a}ger}, C., {Dorschner}, J., {Mutschke}, H., {Posch}, T., \& {Henning}, T. 2003{\natexlab{a}}, \aap, 408, 193, \dodoi{10.1051/0004-6361:20030916}

\bibitem[{{J{\"a}ger} {et~al.}(2003{\natexlab{b}}){J{\"a}ger}, {Fabian}, {Schrempel}, {Dorschner}, {Henning}, \& {Wesch}}]{2003A&A...401...57J}
{J{\"a}ger}, C., {Fabian}, D., {Schrempel}, F., {et~al.} 2003{\natexlab{b}}, \aap, 401, 57, \dodoi{10.1051/0004-6361:20030002}

\bibitem[{{J{\"a}ger} {et~al.}(2003{\natexlab{c}}){J{\"a}ger}, {Il'in}, {Henning}, {Mutschke}, {Fabian}, {Semenov}, \& {Voshchinnikov}}]{2003JQSRT..79..765J}
{J{\"a}ger}, C., {Il'in}, V.~B., {Henning}, T., {et~al.} 2003{\natexlab{c}}, \jqsrt, 79-80, 765, \dodoi{10.1016/S0022-4073(02)00301-1}

\bibitem[{{Juh{\'a}sz} {et~al.}(2009){Juh{\'a}sz}, {Henning}, {Bouwman}, {Dullemond}, {Pascucci}, \& {Apai}}]{2009ApJ...695.1024J}
{Juh{\'a}sz}, A., {Henning}, T., {Bouwman}, J., {et~al.} 2009, \apj, 695, 1024, \dodoi{10.1088/0004-637X/695/2/1024}

\bibitem[{{Juh{\'a}sz} {et~al.}(2010){Juh{\'a}sz}, {Bouwman}, {Henning}, {Acke}, {van den Ancker}, {Meeus}, {Dominik}, {Min}, {Tielens}, \& {Waters}}]{2010ApJ...721..431J}
{Juh{\'a}sz}, A., {Bouwman}, J., {Henning}, T., {et~al.} 2010, \apj, 721, 431, \dodoi{10.1088/0004-637X/721/1/431}

\bibitem[{{Kaeufer} {et~al.}(2023){Kaeufer}, {Woitke}, {Min}, {Kamp}, \& {Pinte}}]{2023A&A...672A..30K}
{Kaeufer}, T., {Woitke}, P., {Min}, M., {Kamp}, I., \& {Pinte}, C. 2023, \aap, 672, A30, \dodoi{10.1051/0004-6361/202245461}

\bibitem[{{Kastner} {et~al.}(2010{\natexlab{a}}){Kastner}, {Buchanan}, {Sahai}, {Forrest}, \& {Sargent}}]{2010AJ....139.1993K}
{Kastner}, J.~H., {Buchanan}, C., {Sahai}, R., {Forrest}, W.~J., \& {Sargent}, B.~A. 2010{\natexlab{a}}, \aj, 139, 1993, \dodoi{10.1088/0004-6256/139/5/1993}

\bibitem[{{Kastner} {et~al.}(2006){Kastner}, {Buchanan}, {Sargent}, \& {Forrest}}]{2006ApJ...638L..29K}
{Kastner}, J.~H., {Buchanan}, C.~L., {Sargent}, B., \& {Forrest}, W.~J. 2006, \apj, 638, L29, \dodoi{10.1086/500804}

\bibitem[{{Kastner} {et~al.}(2010{\natexlab{b}}){Kastner}, {Hily-Blant}, {Sacco}, {Forveille}, \& {Zuckerman}}]{2010ApJ...723L.248K}
{Kastner}, J.~H., {Hily-Blant}, P., {Sacco}, G.~G., {Forveille}, T., \& {Zuckerman}, B. 2010{\natexlab{b}}, \apjl, 723, L248, \dodoi{10.1088/2041-8205/723/2/L248}

\bibitem[{{Kim} {et~al.}(2016){Kim}, {Watson}, {Manoj}, {Forrest}, {Furlan}, {Najita}, {Sargent}, {Hern{\'a}ndez}, {Calvet}, {Adame}, {Espaillat}, {Megeath}, {Muzerolle}, \& {McClure}}]{2016ApJS..226....8K}
{Kim}, K.~H., {Watson}, D.~M., {Manoj}, P., {et~al.} 2016, \apjs, 226, 8, \dodoi{10.3847/0067-0049/226/1/8}

\bibitem[{{Lebouteiller} {et~al.}(2011){Lebouteiller}, {Barry}, {Spoon}, {Bernard-Salas}, {Sloan}, {Houck}, \& {Weedman}}]{2011ApJS..196....8L}
{Lebouteiller}, V., {Barry}, D.~J., {Spoon}, H.~W.~W., {et~al.} 2011, The Astrophysical Journal Supplement Series, 196, 8, \dodoi{10.1088/0067-0049/196/1/8}

\bibitem[{{Lindegren} {et~al.}(2021){Lindegren}, {Klioner}, {Hern{\'a}ndez}, {Bombrun}, {Ramos-Lerate}, {Steidelm{\"u}ller}, {Bastian}, {Biermann}, {de Torres}, {Gerlach}, {Geyer}, {Hilger}, {Hobbs}, {Lammers}, {McMillan}, {Stephenson}, {Casta{\~n}eda}, {Davidson}, {Fabricius}, {Gracia-Abril}, {Portell}, {Rowell}, {Teyssier}, {Torra}, {Bartolom{\'e}}, {Clotet}, {Garralda}, {Gonz{\'a}lez-Vidal}, {Torra}, {Abbas}, {Altmann}, {Anglada Varela}, {Balaguer-N{\'u}{\~n}ez}, {Balog}, {Barache}, {Becciani}, {Bernet}, {Bertone}, {Bianchi}, {Bouquillon}, {Brown}, {Bucciarelli}, {Busonero}, {Butkevich}, {Buzzi}, {Cancelliere}, {Carlucci}, {Charlot}, {Cioni}, {Crosta}, {Crowley}, {del Peloso}, {del Pozo}, {Drimmel}, {Esquej}, {Fienga}, {Fraile}, {Gai}, {Garcia-Reinaldos}, {Guerra}, {Hambly}, {Hauser}, {Jan{\ss}en}, {Jordan}, {Kostrzewa-Rutkowska}, {Lattanzi}, {Liao}, {Licata}, {Lister}, {L{\"o}ffler}, {Marchant}, {Masip}, {Mignard}, {Mints}, {Molina}, {Mora}, {Morbidelli}, {Murphy}, {Pagani}, {Panuzzo}, {Pe{\~n}alosa
  Esteller}, {Poggio}, {Re Fiorentin}, {Riva}, {Sagrist{\`a} Sell{\'e}s}, {Sanchez Gimenez}, {Sarasso}, {Sciacca}, {Siddiqui}, {Smart}, {Souami}, {Spagna}, {Steele}, {Taris}, {Utrilla}, {van Reeven}, \& {Vecchiato}}]{2021A&A...649A...2L}
{Lindegren}, L., {Klioner}, S.~A., {Hern{\'a}ndez}, J., {et~al.} 2021, \aap, 649, A2, \dodoi{10.1051/0004-6361/202039709}

\bibitem[{{Lisse} {et~al.}(1998){Lisse}, {A'Hearn}, {Hauser}, {Kelsall}, {Lien}, {Moseley}, {Reach}, \& {Silverberg}}]{1998ApJ...496..971L}
{Lisse}, C.~M., {A'Hearn}, M.~F., {Hauser}, M.~G., {et~al.} 1998, \apj, 496, 971, \dodoi{10.1086/305397}

\bibitem[{{Liu} {et~al.}(2019){Liu}, {Pascucci}, \& {Henning}}]{2019A&A...623A.106L}
{Liu}, Y., {Pascucci}, I., \& {Henning}, T. 2019, \aap, 623, A106, \dodoi{10.1051/0004-6361/201834418}

\bibitem[{{Lu} {et~al.}(2022){Lu}, {Chen}, {Sargent}, {Watson}, {Lisse}, {Green}, {Sitko}, {Mittal}, {Lebouteiller}, {Sloan}, {Rebollido}, {Hines}, {Girard}, {Werner}, {Stapelfeldt}, {Wu}, \& {Worthen}}]{2022ApJ...933...54L}
{Lu}, C.~X., {Chen}, C.~H., {Sargent}, B.~A., {et~al.} 2022, \apj, 933, 54, \dodoi{10.3847/1538-4357/ac70d1}

\bibitem[{{Mamajek} {et~al.}(2002){Mamajek}, {Meyer}, \& {Liebert}}]{2002AJ....124.1670M}
{Mamajek}, E.~E., {Meyer}, M.~R., \& {Liebert}, J. 2002, \aj, 124, 1670, \dodoi{10.1086/341952}

\bibitem[{{McClure}(2009)}]{2009ApJ...693L..81M}
{McClure}, M. 2009, \apj, 693, L81, \dodoi{10.1088/0004-637X/693/2/L81}

\bibitem[{{McClure} {et~al.}(2012){McClure}, {Manoj}, {Calvet}, {Adame}, {Espaillat}, {Watson}, {Sargent}, {Forrest}, \& {D'Alessio}}]{2012ApJ...759L..10M}
{McClure}, M.~K., {Manoj}, P., {Calvet}, N., {et~al.} 2012, \apjl, 759, L10, \dodoi{10.1088/2041-8205/759/1/L10}

\bibitem[{{Meyer} {et~al.}(2004){Meyer}, {Backman}, {Beckwith}, {Carpenter}, {Cohen}, {Henning}, {Hillenbrand}, {Hines}, {Hollenbach}, {Lunine}, {Malhotra}, {Najita}, {Padgett}, {Soderblom}, {Stauffer}, {Strom}, {Watson}, {Weidenschilling}, \& {Young}}]{2004sptz.prop..148M}
{Meyer}, M.~R., {Backman}, D., {Beckwith}, S. V.~W., {et~al.} 2004, {The Formation and Evolution of Planetary Systems: Placing Our Solar System in Context}, Spitzer Proposal

\bibitem[{{Min} {et~al.}(2003){Min}, {Hovenier}, \& {de Koter}}]{2003A&A...404...35M}
{Min}, M., {Hovenier}, J.~W., \& {de Koter}, A. 2003, \aap, 404, 35, \dodoi{10.1051/0004-6361:20030456}

\bibitem[{Oliphant(2006)}]{oliphant2006guide}
Oliphant, T.~E. 2006, A guide to NumPy, Vol.~1 (Trelgol Publishing USA)

\bibitem[{{Pecaut} \& {Mamajek}(2013)}]{2013ApJS..208....9P}
{Pecaut}, M.~J., \& {Mamajek}, E.~E. 2013, The Astrophysical Journal Supplement Series, 208, 9, \dodoi{10.1088/0067-0049/208/1/9}

\bibitem[{{Perotti} {et~al.}(2023){Perotti}, {Christiaens}, {Henning}, {Tabone}, {Waters}, {Kamp}, {Olofsson}, {Grant}, {Gasman}, {Bouwman}, {Samland}, {Franceschi}, {van Dishoeck}, {Schwarz}, {G{\"u}del}, {Lagage}, {Ray}, {Vandenbussche}, {Abergel}, {Absil}, {Arabhavi}, {Argyriou}, {Barrado}, {Boccaletti}, {Caratti o Garatti}, {Geers}, {Glauser}, {Justannont}, {Lahuis}, {Mueller}, {Nehm{\'e}}, {Pantin}, {Scheithauer}, {Waelkens}, {Guadarrama}, {Jang}, {Kanwar}, {Morales-Calder{\'o}n}, {Pawellek}, {Rodgers-Lee}, {Schreiber}, {Colina}, {Greve}, {{\"O}stlin}, \& {Wright}}]{2023Natur.620..516P}
{Perotti}, G., {Christiaens}, V., {Henning}, T., {et~al.} 2023, \nat, 620, 516, \dodoi{10.1038/s41586-023-06317-9}

\bibitem[{{Pinte}(2022)}]{2022ascl.soft07024P}
{Pinte}, C. 2022, {pymcfost: Python interface to the MCFOST 3D radiative transfer code}, Astrophysics Source Code Library, record ascl:2207.024

\bibitem[{{Pinte} {et~al.}(2009){Pinte}, {Harries}, {Min}, {Watson}, {Dullemond}, {Woitke}, {M{\'e}nard}, \& {Dur{\'a}n-Rojas}}]{2009A&A...498..967P}
{Pinte}, C., {Harries}, T.~J., {Min}, M., {et~al.} 2009, \aap, 498, 967, \dodoi{10.1051/0004-6361/200811555}

\bibitem[{{Pinte} {et~al.}(2006){Pinte}, {M{\'e}nard}, {Duch{\^e}ne}, \& {Bastien}}]{2006A&A...459..797P}
{Pinte}, C., {M{\'e}nard}, F., {Duch{\^e}ne}, G., \& {Bastien}, P. 2006, \aap, 459, 797, \dodoi{10.1051/0004-6361:20053275}

\bibitem[{{Pinte} {et~al.}(2023){Pinte}, {Teague}, {Flaherty}, {Hall}, {Facchini}, \& {Casassus}}]{2023ASPC..534..645P}
{Pinte}, C., {Teague}, R., {Flaherty}, K., {et~al.} 2023, in Astronomical Society of the Pacific Conference Series, Vol. 534, Protostars and Planets VII, ed. S.~{Inutsuka}, Y.~{Aikawa}, T.~{Muto}, K.~{Tomida}, \& M.~{Tamura}, 645, \dodoi{10.48550/arXiv.2203.09528}

\bibitem[{{Pinte} {et~al.}(2022){Pinte}, {M{\'e}nard}, {Duch{\^e}ne}, {Bastien}, {Harries}, {Min}, {Watson}, {Dullemond}, {Woitke}, \& {Dur{\'a}n-Rojas}}]{2022ascl.soft07023P}
{Pinte}, C., {M{\'e}nard}, F., {Duch{\^e}ne}, G., {et~al.} 2022, {MCFOST: Radiative transfer code}, Astrophysics Source Code Library, record ascl:2207.023.
\newblock \doeprint{2207.023}

\bibitem[{{Press} {et~al.}(1992){Press}, {Teukolsky}, {Vetterling}, \& {Flannery}}]{1992nrfa.book.....P}
{Press}, W.~H., {Teukolsky}, S.~A., {Vetterling}, W.~T., \& {Flannery}, B.~P. 1992, {Numerical recipes in FORTRAN. The art of scientific computing}

\bibitem[{{Rapson} {et~al.}(2015){Rapson}, {Sargent}, {Germano Sacco}, {Kastner}, {Wilner}, {Rosenfeld}, {Andrews}, {Herczeg}, \& {van der Marel}}]{2015ApJ...810...62R}
{Rapson}, V.~A., {Sargent}, B., {Germano Sacco}, G., {et~al.} 2015, \apj, 810, 62, \dodoi{10.1088/0004-637X/810/1/62}

\bibitem[{{Ribas} {et~al.}(2020){Ribas}, {Espaillat}, {Mac{\'\i}as}, \& {Sarro}}]{2020A&A...642A.171R}
{Ribas}, {\'A}., {Espaillat}, C.~C., {Mac{\'\i}as}, E., \& {Sarro}, L.~M. 2020, \aap, 642, A171, \dodoi{10.1051/0004-6361/202038352}

\bibitem[{{Ribas} {et~al.}(2023){Ribas}, {Mac{\'\i}as}, {Weber}, {P{\'e}rez}, {Cuello}, {Dong}, {Aguayo}, {C{\'a}ceres}, {Carpenter}, {Dent}, {de Gregorio-Monsalvo}, {Duch{\^e}ne}, {Espaillat}, {Riviere-Marichalar}, \& {Villenave}}]{2023A&A...673A..77R}
{Ribas}, {\'A}., {Mac{\'\i}as}, E., {Weber}, P., {et~al.} 2023, \aap, 673, A77, \dodoi{10.1051/0004-6361/202245637}

\bibitem[{{Rosotti}(2023)}]{2023NewAR..9601674R}
{Rosotti}, G.~P. 2023, \nar, 96, 101674, \dodoi{10.1016/j.newar.2023.101674}

\bibitem[{{Rosotti} {et~al.}(2019){Rosotti}, {Tazzari}, {Booth}, {Testi}, {Lodato}, \& {Clarke}}]{2019MNRAS.486.4829R}
{Rosotti}, G.~P., {Tazzari}, M., {Booth}, R.~A., {et~al.} 2019, \mnras, 486, 4829, \dodoi{10.1093/mnras/stz1190}

\bibitem[{{Sacco} {et~al.}(2014){Sacco}, {Kastner}, {Forveille}, {Principe}, {Montez}, {Zuckerman}, \& {Hily-Blant}}]{2014A&A...561A..42S}
{Sacco}, G.~G., {Kastner}, J.~H., {Forveille}, T., {et~al.} 2014, \aap, 561, A42, \dodoi{10.1051/0004-6361/201322273}

\bibitem[{{Sargent} {et~al.}(2006){Sargent}, {Forrest}, {D'Alessio}, {Li}, {Najita}, {Watson}, {Calvet}, {Furlan}, {Green}, {Kim}, {Sloan}, {Chen}, {Hartmann}, \& {Houck}}]{2006ApJ...645..395S}
{Sargent}, B., {Forrest}, W.~J., {D'Alessio}, P., {et~al.} 2006, \apj, 645, 395, \dodoi{10.1086/504283}

\bibitem[{{Sargent} {et~al.}(2009{\natexlab{a}}){Sargent}, {Forrest}, {Tayrien}, {McClure}, {Li}, {Basu}, {Manoj}, {Watson}, {Bohac}, {Furlan}, {Kim}, {Green}, \& {Sloan}}]{2009ApJ...690.1193S}
{Sargent}, B.~A., {Forrest}, W.~J., {Tayrien}, C., {et~al.} 2009{\natexlab{a}}, \apj, 690, 1193, \dodoi{10.1088/0004-637X/690/2/1193}

\bibitem[{{Sargent} {et~al.}(2009{\natexlab{b}}){Sargent}, {Forrest}, {Tayrien}, {McClure}, {Watson}, {Sloan}, {Li}, {Manoj}, {Bohac}, {Furlan}, {Kim}, \& {Green}}]{2009ApJS..182..477S}
---. 2009{\natexlab{b}}, The Astrophysical Journal Supplement Series, 182, 477, \dodoi{10.1088/0067-0049/182/2/477}

\bibitem[{{Schneider} {et~al.}(2014){Schneider}, {Grady}, {Hines}, {Stark}, {Debes}, {Carson}, {Kuchner}, {Perrin}, {Weinberger}, {Wisniewski}, {Silverstone}, {Jang-Condell}, {Henning}, {Woodgate}, {Serabyn}, {Moro-Martin}, {Tamura}, {Hinz}, \& {Rodigas}}]{2014AJ....148...59S}
{Schneider}, G., {Grady}, C.~A., {Hines}, D.~C., {et~al.} 2014, \aj, 148, 59, \dodoi{10.1088/0004-6256/148/4/59}

\bibitem[{{Sch{\"u}tz} {et~al.}(2005){Sch{\"u}tz}, {Meeus}, \& {Sterzik}}]{2005A&A...431..165S}
{Sch{\"u}tz}, O., {Meeus}, G., \& {Sterzik}, M.~F. 2005, \aap, 431, 165, \dodoi{10.1051/0004-6361:20041489}

\bibitem[{{Skrutskie} {et~al.}(2006){Skrutskie}, {Cutri}, {Stiening}, {Weinberg}, {Schneider}, {Carpenter}, {Beichman}, {Capps}, {Chester}, {Elias}, {Huchra}, {Liebert}, {Lonsdale}, {Monet}, {Price}, {Seitzer}, {Jarrett}, {Kirkpatrick}, {Gizis}, {Howard}, {Evans}, {Fowler}, {Fullmer}, {Hurt}, {Light}, {Kopan}, {Marsh}, {McCallon}, {Tam}, {Van Dyk}, \& {Wheelock}}]{2006AJ....131.1163S}
{Skrutskie}, M.~F., {Cutri}, R.~M., {Stiening}, R., {et~al.} 2006, \aj, 131, 1163, \dodoi{10.1086/498708}

\bibitem[{{Sogawa} {et~al.}(2006){Sogawa}, {Koike}, {Chihara}, {Suto}, {Tachibana}, {Tsuchiyama}, \& {Kozasa}}]{2006A&A...451..357S}
{Sogawa}, H., {Koike}, C., {Chihara}, H., {et~al.} 2006, \aap, 451, 357, \dodoi{10.1051/0004-6361:20041538}

\bibitem[{{Storn} \& {Price}(1997)}]{1997JGOpt..11..341S}
{Storn}, R., \& {Price}, K. 1997, Journal of Global Optimization, 11, 341, \dodoi{10.1023/A:1008202821328}

\bibitem[{{Torres} {et~al.}(2006){Torres}, {Quast}, {da Silva}, {de La Reza}, {Melo}, \& {Sterzik}}]{2006A&A...460..695T}
{Torres}, C.~A.~O., {Quast}, G.~R., {da Silva}, L., {et~al.} 2006, \aap, 460, 695, \dodoi{10.1051/0004-6361:20065602}

\bibitem[{{van Boekel} {et~al.}(2005){van Boekel}, {Min}, {Waters}, {de Koter}, {Dominik}, {van den Ancker}, \& {Bouwman}}]{2005A&A...437..189V}
{van Boekel}, R., {Min}, M., {Waters}, L.~B.~F.~M., {et~al.} 2005, \aap, 437, 189, \dodoi{10.1051/0004-6361:20042339}

\bibitem[{{van der Walt} {et~al.}(2011){van der Walt}, {Colbert}, \& {Varoquaux}}]{2011CSE....13b..22V}
{van der Walt}, S., {Colbert}, S.~C., \& {Varoquaux}, G. 2011, Computing in Science and Engineering, 13, 22, \dodoi{10.1109/MCSE.2011.37}

\bibitem[{{van Holstein} {et~al.}(2020){van Holstein}, {Girard}, {de Boer}, {Snik}, {Milli}, {Stam}, {Ginski}, {Mouillet}, {Wahhaj}, {Schmid}, {Keller}, {Langlois}, {Dohlen}, {Vigan}, {Pohl}, {Carbillet}, {Fantinel}, {Maurel}, {Orign{\'e}}, {Petit}, {Ramos}, {Rigal}, {Sevin}, {Boccaletti}, {Le Coroller}, {Dominik}, {Henning}, {Lagadec}, {M{\'e}nard}, {Turatto}, {Udry}, {Chauvin}, {Feldt}, \& {Beuzit}}]{2020A&A...633A..64V}
{van Holstein}, R.~G., {Girard}, J.~H., {de Boer}, J., {et~al.} 2020, \aap, 633, A64, \dodoi{10.1051/0004-6361/201834996}

\bibitem[{{Virtanen} {et~al.}(2020){Virtanen}, {Gommers}, {Oliphant}, {Haberland}, {Reddy}, {Cournapeau}, {Burovski}, {Peterson}, {Weckesser}, {Bright}, {van der Walt}, {Brett}, {Wilson}, {Jarrod Millman}, {Mayorov}, {Nelson}, {Jones}, {Kern}, {Larson}, {Carey}, {Polat}, {Feng}, {Moore}, {Vand erPlas}, {Laxalde}, {Perktold}, {Cimrman}, {Henriksen}, {Quintero}, {Harris}, {Archibald}, {Ribeiro}, {Pedregosa}, {van Mulbregt}, \& {Contributors}}]{2020SciPy-NMeth}
{Virtanen}, P., {Gommers}, R., {Oliphant}, T.~E., {et~al.} 2020, Nature Methods, 17, 261, \dodoi{https://doi.org/10.1038/s41592-019-0686-2}

\bibitem[{{Watson}(2009)}]{2009ASPC..414...77W}
{Watson}, D. 2009, in Astronomical Society of the Pacific Conference Series, Vol. 414, Cosmic Dust - Near and Far, ed. T.~{Henning}, E.~{Gr{\"u}n}, \& J.~{Steinacker}, 77, \dodoi{10.48550/arXiv.0902.2744}

\bibitem[{{Watson} {et~al.}(2009){Watson}, {Leisenring}, {Furlan}, {Bohac}, {Sargent}, {Forrest}, {Calvet}, {Hartmann}, {Nordhaus}, {Green}, {Kim}, {Sloan}, {Chen}, {Keller}, {d'Alessio}, {Najita}, {Uchida}, \& {Houck}}]{2009ApJS..180...84W}
{Watson}, D.~M., {Leisenring}, J.~M., {Furlan}, E., {et~al.} 2009, The Astrophysical Journal Supplement Series, 180, 84, \dodoi{10.1088/0067-0049/180/1/84}

\bibitem[{{Weingartner} \& {Draine}(2001)}]{2001ApJ...548..296W}
{Weingartner}, J.~C., \& {Draine}, B.~T. 2001, \apj, 548, 296, \dodoi{10.1086/318651}

\bibitem[{{Wells} {et~al.}(2015){Wells}, {Pel}, {Glasse}, {Wright}, {Aitink-Kroes}, {Azzollini}, {Beard}, {Brandl}, {Gallie}, {Geers}, {Glauser}, {Hastings}, {Henning}, {Jager}, {Justtanont}, {Kruizinga}, {Lahuis}, {Lee}, {Martinez-Delgado}, {Mart{\'\i}nez-Galarza}, {Meijers}, {Morrison}, {M{\"u}ller}, {Nakos}, {O'Sullivan}, {Oudenhuysen}, {Parr-Burman}, {Pauwels}, {Rohloff}, {Schmalzl}, {Sykes}, {Thelen}, {van Dishoeck}, {Vandenbussche}, {Venema}, {Visser}, {Waters}, \& {Wright}}]{2015PASP..127..646W}
{Wells}, M., {Pel}, J.~W., {Glasse}, A., {et~al.} 2015, \pasp, 127, 646, \dodoi{10.1086/682281}

\bibitem[{{Woitke} {et~al.}(2016){Woitke}, {Min}, {Pinte}, {Thi}, {Kamp}, {Rab}, {Anthonioz}, {Antonellini}, {Baldovin-Saavedra}, {Carmona}, {Dominik}, {Dionatos}, {Greaves}, {G{\"u}del}, {Ilee}, {Liebhart}, {M{\'e}nard}, {Rigon}, {Waters}, {Aresu}, {Meijerink}, \& {Spaans}}]{2016A&A...586A.103W}
{Woitke}, P., {Min}, M., {Pinte}, C., {et~al.} 2016, \aap, 586, A103, \dodoi{10.1051/0004-6361/201526538}

\bibitem[{{Wolff} {et~al.}(2016){Wolff}, {Perrin}, {Millar-Blanchaer}, {Nielsen}, {Wang}, {Cardwell}, {Chilcote}, {Dong}, {Draper}, {Duch{\^e}ne}, {Fitzgerald}, {Goodsell}, {Grady}, {Graham}, {Greenbaum}, {Hartung}, {Hibon}, {Hines}, {Hung}, {Kalas}, {Macintosh}, {Marchis}, {Marois}, {Pueyo}, {Rantakyr{\"o}}, {Schneider}, {Sivaramakrishnan}, \& {Wiktorowicz}}]{2016ApJ...818L..15W}
{Wolff}, S.~G., {Perrin}, M., {Millar-Blanchaer}, M.~A., {et~al.} 2016, \apjl, 818, L15, \dodoi{10.3847/2041-8205/818/1/L15}

\bibitem[{{Zeidler} {et~al.}(2013){Zeidler}, {Posch}, \& {Mutschke}}]{2013A&A...553A..81Z}
{Zeidler}, S., {Posch}, T., \& {Mutschke}, H. 2013, \aap, 553, A81, \dodoi{10.1051/0004-6361/201220459}

\end{thebibliography}
